\documentclass[12pt,letterpaper]{article}
\usepackage{amsmath}
\usepackage{graphicx}
\usepackage{setspace}   
\usepackage[colon, sort&compress]{natbib}
\usepackage{textcomp} 
\usepackage{multirow} 

\newcommand{\tx}[1]{\text{#1}}
\newcommand{\mb}[1]{\mathbf{#1}}
\newcommand{\nin}{\noindent}
\newcommand{\non}{\nonumber}

\usepackage{titlesec}
\titlespacing{\section}{0pt}{12pt}{0pt}
\titlespacing{\subsection}{0pt}{12pt}{0pt}
\titleformat{\section}
  {\large\bfseries}
  {\thesection}{1em}{}
\titleformat{\subsection}
  {\normalsize\bfseries}
  {\thesubsection}{1em}{}

\usepackage{caption}
\usepackage{fancyhdr}
\begin{document}

\title{Geostatistical Model Resolution Enhancement in the Context of Multiple-Point Statistics}
\author{Saina Lajevardi and Clayton V. Deutsch}
\date{}
\maketitle

\setcounter{page}{1}
\nin Current multiple-point based simulations implementations generate geostatistical models at the scale of the training image; there is an assumption that the categories are exclusive at smaller scales. The goal of this paper is to generate models with multiple-point statistics (MPS) at a higher resolution than that of the available training image.
This paper addresses model resolution enhancement by studying the scale-dependence of spatial structure in MPS based models---extrapolating the smaller scale MPS from the larger scale MPS, and (2) rescaling the training image directly to the smaller scale. The first approach investigates the MPS probabilities. A number of challenges in characterizing smaller scale variability using high-order statistics are documented. The paper concludes by advocating the direct rescaling of the training image to generate models at higher resolution.

\section{Introduction}
\nin MPS is a cell-based simulation technique that characterizes complex geological features with a training image. MPS was introduced a few decades ago by \citep{fg:93a} and has been successfully developed for practical purposes by \citep{ss:00}. In MPS, information regarding spatial heterogeneity is inferred from a training image that is a rasterized depiction of the properties of interest \citep{ab:09}. The training image is a conceptual model representing a particular phenomena \citep{jbb:07,sjl:09}. MPS simulation benefits from high order statistics captured from the training image as opposed to conventional variogram techniques \citep{gm:14}. Enforcement of complex geological patterns is the main reason that MPS has become appealing in modern geostatistics.

In MPS, the relationship between categories at different locations is provided through high-order statistics inferred from a training image that represents the type of structural characteristics that are expected in the geological setting of the deposit. The training image must be representative of the geology under study in terms of its complexity and scale. Numerical models at higher resolution are often required as the study progresses; precise assessment of connectivity for instance may warrant models at higher resolution \citep{dkl:06,dkl:07}.
A training image, however, is often available only at a fixed scale.

\begin{figure}[h!]\centering
\includegraphics[scale=1.2]{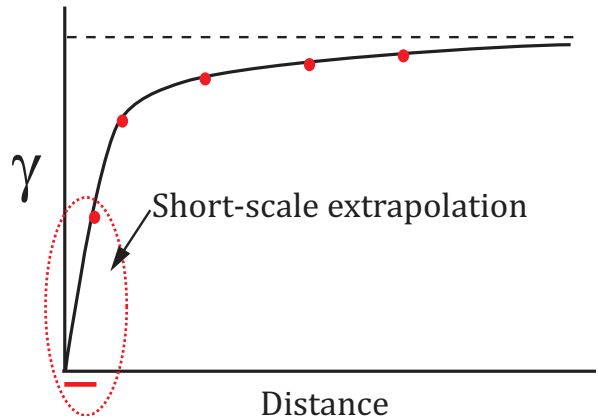}
\caption{\small A schematic of extrapolating the variogram to smaller scale, with the intention of modeling short scale variability. The red dots represent experimental variogram values at specific distances. The red horizontal line underneath the distance axis shows that the extrapolation of first experimental variogram value to zero value permits the variogram determination for distances smaller than the data spacing.}
\label{fig:extrapolation}
\end{figure}

Extrapolation of spatial statistics to different scales than that of the conditioning data has been considered in many conventional geostatistical technique. In terms of simulation, data reconciliation at different scales is often addressed through simulation at the point scale of the data then averaging to larger scales \citep{pf:99}. Variogram extrapolation from the smallest lag distance to the nugget effect permits specifying statistics at a higher resolution than the data spacing \citep{hd:98,pf:99} (see Figure~\ref{fig:extrapolation}). MPS techniques, however, do not have this flexibility as the minimum scale of information is the resolution of the training image. 

The most prominent work on the generation of high resolution models from the coarse resolution ones includes super-resolution land cover mapping \citep{ab:08}, remote sensing \citep{ab:09}, and global climate modeling (GCM). The coarse resolution regional model is not capable of capturing variability in local areas. Downscaling methods are developed to capture small scale variability through integration of coarse resolution models and several local data sources that are not considered in simulating the global models.
Empirical techniques are then utilized to extrapolate spatial continuity to the higher resolution. Reference to large or small scale in GCM applications mostly implies the model covering global or local regions. For example, \citep{ab:09} utilizes training image high-order statistics to generate super-resolution maps from coarse-scale satellite sensor information. GCM provides models for regional domains with coarse grid cells of hundreds of kilometers whereas the downscaled models are implemented to provide for local weather forecasting. 

In earlier geostatistics literature, fractals were sometimes considered to characterize reservoir heterogeneity; fractals inherently deal with different scales of simulation \citep{tah:93}. This is, however, different from the problem that is going to be addressed in this work. The problem is to generate high resolution models from coarse resolution ones with no access to small scale data. Constructing a high resolution model based on the scale being depicted in the training image (coarse scale) would not be necessarily accurate unless a strong self-similarity assumption is considered \citep{gm:11}. Thus, the resolution enhancement in this work addresses model regridding and MPS downscaling.

A comprehensive literature review regarding downscaling  in different fields is provided by \citep{wh:12} that categorize downscaling methods into (1) statistical downscaling, and (2) other methods.
However, these approaches require having access to training images at finer scale. The training images at finer scales are utilized as the reference for the downscaling process and interpolating from coarse scale models to fine scale (high resolution) models. The large scale data are basically considered as the conditioning data to the resultant high resolution maps (models) \citep{ab:08}. Geostatistical interpolation/simulation techniques could effectively be employed to generate plausible realizations of high resolution maps \citep{ab:08}.

The remainder of this paper is organized as follows: the next section defines the frequency of occurrence of patterns (FOP) and its utilization to understand high-order statistics in terms of lag distance. The next section investigates the challenges of inferring small scale information from the proposed FOP plots. The last section proposes the direct manipulation of the training image as a practical alternative.

\section{Frequency of Patterns}
\begin{figure}[h!]\centering
\includegraphics[scale=0.9]{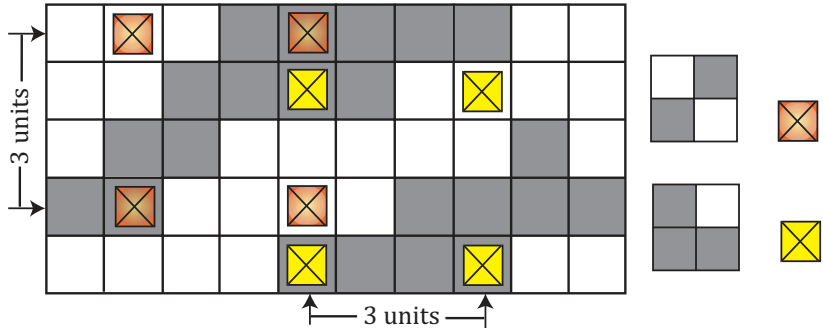}
\caption{\small Example on pattern extraction in training image in terms of lag distance. The two patterns on the right are extracted on the map with 3 units distance lag shown in yellow and orange.}
\label{fig:example}
\end{figure}

\nin High-order statistics in MPS simulations could be summarized by occurrences of patterns of $n$-point (limited to four, six and nine in this paper) template configurations. The number of possible patterns for a particular template configuration is $K^N$, where $K$ is the number of categories and $N$ is the number of locations. For example, five categories ($K$=5), and six locations ($N$ = 6) will result in 15625 (= $5^6$) possible configurations. The possibility that all these combinations are found in a training image is low. 
Similar to the variogram that determines the variability of two-point data statistics as a function of lag distance, the increase or decrease of the frequency of occurrence of high-order patterns is considered to characterize the variability of the training image in terms of lag distance.
A concept is developed to predict the frequency of patterns (FOP) at smaller scales by extrapolating an FOP plot to shorter lag distances as done for the variogram. FOP is the proportion of occurrences of a pattern at a specific lag distance relative to the total number of patterns. For every specific lag distance, the grid cell data is extracted and recorded in the form of template configurations. For instance, if a lag distance of 3 units is of interest, the grid cells are scanned for distances of 3 (see example depicted in Figure~\ref{fig:example}) in the form of 2x2 template configuration.

The variations in FOP depend on the training image, the continuity, and the scale of the structures. Some patterns are expected to be more common at a particular scale, while some others would appear infrequently.
If one pattern occurs more, the remaining patterns of the same configuration must occur less. Patterns that appear more frequently at a unit lag distance represent the structuredness of the training image. With the increase of lag distance, patterns become more random.

\subsection{Quantification of Structuredness/Randomness}
\nin The frequency or proportion of patterns is the ratio of the number of a specific pattern to the total number of patterns of the same configuration. This ratio depends on the global proportion of the categories and the continuity and spatial structure of the model \citep{sl:15a}.
For instance, if there is more black than white, the frequency of patterns that contain more black will be larger than those that have more white.

\begin{figure}[h!]\centering
\includegraphics[scale=0.45]{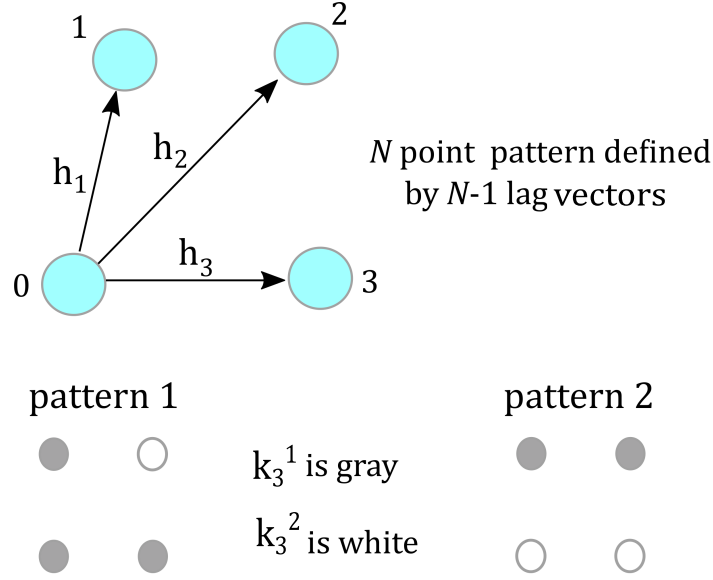}
\caption{\small Example of 4 point configuration by three lag vectors of $h_1$, $h_2$, and $h_3$. Every location in the pattern is indicated by ${k_j}^{(i)}$, where k denotes the category, and $j$ refers to the location in pattern $(i)$.}
\label{fig:conf}
\end{figure}

Standardized frequency of pattern or SFOP is defined to summarize the occurrence of a pattern independent of the univariate proportions of categories. The FOP is scaled by the product of the global proportions for all the locations in the pattern. 

\begin{align}
&\tx{FOP}^{(i)} = \frac{\tx{occurrence of pattern $(i)$}}{\tx{total \# of patterns}}, \;\;\;\; i = 1, \ldots , K^N \non\\
&\tx{FOP}^{(i)}_{\tx{rand}} = \prod_{j=1}^{N} \tx{P}_{{k_{j}}^{(i)}}, \non \\
&\tx{SFOP}^{(i)}  = \tx{FOP}^{(i)} / \prod_{j=1}^{N} \tx{P}_{{k_{j}}^{(i)}},  \;\;\;\;\;\;\; {k_{j}}^{(i)} \in [1, \ldots , K]  \non\\
&\tx{SFOP}^{(i)} = \tx{FOP}^{(i)} / {\tx{FOP}^{(i)}_{\tx{rand}}}.
\label{eq:max}
\end{align}

\nin where $\tx{P}_{{k_{j}}^{(i)}}$ indicates the global proportion of category $k$ of location $j$ in pattern $i$, see Figure~\ref{fig:conf}. 
The product of this proportion, $\prod_{j=1}^{N} \tx{P}_{{k_{j}}^{(i)}}$, for all $N$ locations in pattern $(i)$ represents the frequency of occurrence of pattern $(i)$ if the corresponding map was random.
In case of a noisy map, the occurrence of all patterns are equally likely subject to the proportions of each category. Thus, this product will converge to $1/K^N$ for the random map of equal global proportions of categories. The lower bound of $\tx{SFOP}_{(i)}$ is 0, which occurs when the occurrence of pattern $(i)$ is zero, and its upper bound is $1/\prod_{j=1}^{N} \tx{P}_{{k_{j}}^{(i)}}$ when the pattern occurs with a frequency of 1, which is unlikely in practice. 

%

An example is shown to demonstrate how the global proportions of categories influences the probability of occurrence of specific patterns. The probability of occurrence is evaluated for 16 patterns of binary 2$\times$2 configuration for a noisy map consisting of 60\% black and 40\% white ($\tx{FOP}_\tx{rand}$). The 16 patterns are grouped into 5 categories based on the fraction of black locations. 
The patterns with 3 blacks and 1 white have the FOP$_\tx{rand}$ value of ($0.6^3 \times 0.4^1 = 0.0864$) which is greater than the FOP$_\tx{rand}$ of the same patterns if the global proportions for black and white were 50/50 (=0.0625). It can also be evaluated that the chances for patterns containing 1 black and three white (=0.0384) is less than half the probability of occurrence for patterns with 3 black and 1 white, although the global proportion of black is only 1.5 times that of white. 

\subsection{Odds Ratios for FOP Evaluations}
\nin Odds ratios are another representation of association. The interpretation of probabilities of events relative to other events in the form of odds ratios is quite popular in statistics and medical literature. Odds ratios are mainly used when the study is designed in the form of case-control. To analyze the frequency of patterns independent of the global proportions of category, the model with random occurrence of patterns with same proportions could be selected as the control case \citep{dag:08}. The odds of occurrence of a specific pattern relative to its associated control model could be used to discuss structuredness in MPS models. Therefore, the Odds-FOP is the ratio describing the odds that pattern $n$ occurs relative to the rest of the patterns in the template, which is the probability of the occurrence of pattern $n$ to the probability of occurrence of all other patterns. For example, for a total number of templates in a map ($M$), and $m$ number of occurrence for pattern $n$, the odds of pattern $n$ is evaluated as the proportion of $m/(M-m).$ In other words, \textit{Odds-FOP} explains the probability of pattern $n$ occurs to the probability that it does not occur \citep{jmb:00}.
Also, the odds of occurrence of every pattern in a model of study relative to its occurrence in its corresponding random model (baseline model) could be expressed in the odds ratio:

\begin{align}
& \tx{odds\_FOP}^{(i)} = \tx{FOP}^{(i)}/(1-\tx{FOP}^{(i)}),\non \\
& \tx{odds\_FOP}^{(i)}_\tx{rand}= \tx{FOP}^{(i)}_{\tx{rand}} / (1-\tx{FOP}^{(i)}_{\tx{rand}}) \non\\
& \tx{odds\_ratio}^{(i)} = \tx{odds\_FOP}^{(i)} /\tx{odds\_FOP}^{(i)}_\tx{rand}
\label{eq:ratio}
\end{align}

\nin This ratio has a lower bound of 0 indicating no occurrence of a specific pattern and has no upper bound. For 100\% FOP, the odds ratio is undefined since SFOP reaches FOP$^{-1}_\tx{rand}$. Also, the odds ratio is 1 when FOP is equivalent to $\tx{FOP}_{\tx{rand}}$. Thus, a value of 1 for the odds ratio is for the independent case.

\begin{figure}[t!]\centering
\includegraphics[scale=0.42]{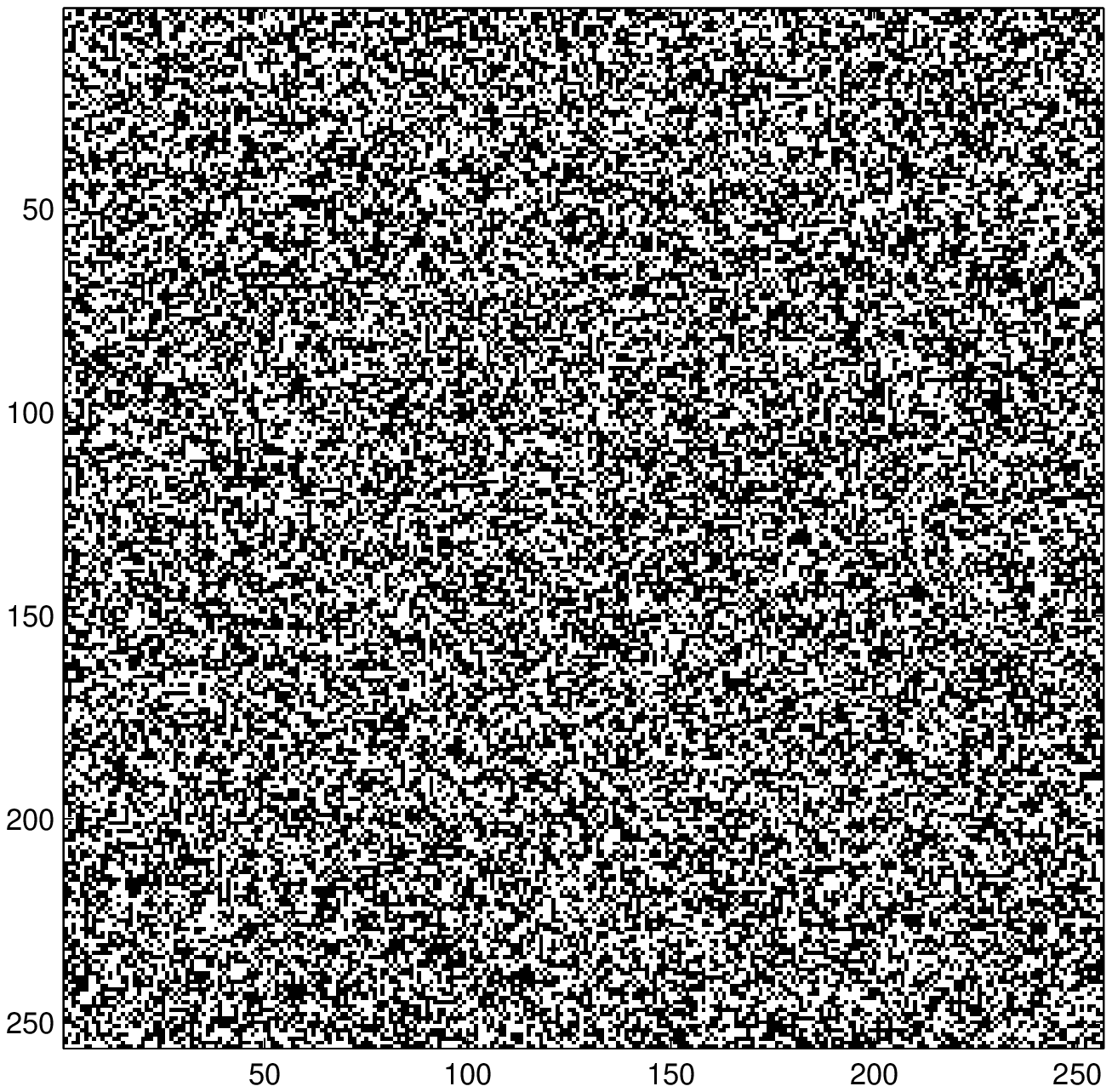}
{\includegraphics[scale=0.42]{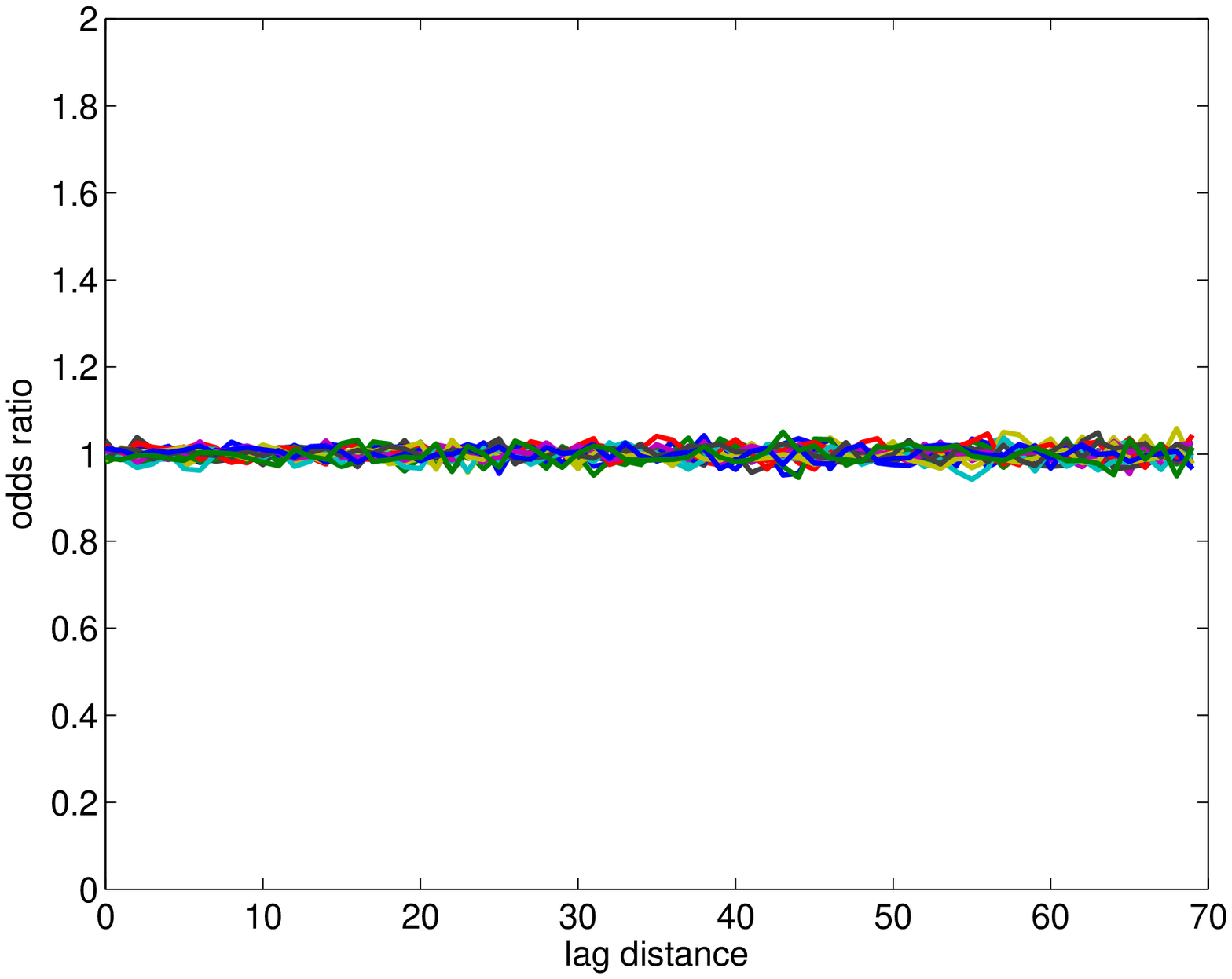}}\\
{\includegraphics[scale=0.36]{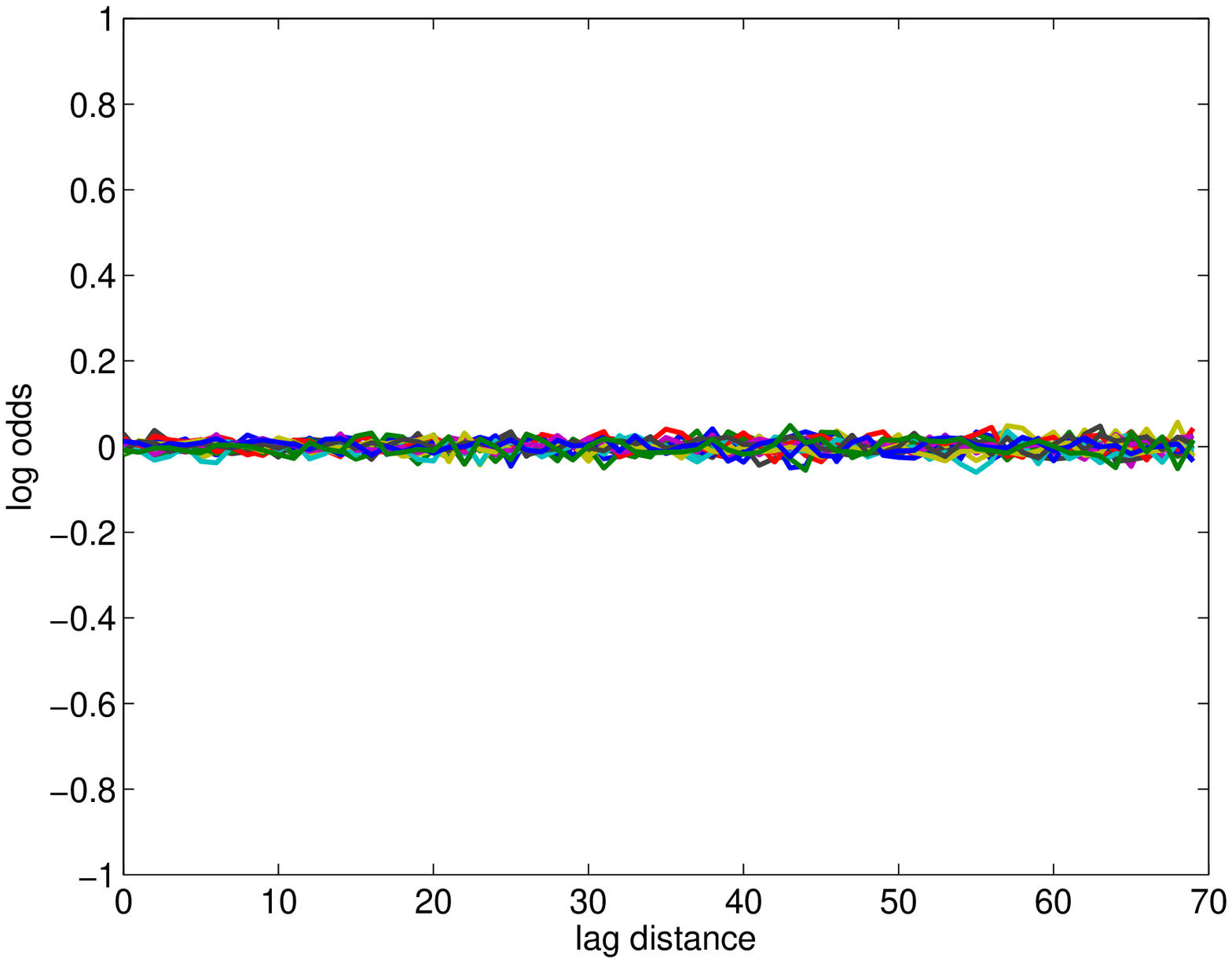}
\includegraphics[scale=0.36]{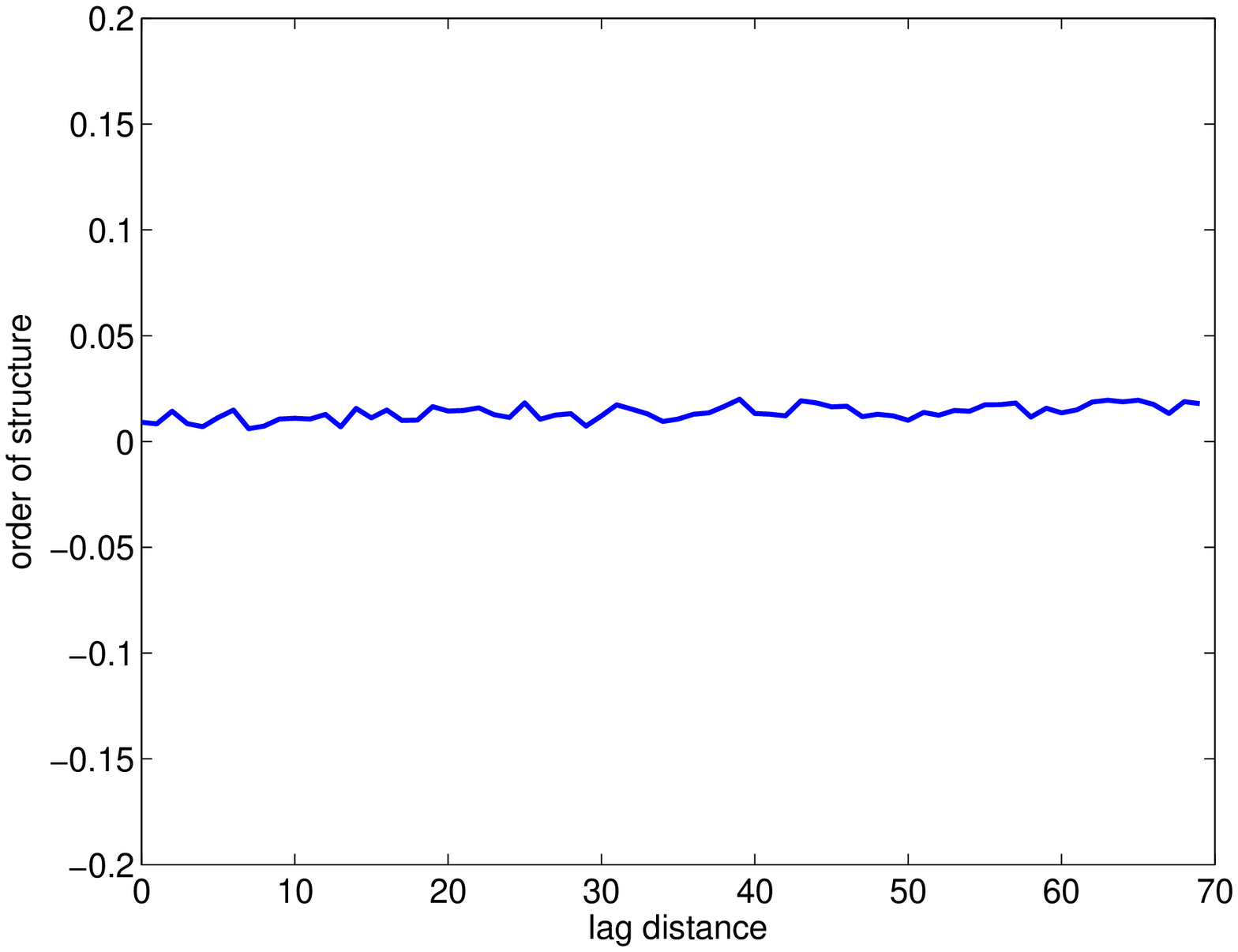}}\\
\caption[A binary training image is shown on the top-left representing a random map with equal proportions of black and white. This leads to equal FOP of 1/16 for all 2$\times$2 patterns.]{\small A binary training image is shown on the top-left representing a random map with equal proportions of black and white. This leads to equal FOP of 1/16 for all 2$\times$2 patterns. The plots for odds ratio (=1) and log odds (=0) are also presented. The order of structure is very small value at all lag distances as shown in bottom right figure.}
\label{fig:rand}
\end{figure}

In two-point statistics, the variogram sill is the data variance. A similar concept could be applied to multiple-point statistics. As discussed above, 1 is the reference value for the random case. 
Thus, in the context of occurrence of patterns, the odds ratio of every pattern converges to a value of 1 at a large lag distance, since the model commonly loses its structure when the lag distances increase (see Figure~\ref{fig:rand}). Therefore, it is reasonable to interpret any deviation from one to be an indication of continuity or structuredness as opposed to randomness.
A summary measure of structure could be calculated by the sum of absolute values of all deviations from the reference value of 1 at every lag distance \[\tx{Order of Structure} = \frac{\sum_{i=1}^{K^N} \mid \tx{odds\_ratio}^{(i)} - 1 \mid}{K^N}.\]
%

\begin{figure}[t!]\centering
\includegraphics[scale=0.38]{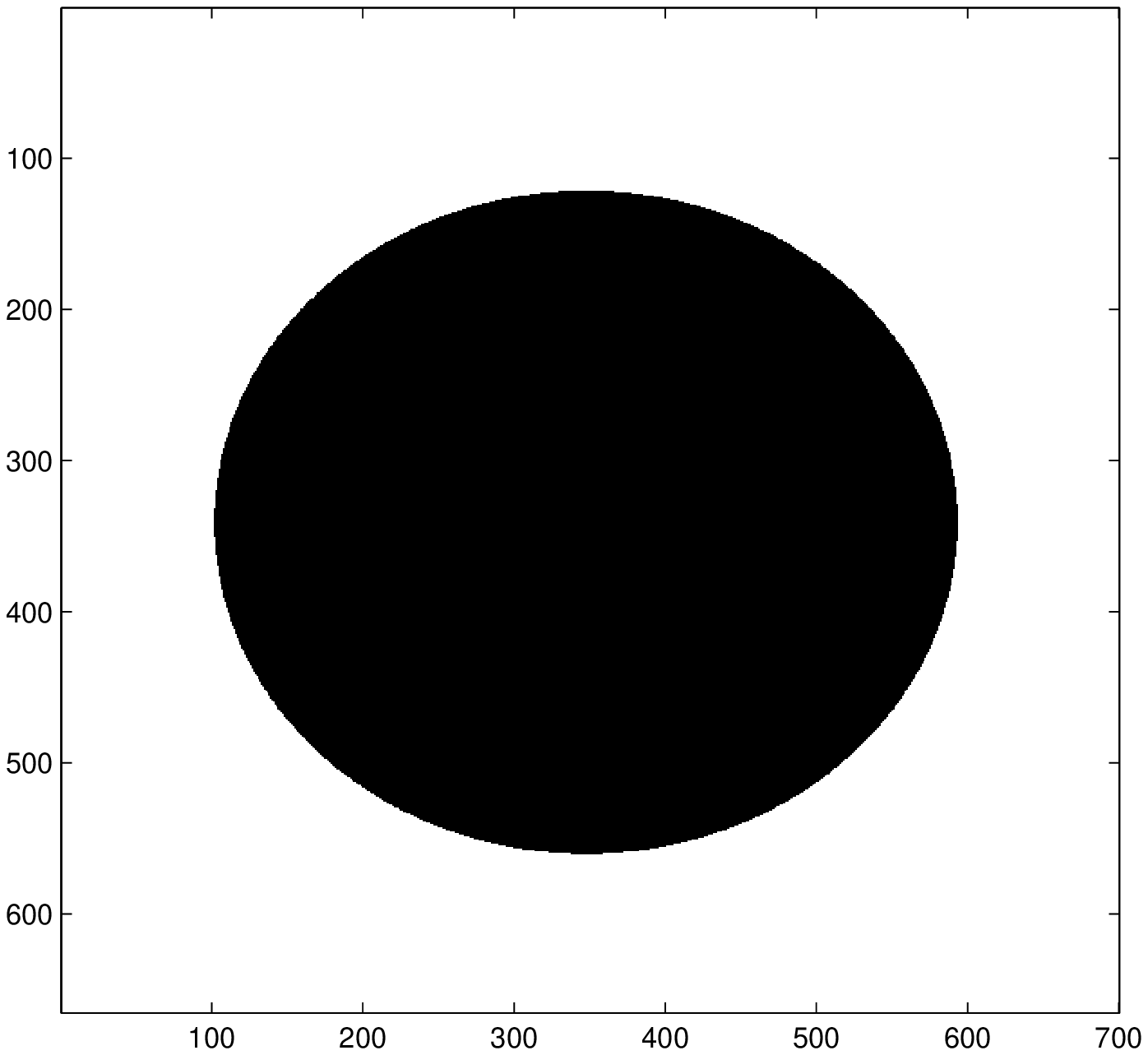}
{\includegraphics[scale=0.38]{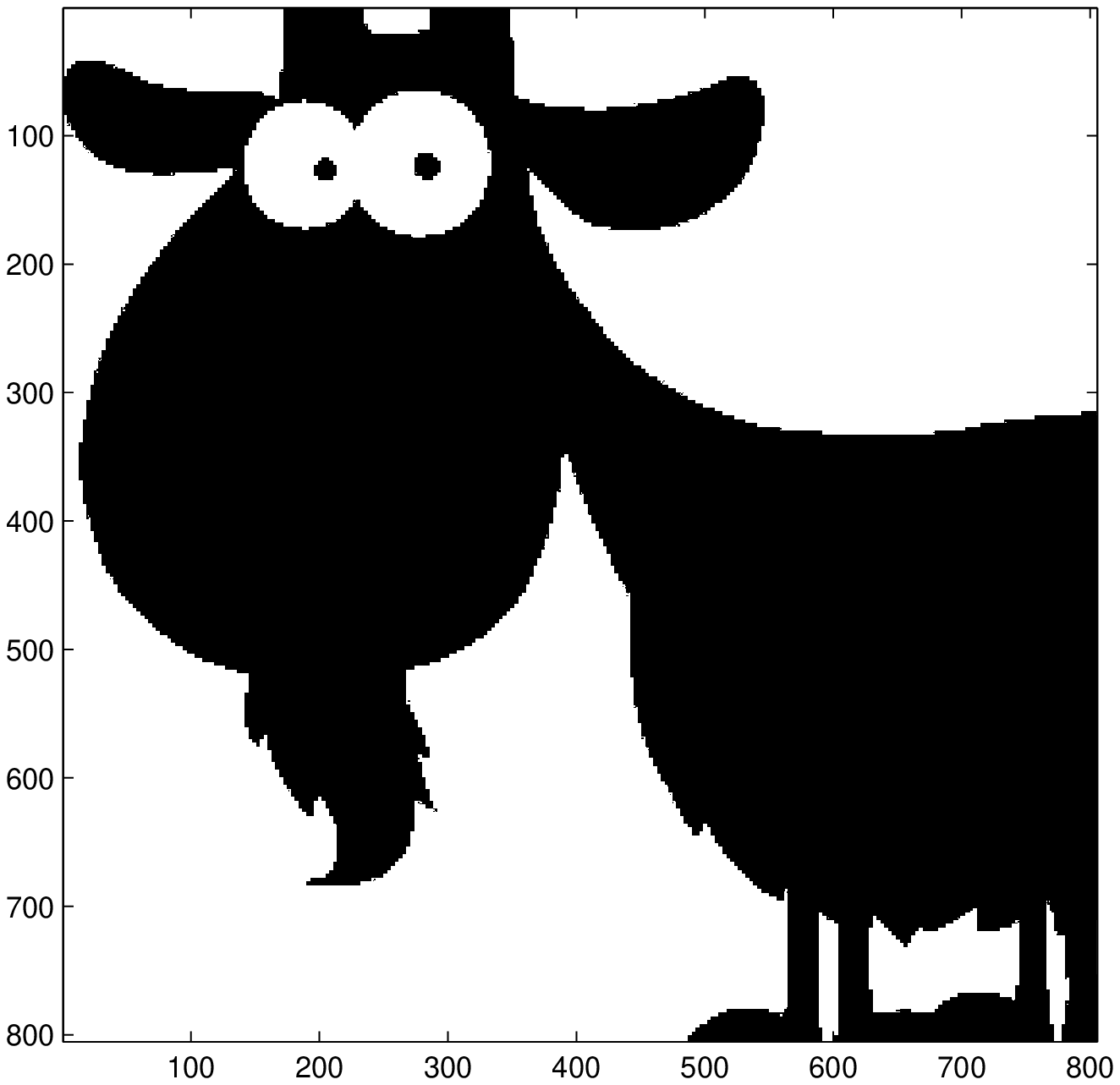}}\\
{\includegraphics[scale=0.38]{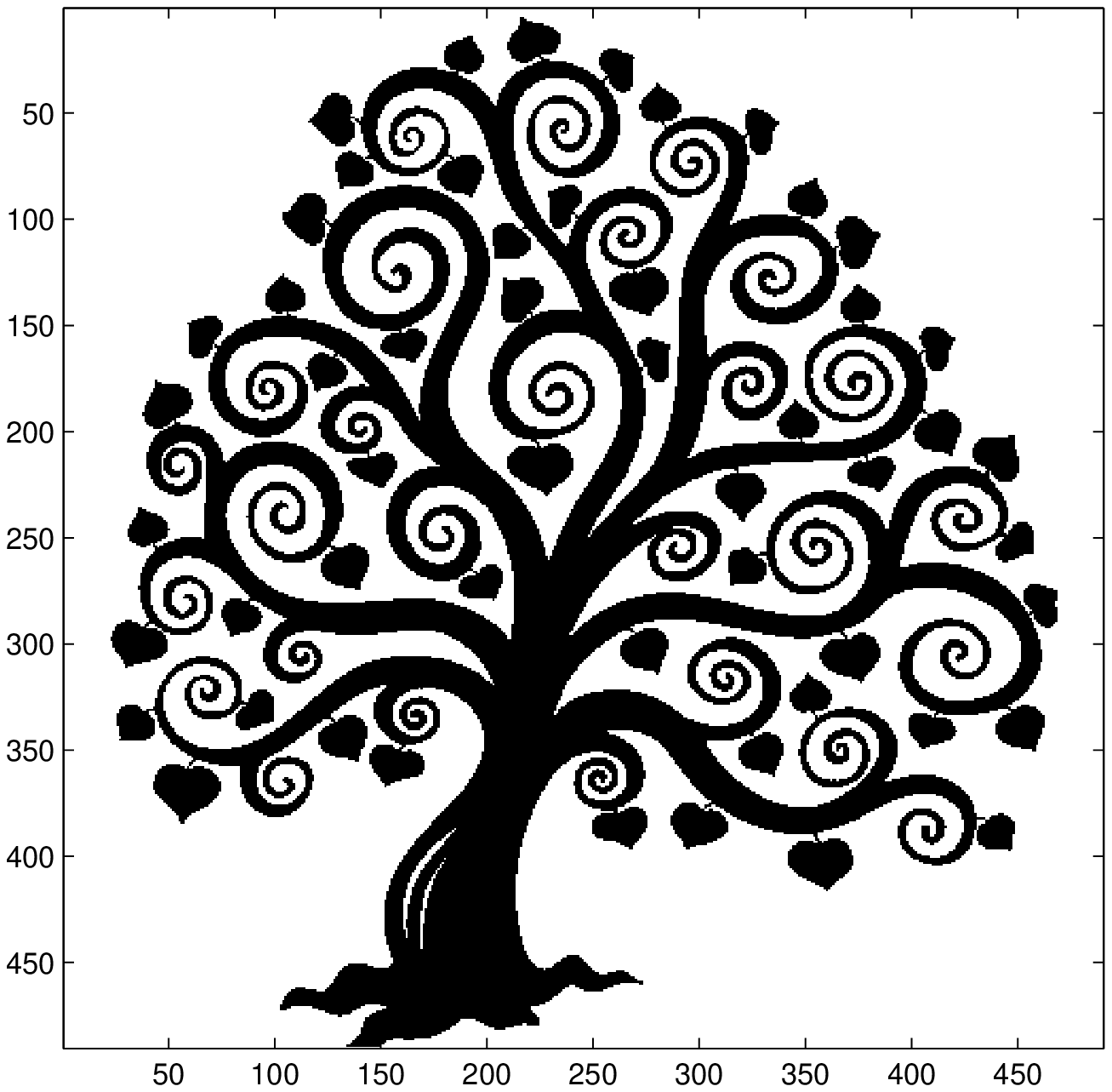}
\includegraphics[scale=0.38]{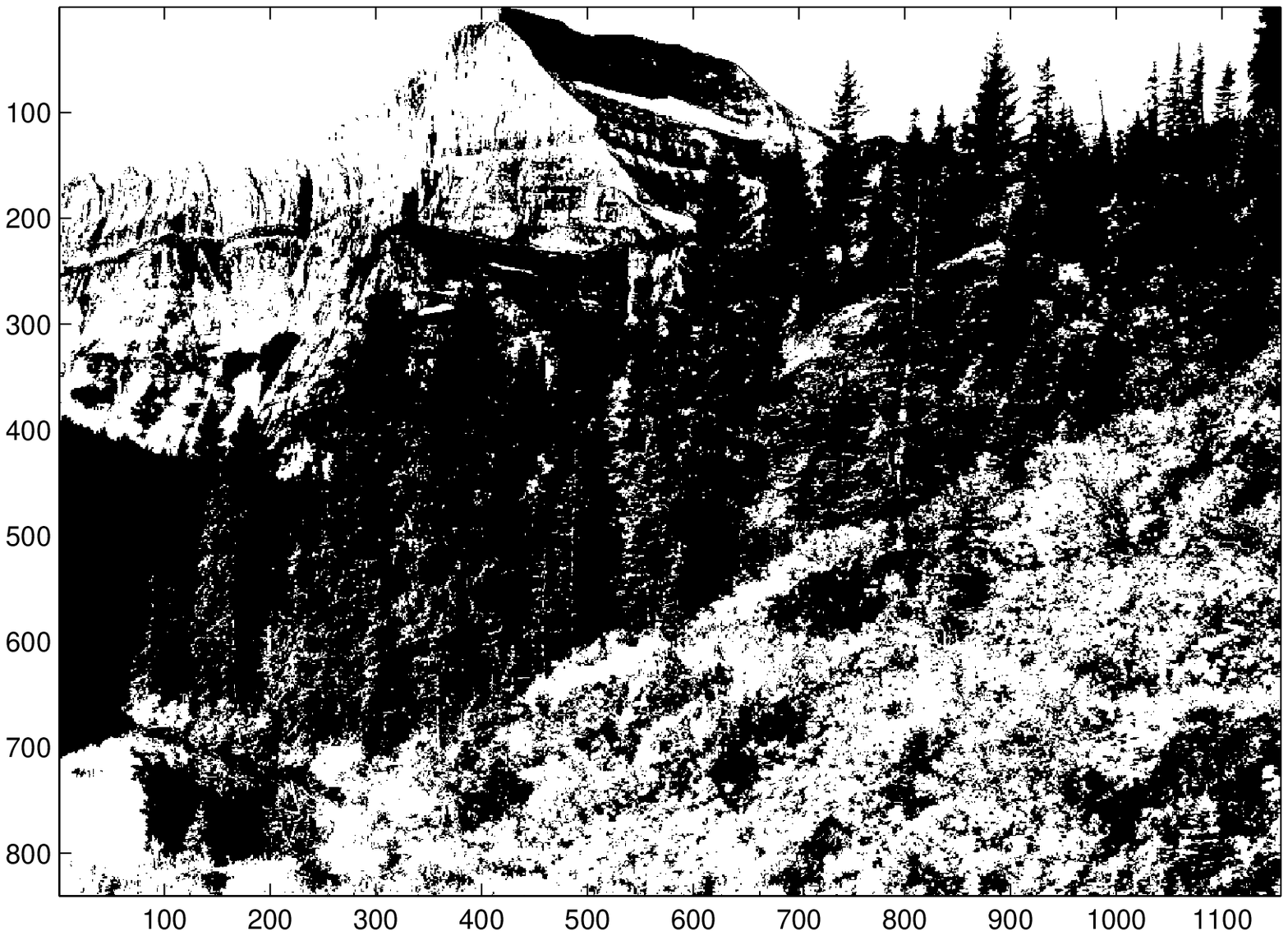}}\\
{\includegraphics[scale=0.38]{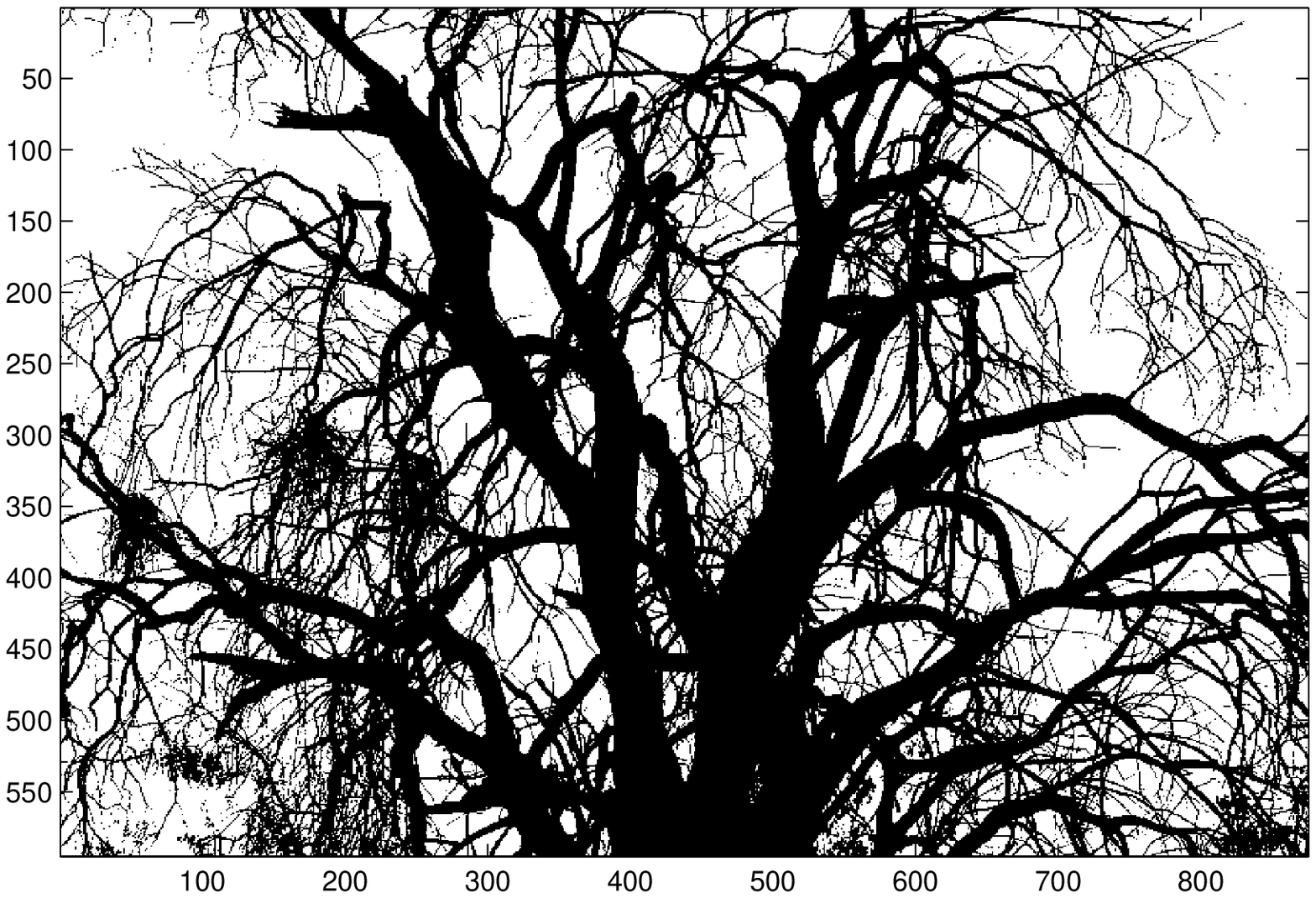}
\includegraphics[scale=0.38]{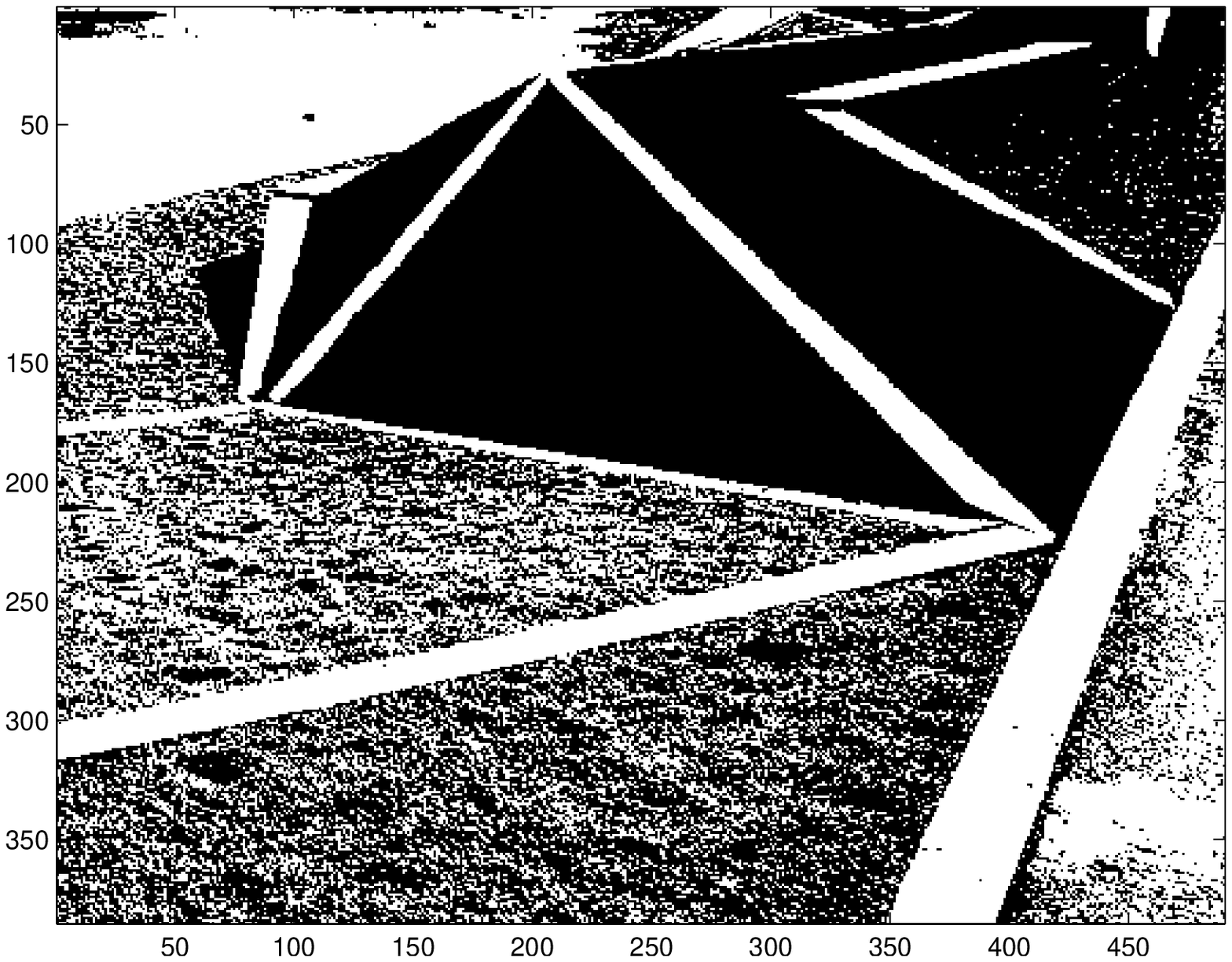}}\\
\caption[The images are listed from left to right and top to bottom from 1 to 6. The results and analysis on every training image have listed in the same order in the next following figures]{\small The images are listed from left to right and up to bottom from 1 to 6 \citep{image2,image3,image4,image5,image6}. The results and analysis on every training image have listed in the same order in the following figures.}
\label{fig:TI}
\end{figure}

\begin{figure}[t!]\centering
{\includegraphics[scale=0.4]{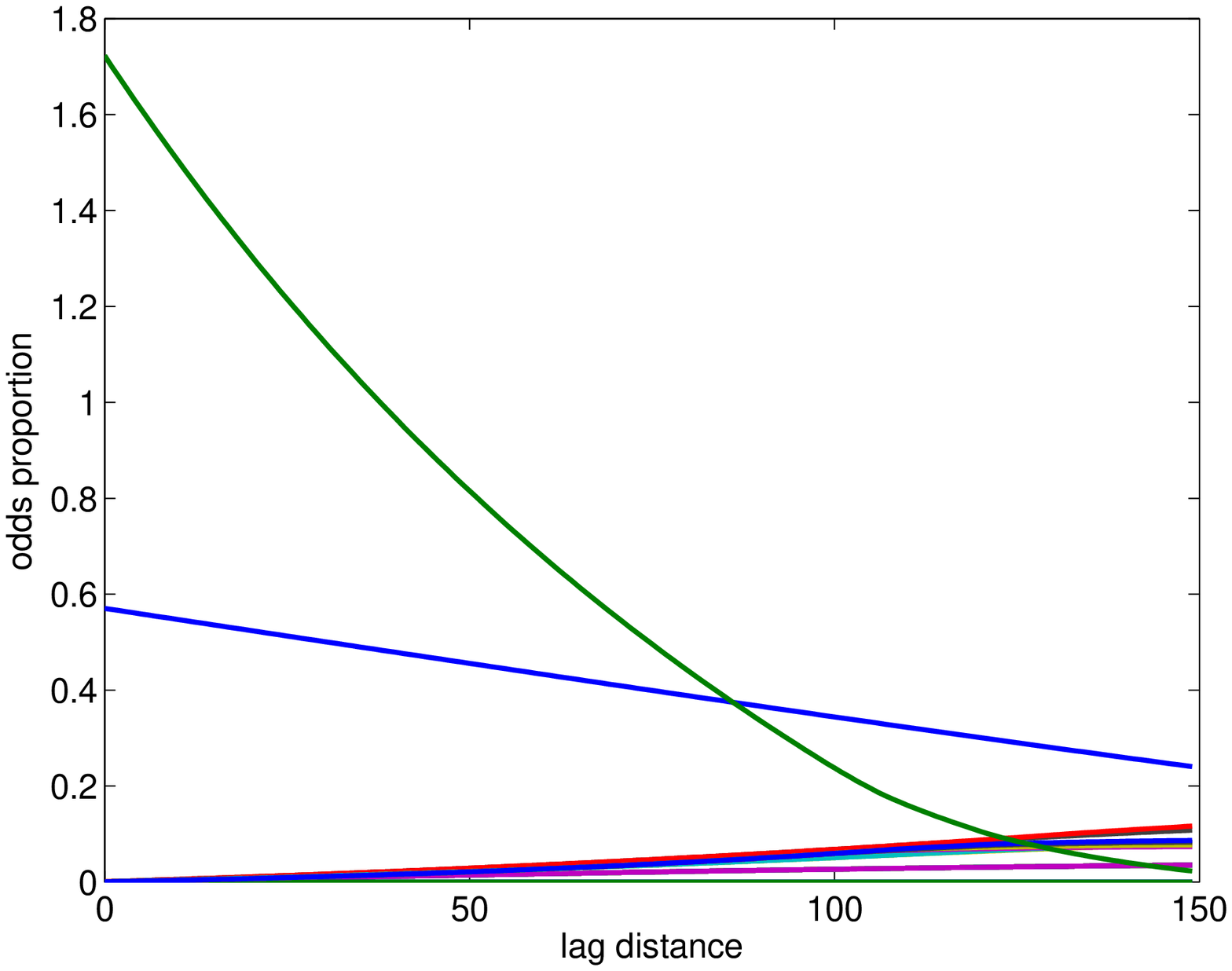}
\includegraphics[scale=0.4]{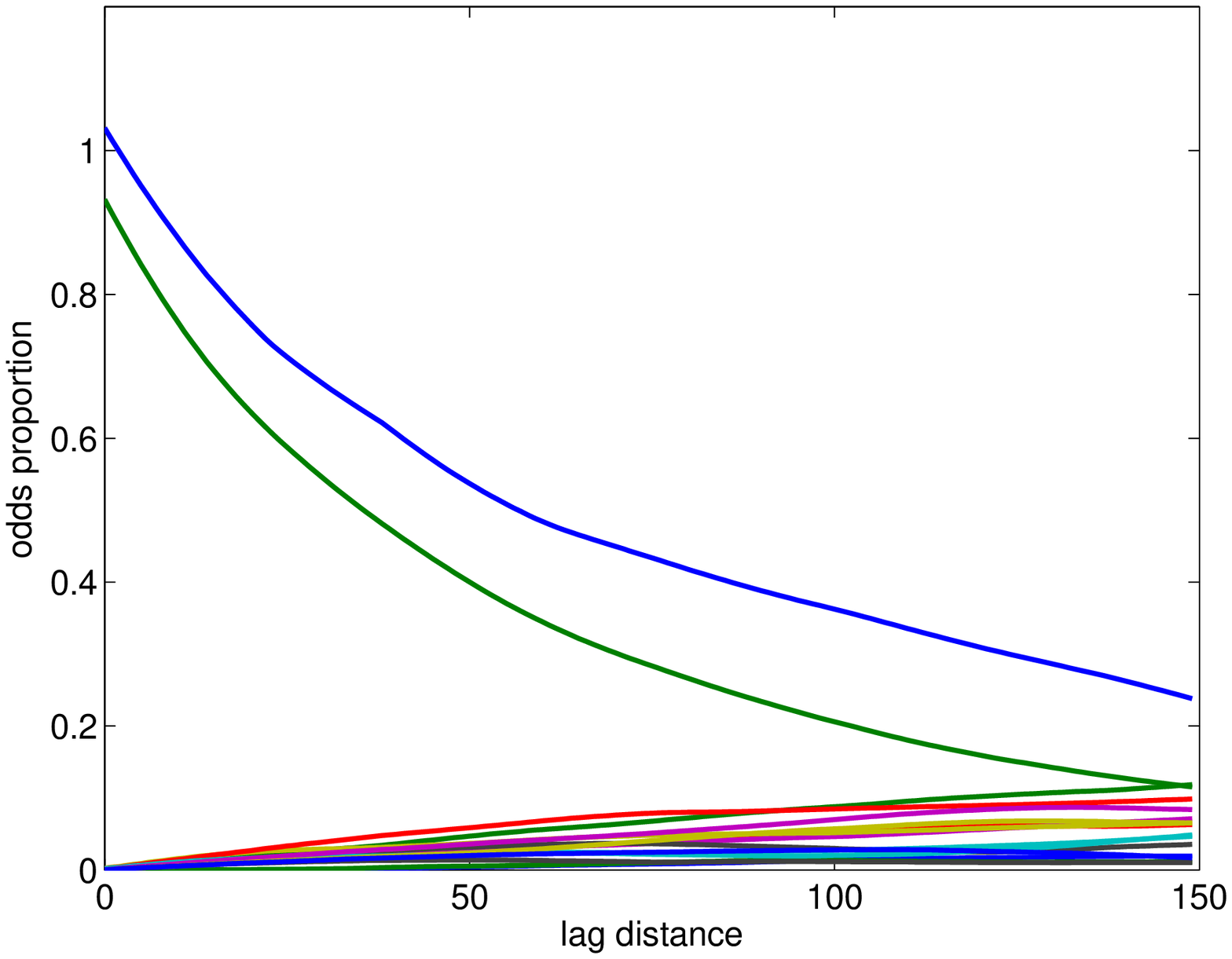}}\\
{\includegraphics[scale=0.4]{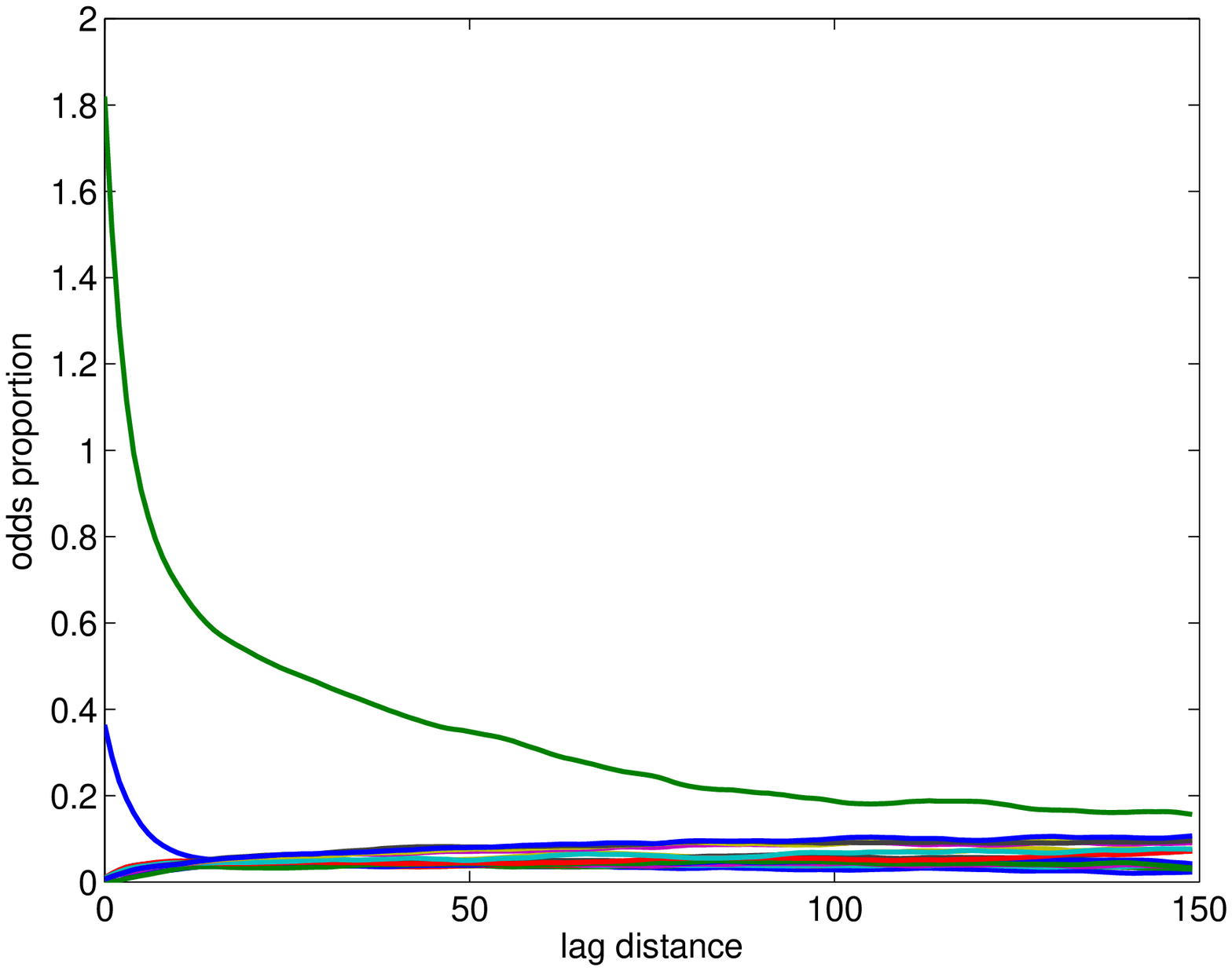}
\includegraphics[scale=0.4]{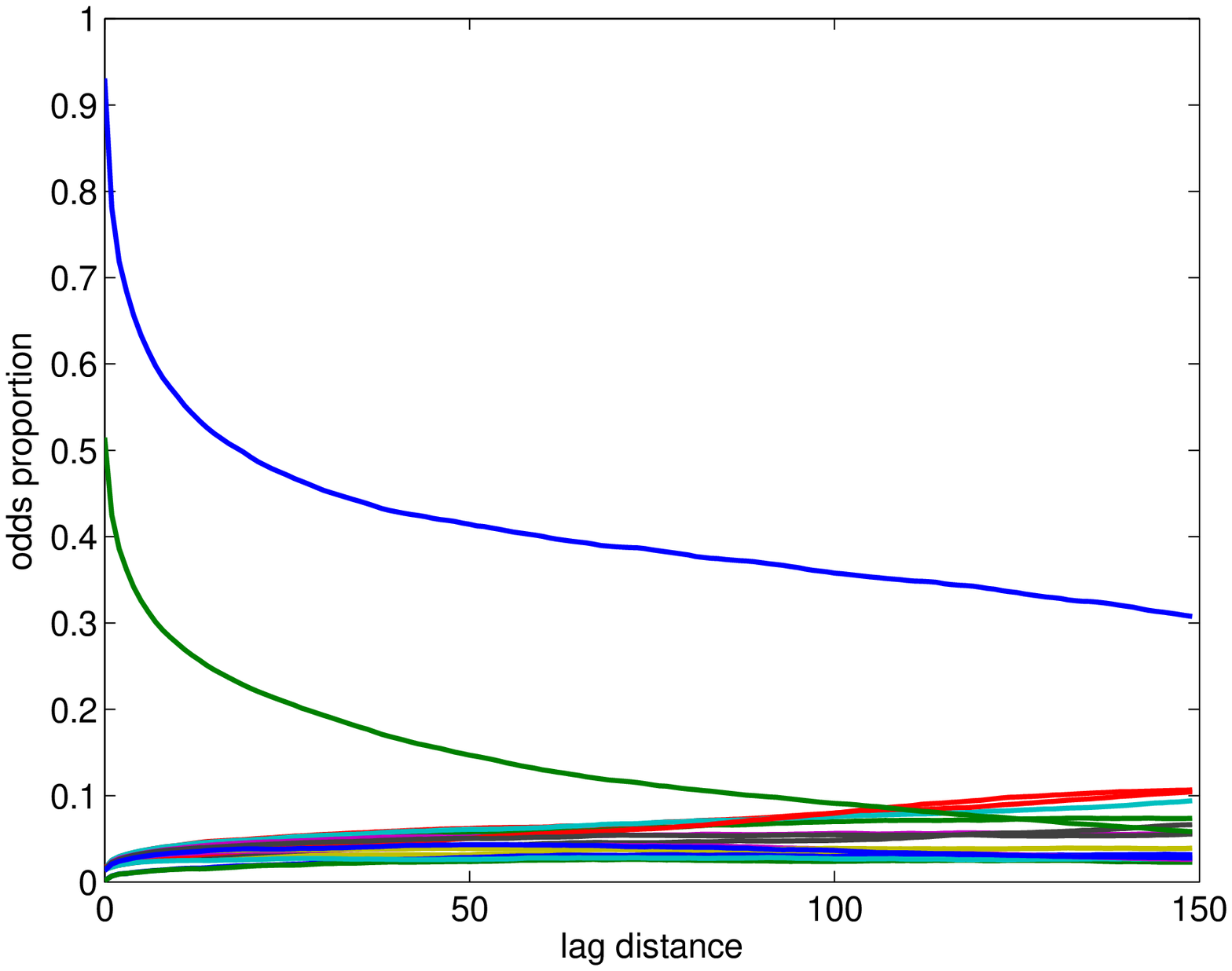}}\\
{\includegraphics[scale=0.4]{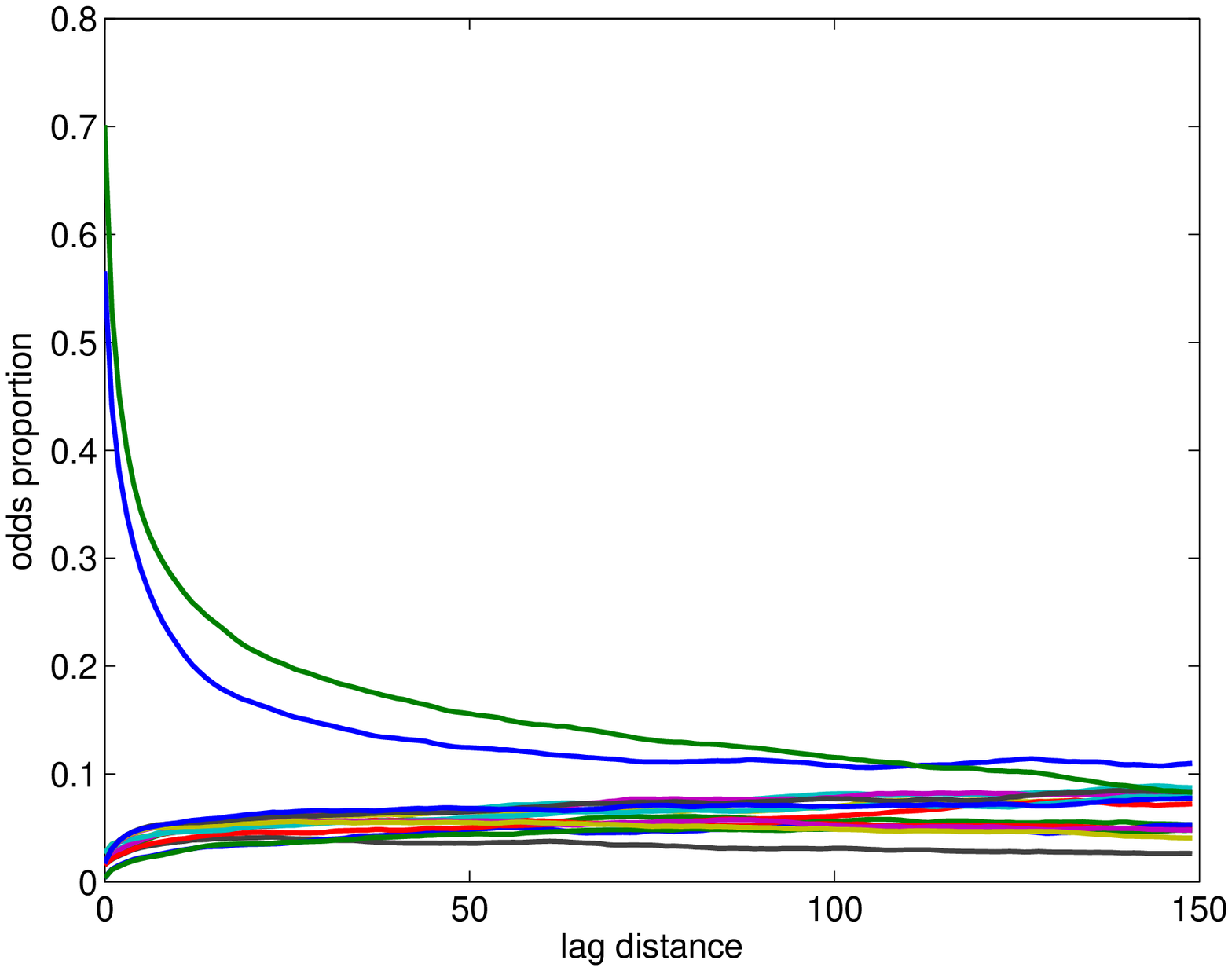}
\includegraphics[scale=0.4]{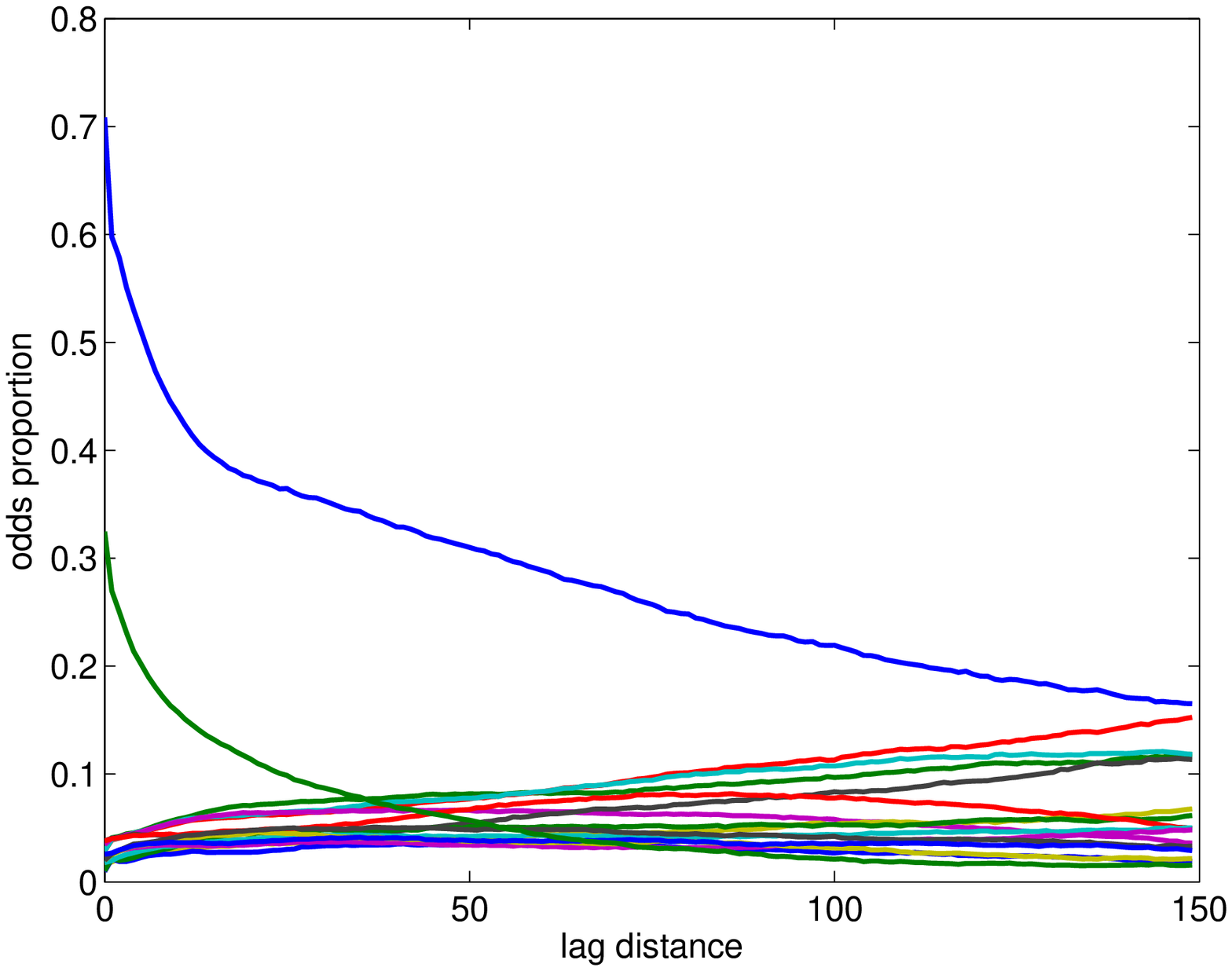}}\\
\caption[The occurrence of all 16 patterns of 2$\times$2 configuration are plotted as a function of lag distance. The dominant patterns of 1 and 16 at zero lag distances are shown in blue and violet]{\small The occurrence of all 16 patterns of 2$\times$2 configuration are plotted as a function of lag distance. The dominant patterns of 1 and 16 at zero lag distances are shown in blue and violet; these two patterns are the most structured as they are at their maximum at lag 0. The other three lines represent group patterns of (1) one white, three black (2) two white, two black (3) one black, three whites. These 14 patterns represent non-structured (noise) in the map. The proportion increase with the lag distance as the continuity of the structure discontinues with the continuity of spatial structure.}
\label{fig:proportion}
\end{figure}

\begin{figure}[t!]\centering
{\includegraphics[scale=0.37]{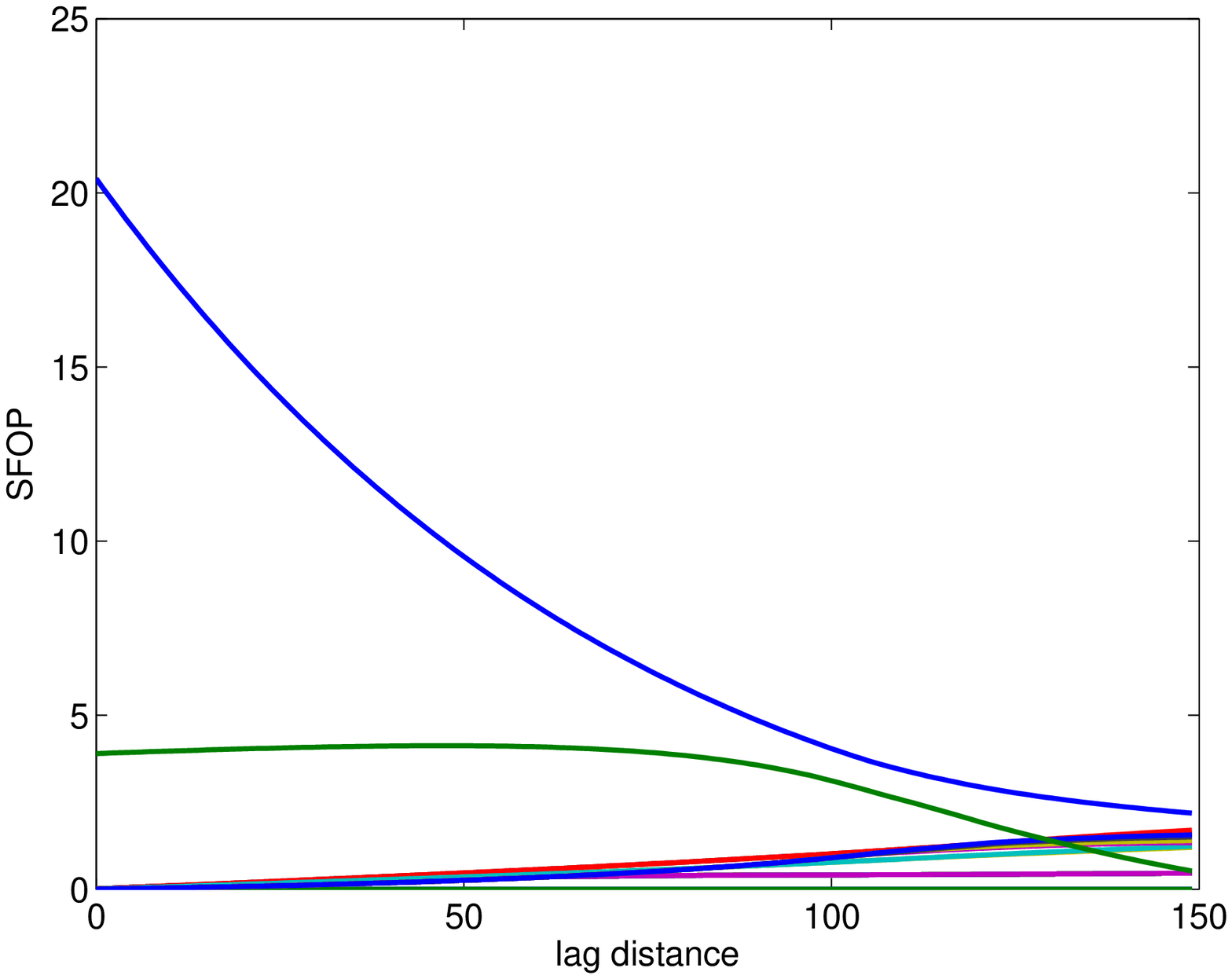}
\includegraphics[scale=0.37]{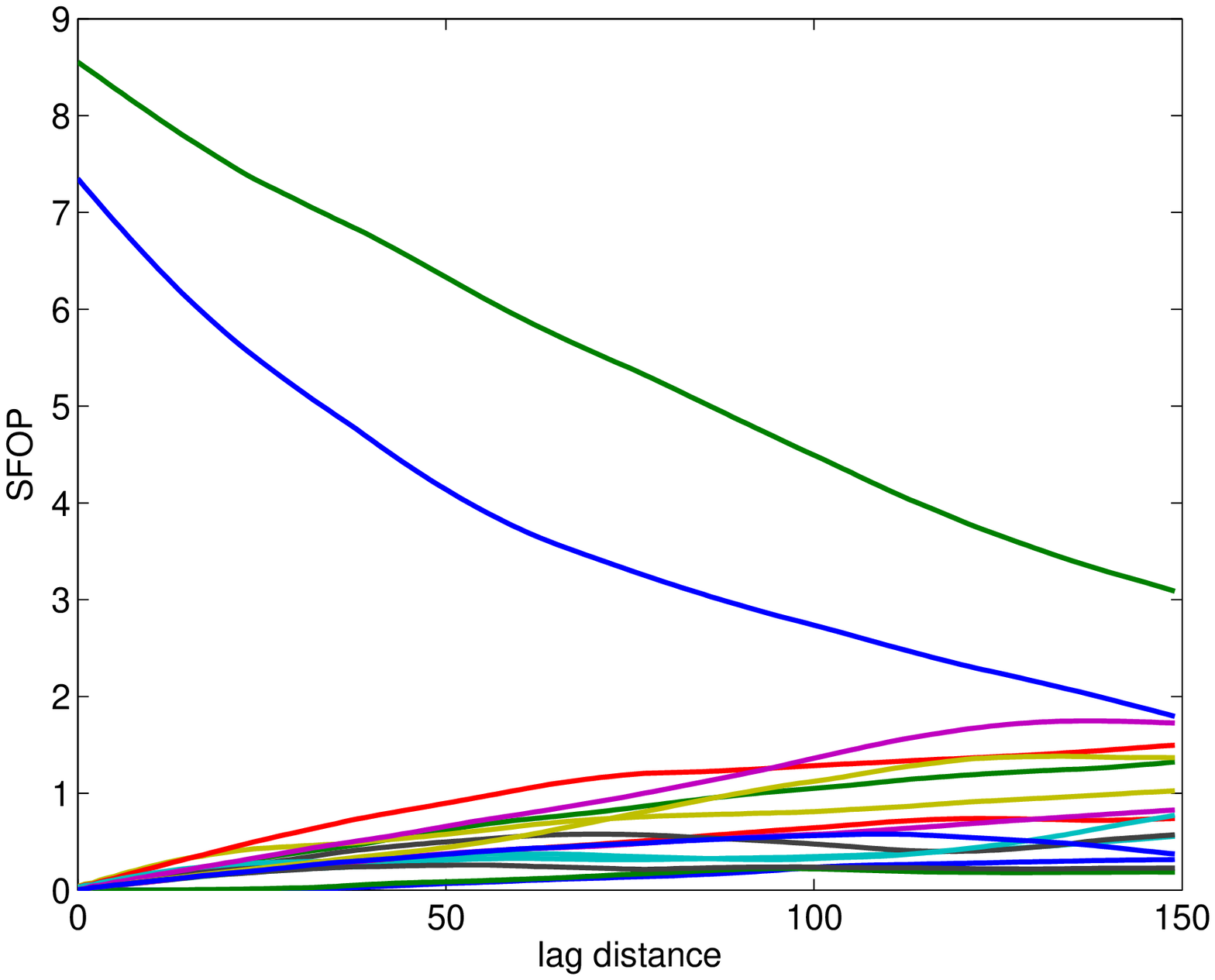}}\\
{\includegraphics[scale=0.37]{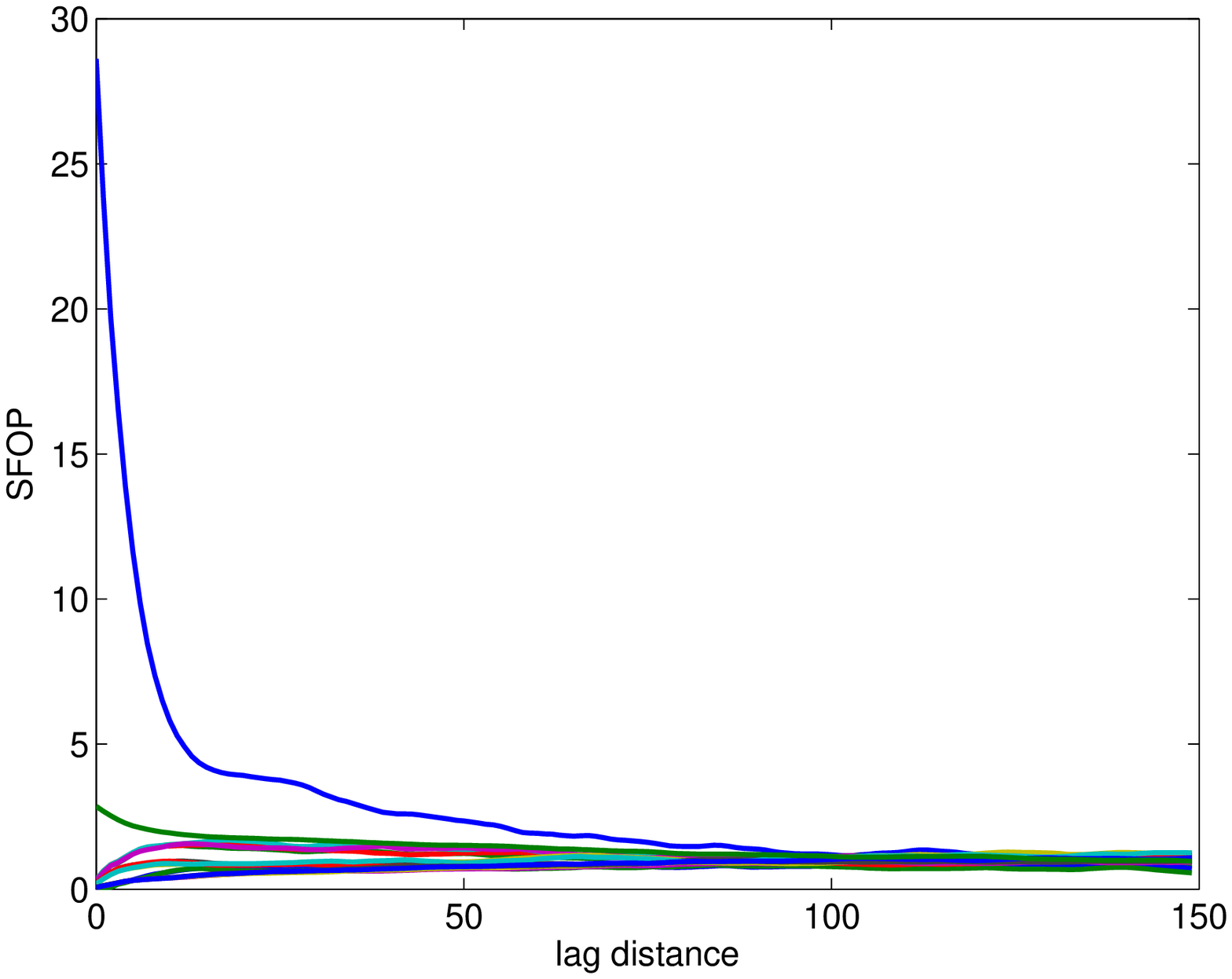}
\includegraphics[scale=0.37]{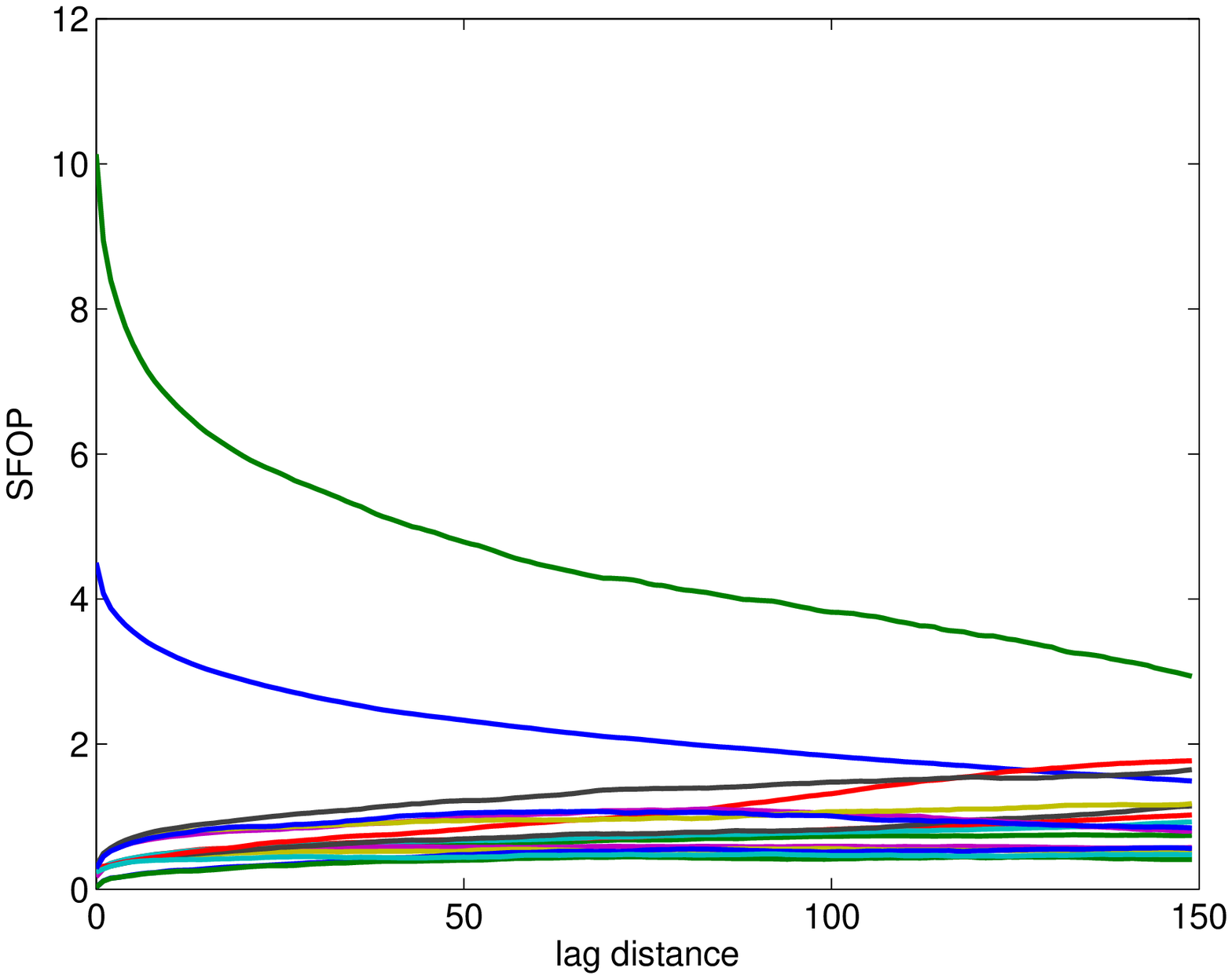}}\\
{\includegraphics[scale=0.37]{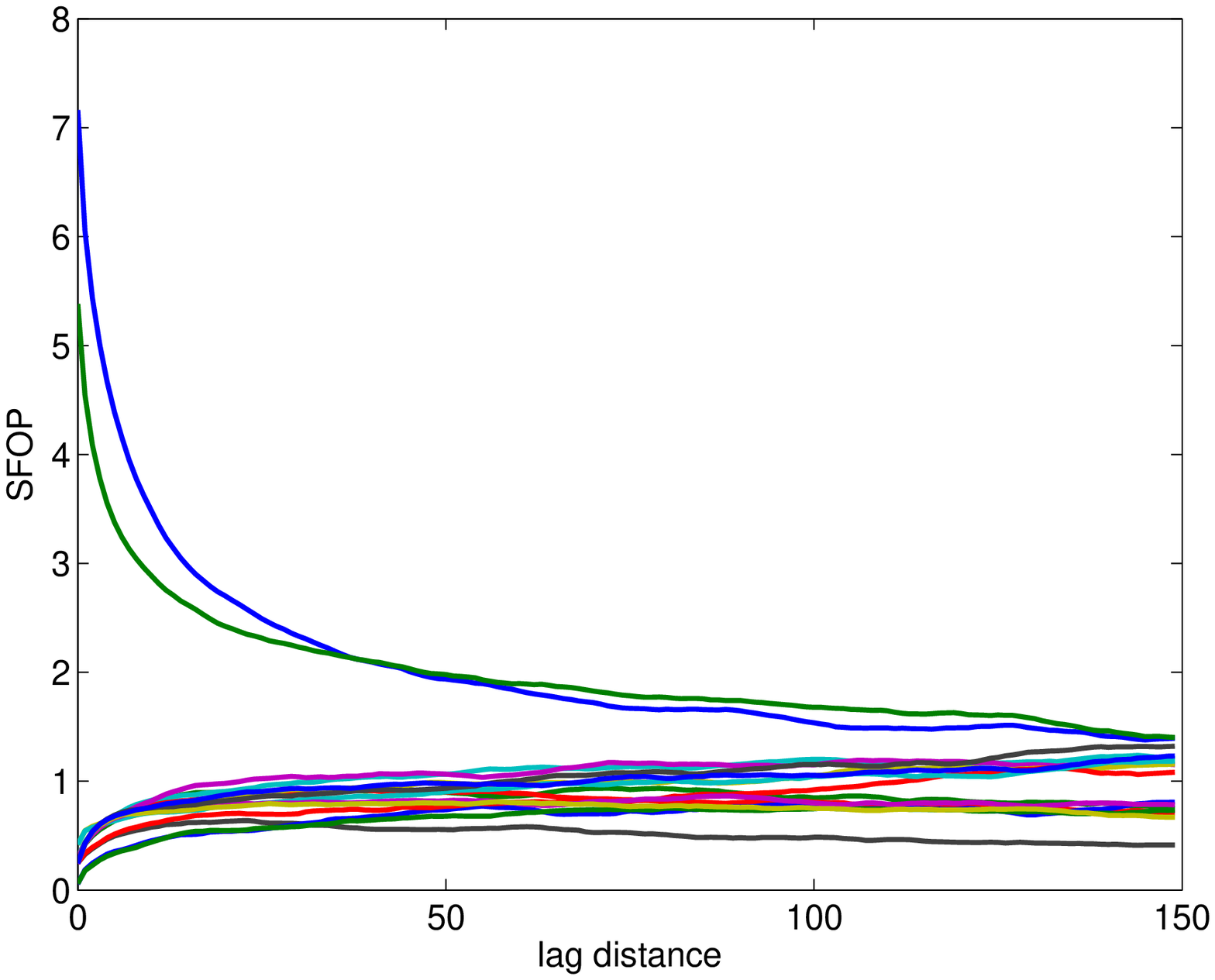}
\includegraphics[scale=0.37]{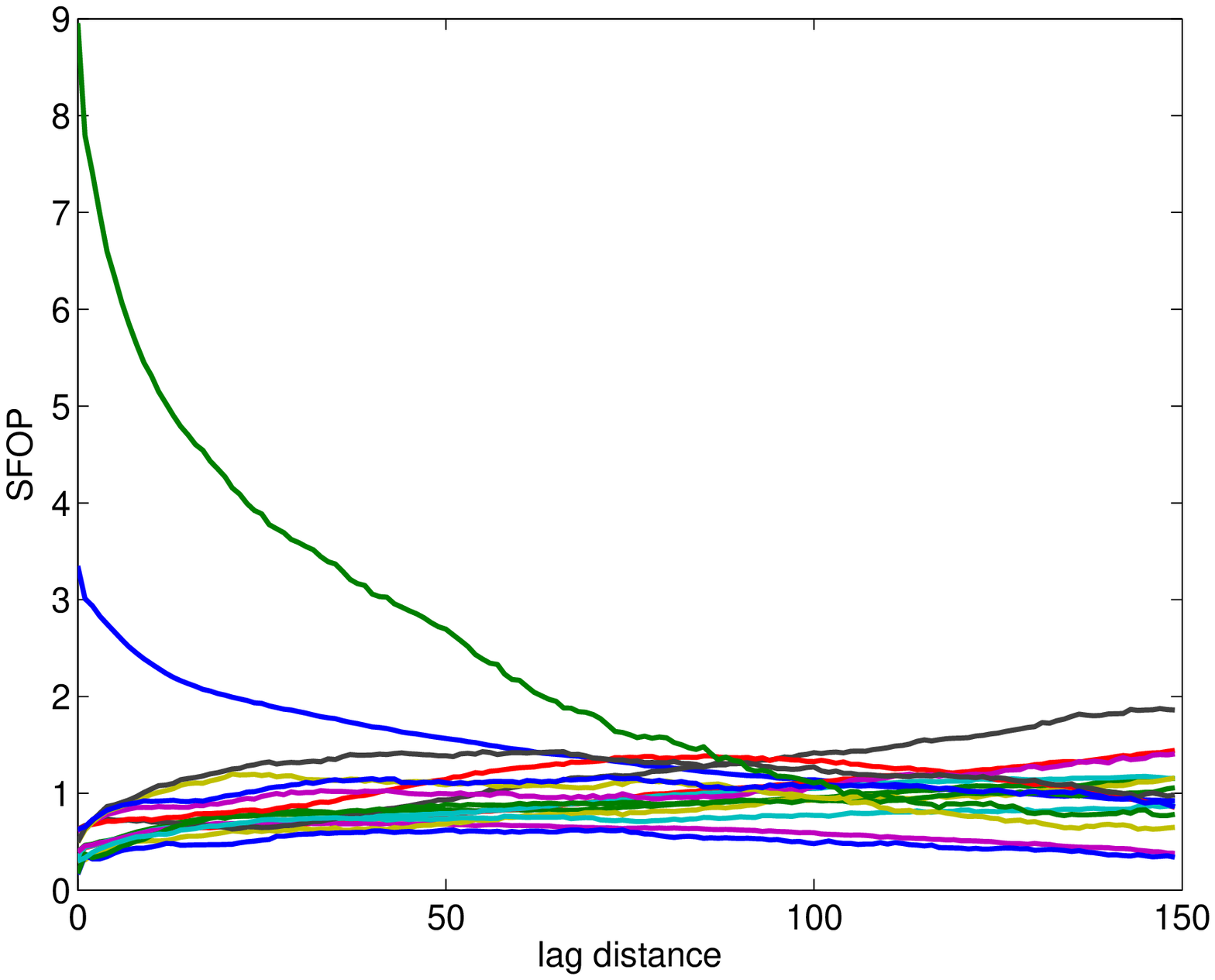}}\\
\caption[The standardize frequency of every pattern are shown for the six images. SFOP has been determined as shown in \eqref{eq:max}]{\small The standardize frequency of every pattern are shown for the six images. SFOP has been determined as shown in \eqref{eq:max}. The two patterns of maximum occurrence or the more structured ones which are shown in blue and green revert position compares to odds proportions of patterns plot. The standardization of the odds proportions account for the available data and probability of occurrence.}
\label{fig:occur}
\end{figure}

\begin{figure}[t!]\centering
{\includegraphics[scale=0.37]{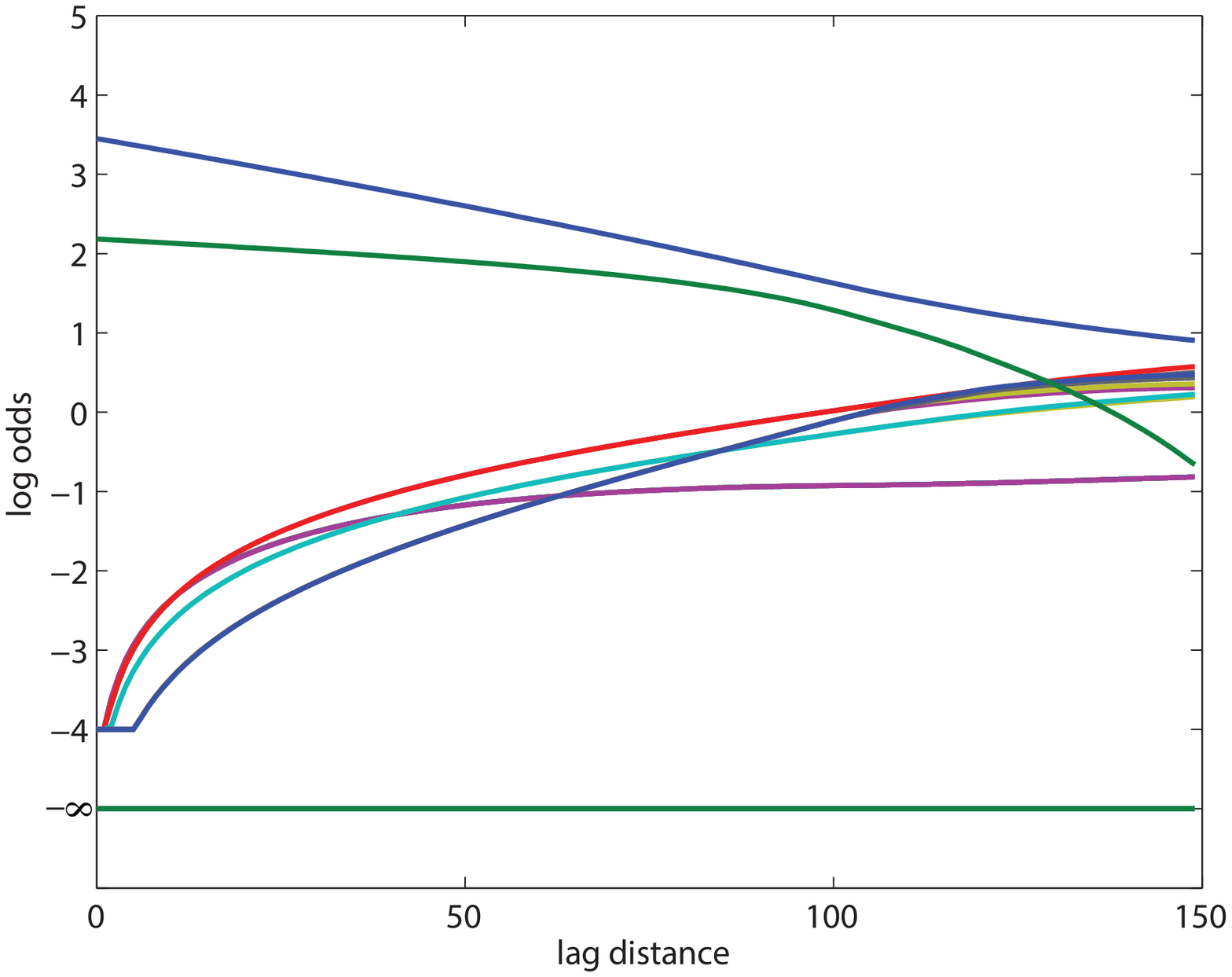}
\includegraphics[scale=0.37]{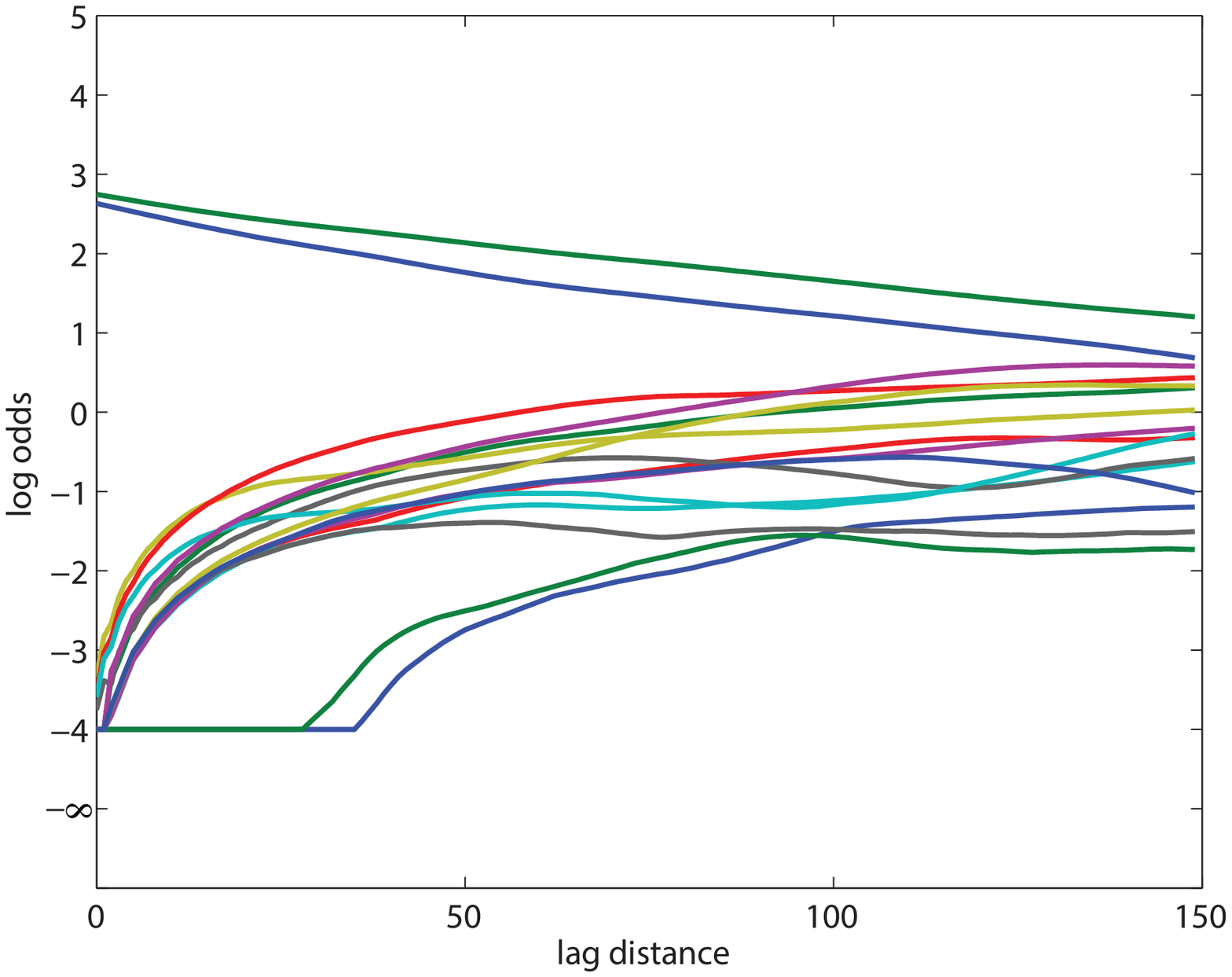}}\\
{\includegraphics[scale=0.37]{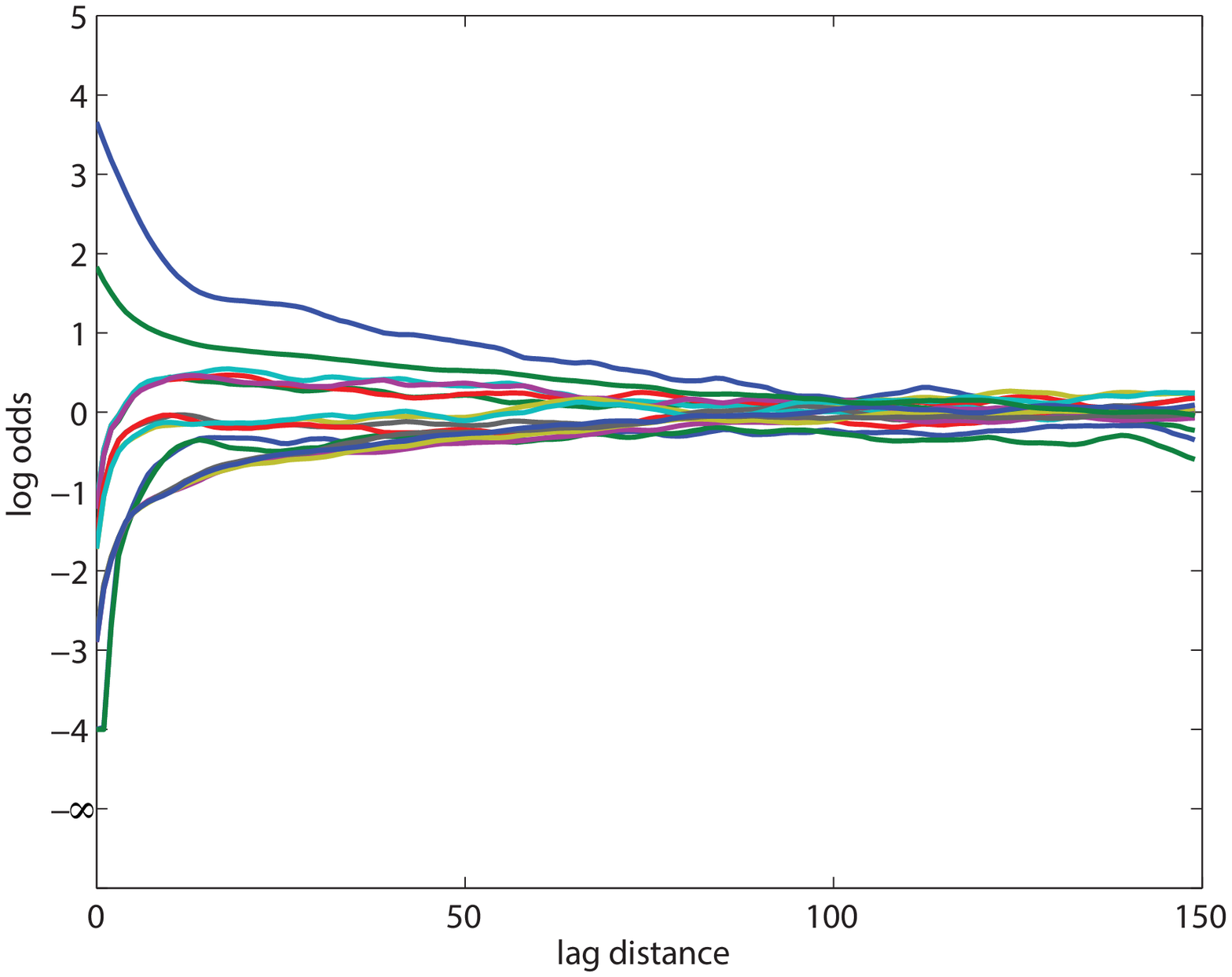}
\includegraphics[scale=0.37]{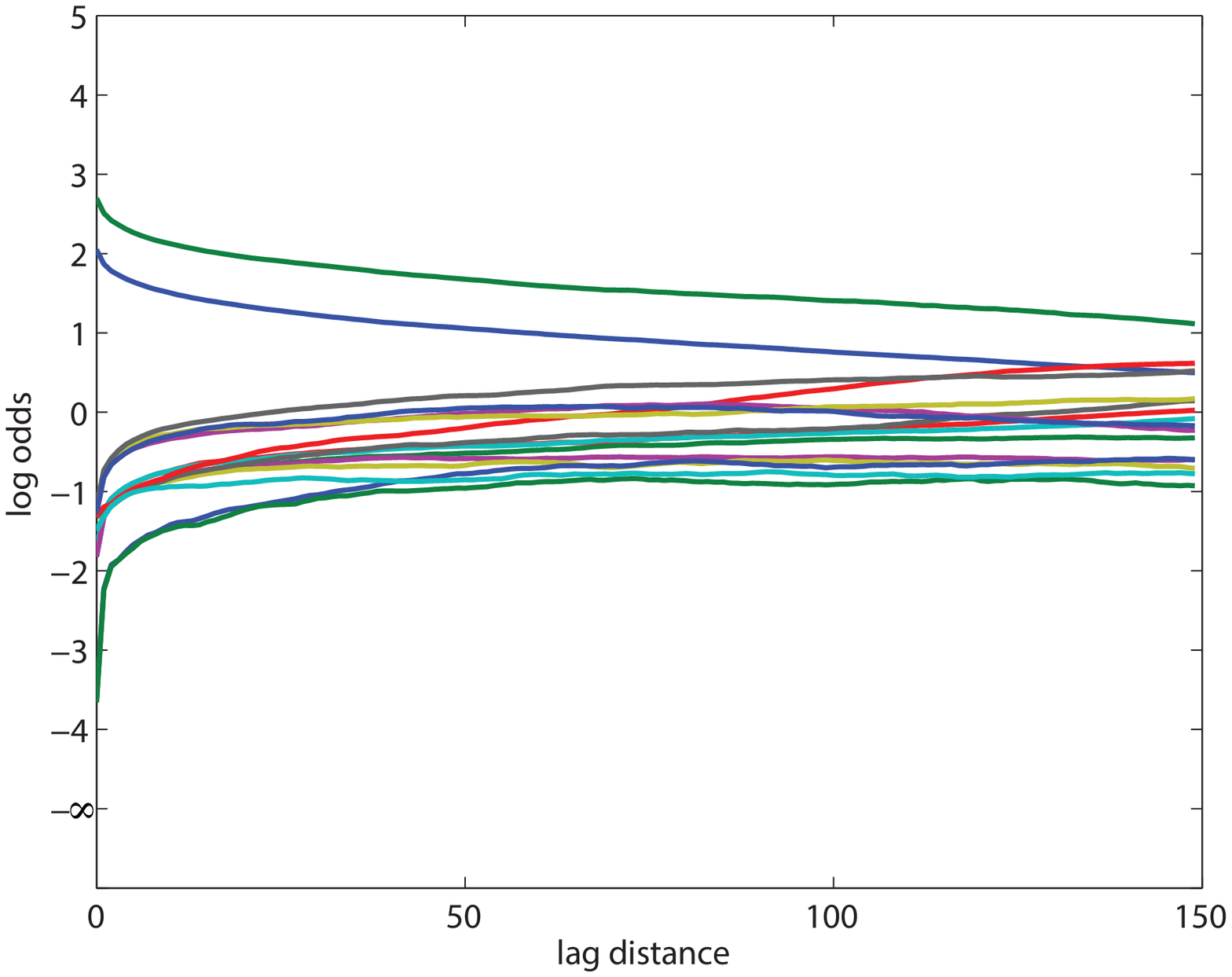}}\\
{\includegraphics[scale=0.37]{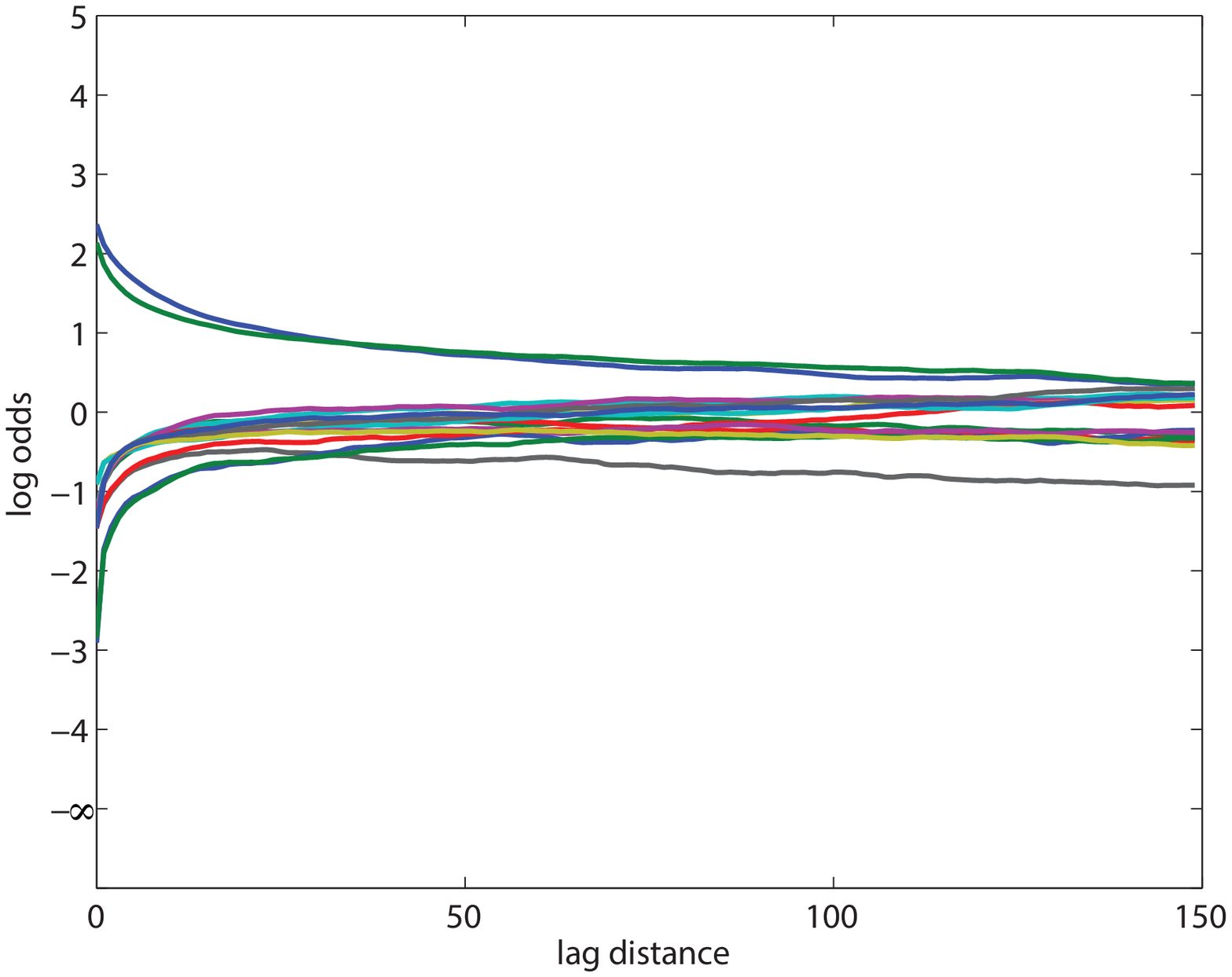}
\includegraphics[scale=0.37]{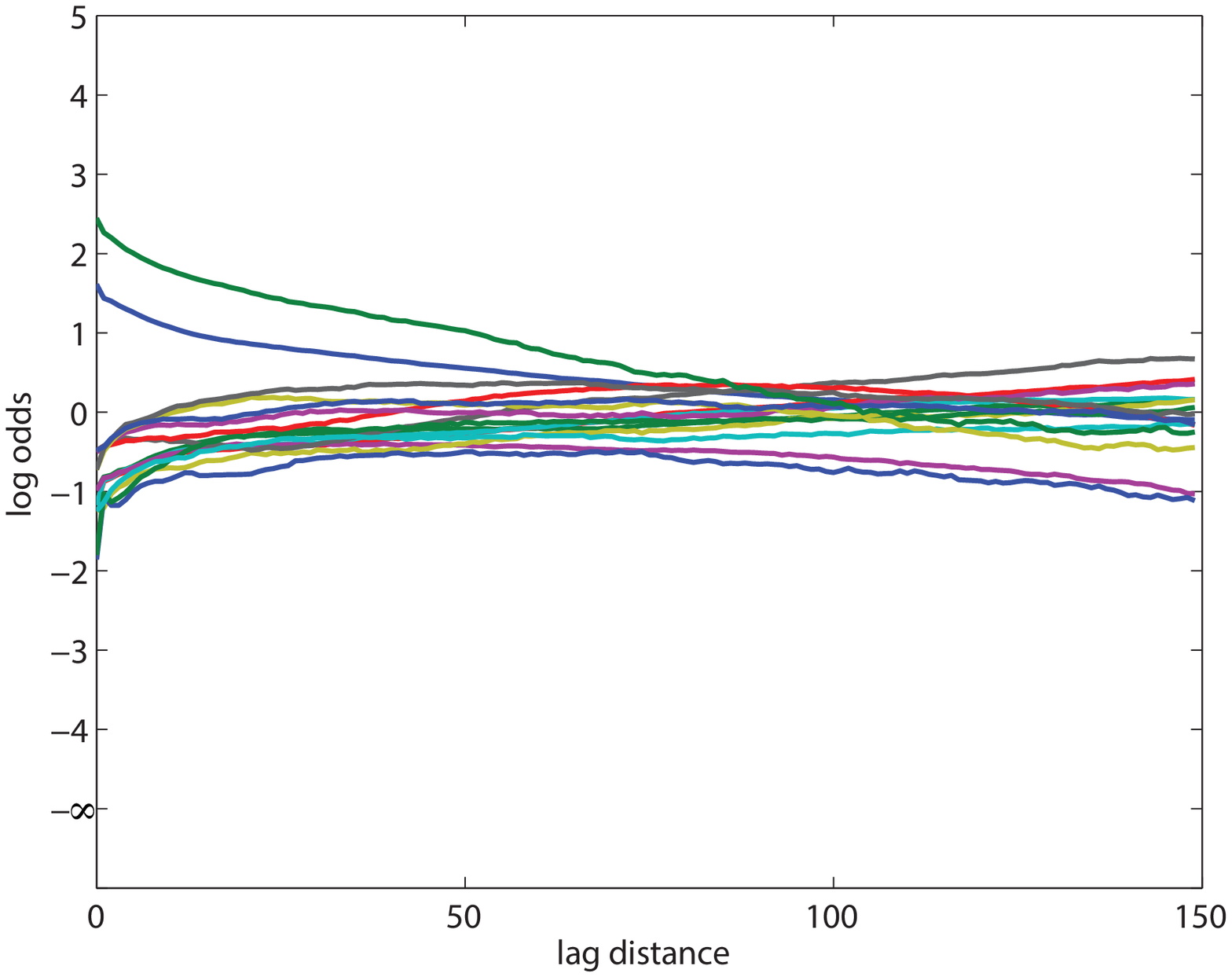}}\\
\caption[The log of odds ratio for 16 patterns of 2$\times$2 configuration are shown for the six images. The lag distance starts at 0 and goes for a range of 150. All log ratios are constrained to the range of -4 and 4. All log ratios are constrained to the range of -4 and 4]{\small The log of odds ratio for 16 patterns of 2$\times$2 configuration are shown for the six images. The lag distance starts at 0 and goes for a range of 150. All log ratios are constrained to the range of -4 and 4. Occurrence of real zero is represented by -$\infty$ in the plot and any log odds less than -4 or greater than 4 would be represented as the minimum of -4 or maximum of 4.}
\label{fig:log}
\end{figure}

\begin{figure}[t!]\centering
{\includegraphics[scale=0.37]{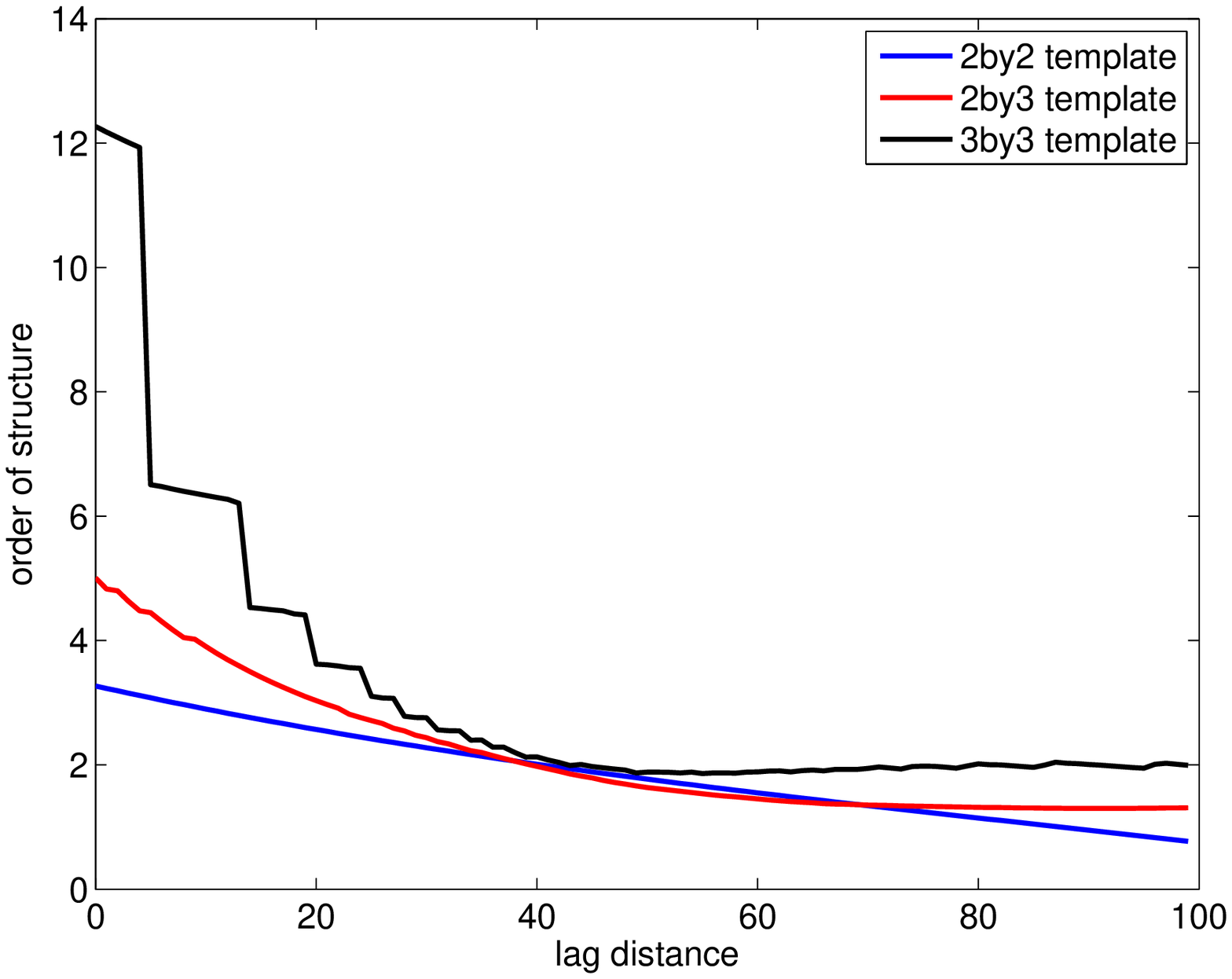}
\includegraphics[scale=0.37]{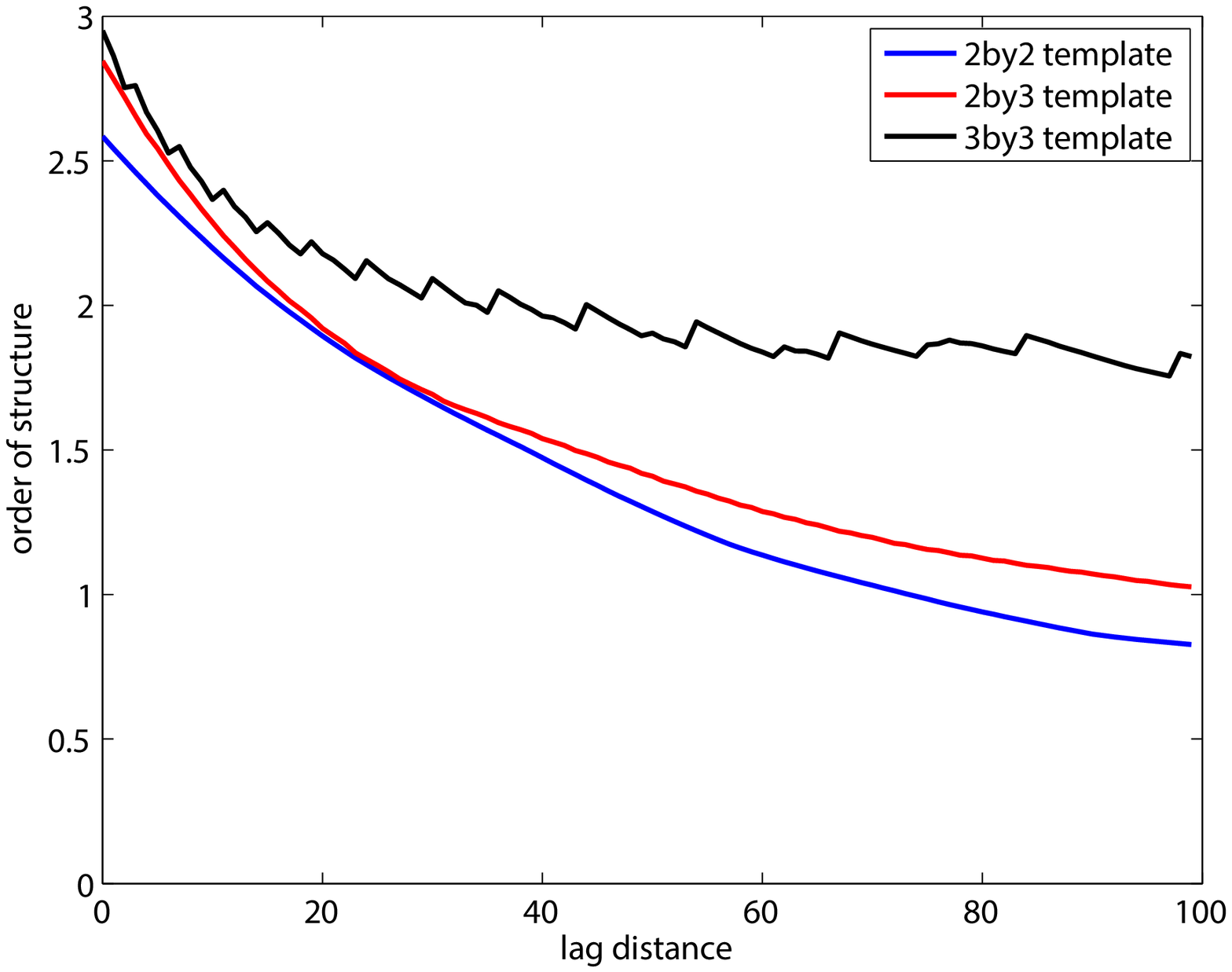}}\\
{\includegraphics[scale=0.37]{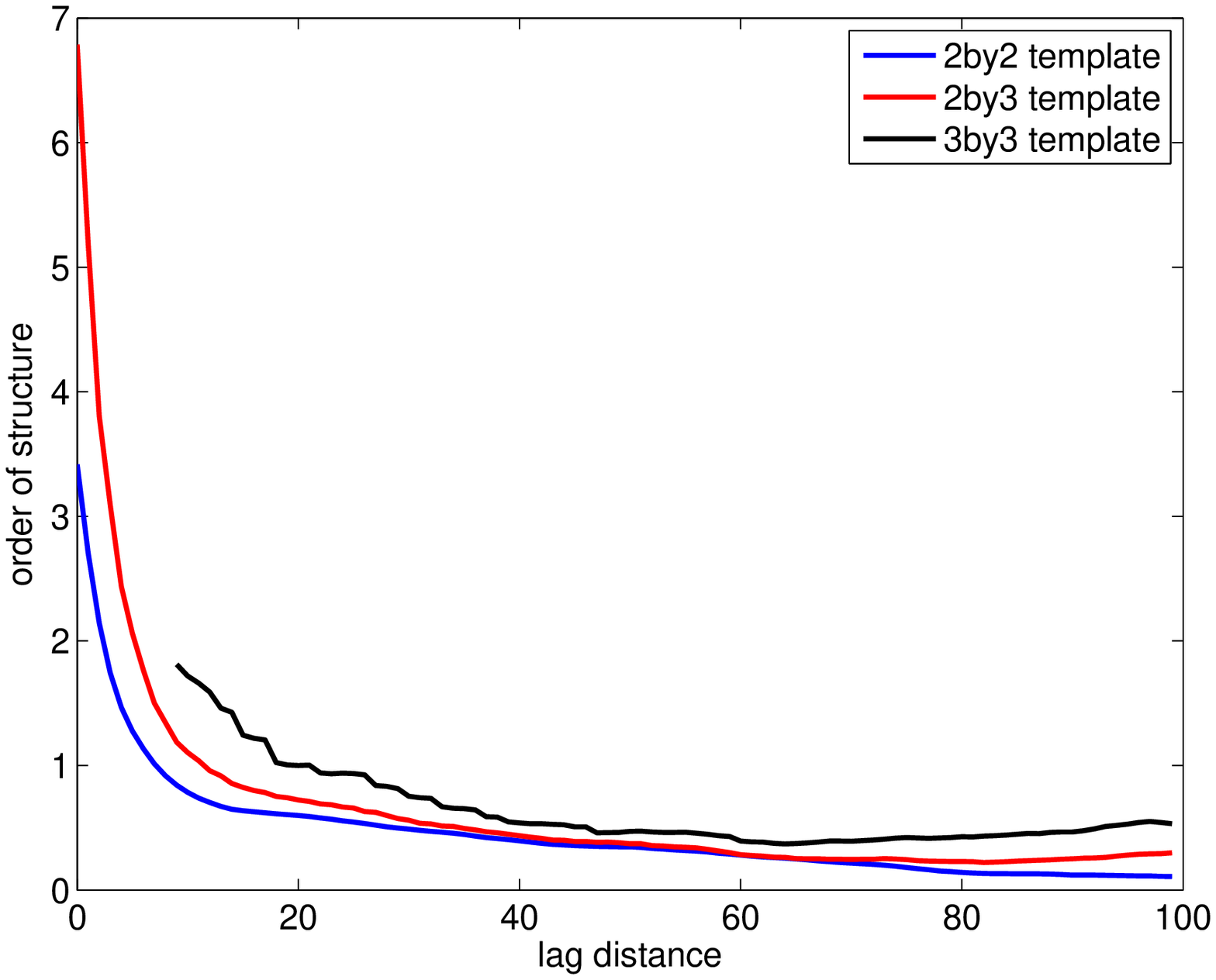}
\includegraphics[scale=0.37]{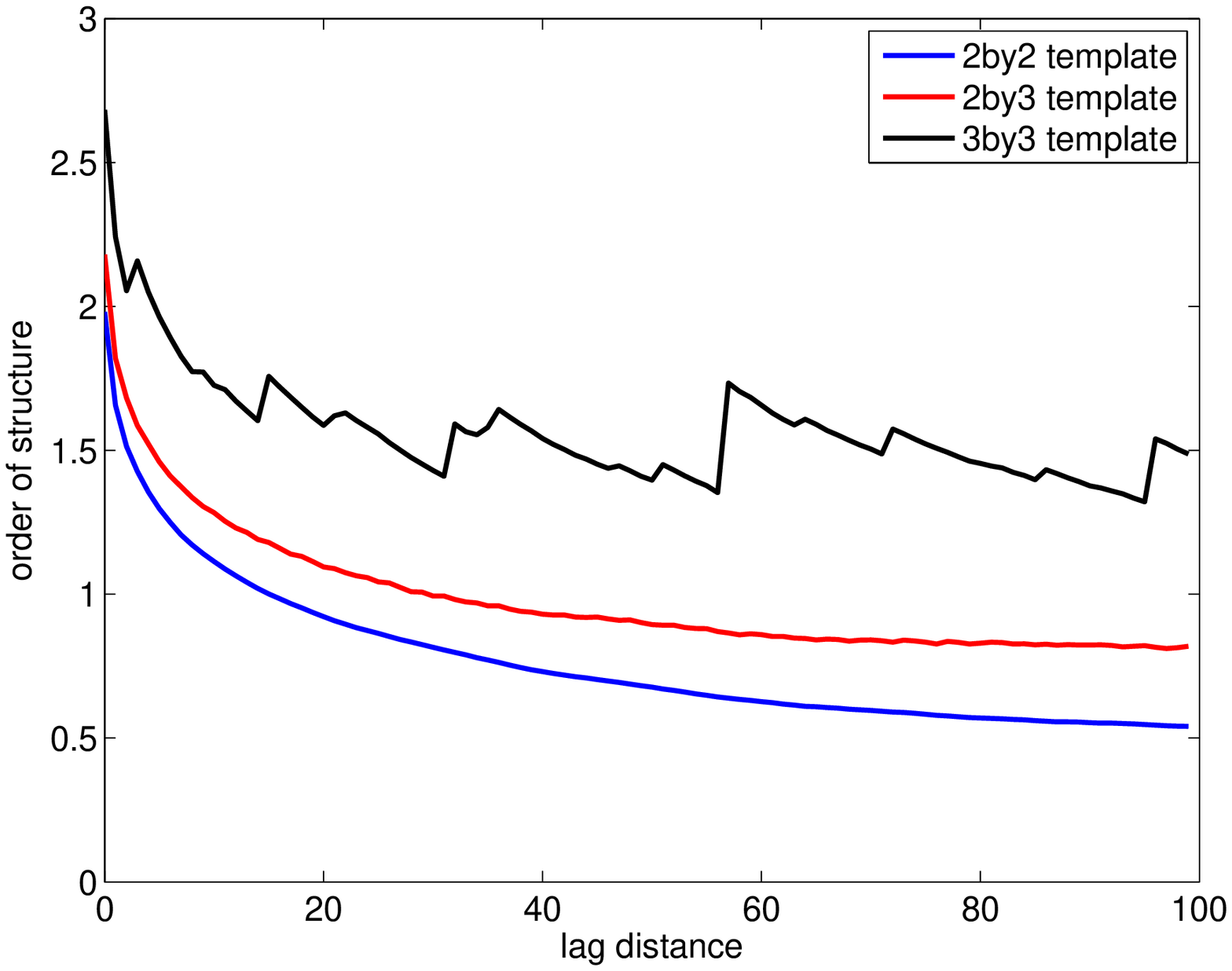}}\\
{\includegraphics[scale=0.37]{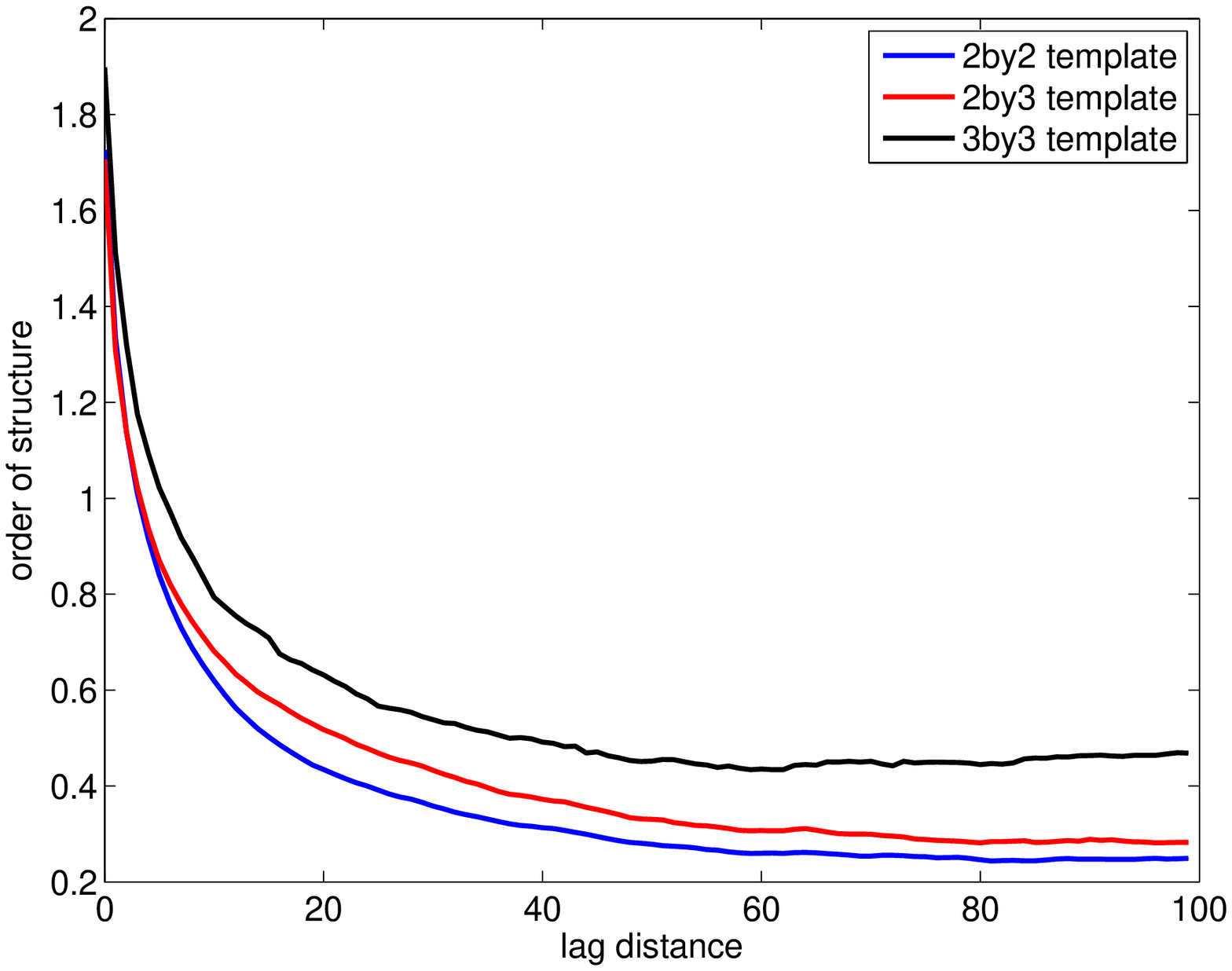}
\includegraphics[scale=0.37]{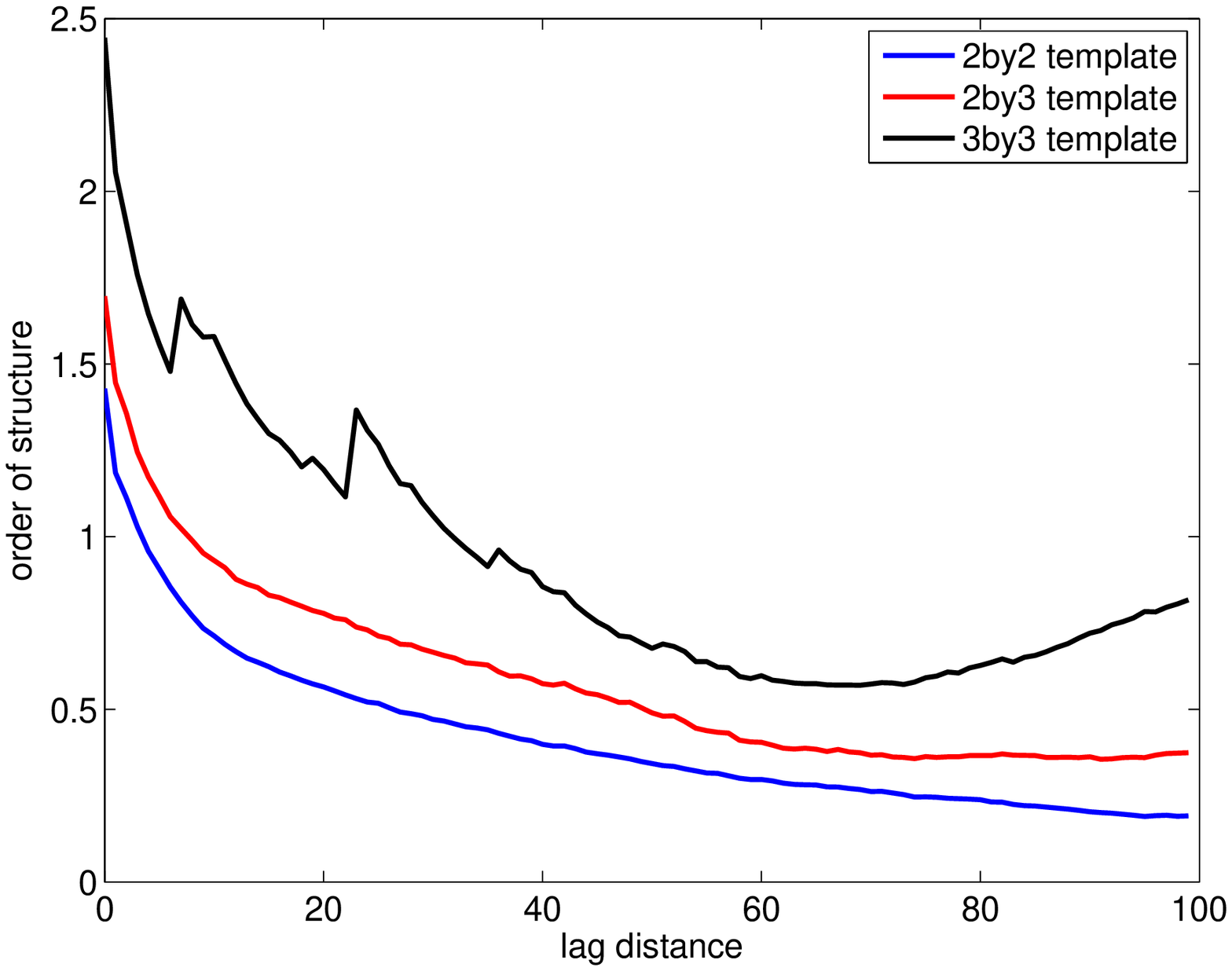}}\\
\caption[An analogy to the standardized variogram plot that represents the spatial continuity within 0 and 1]{\small An analogy to the standardized variogram plot that represents the spatial continuity within 0 and 1. The FOP ratio in $y$-axis are determined by calculating the ratio of non-ordered structure as a function of all patterns FOP. The jump from zero is analogous to nugget effect in variogram plot. The first two maps are the least noisy and have nugget effect of almost zero value.}
\label{fig:stf}
\end{figure}
\begin{figure}[t!]\centering
{\includegraphics[width=8.25cm,height=6.2cm]{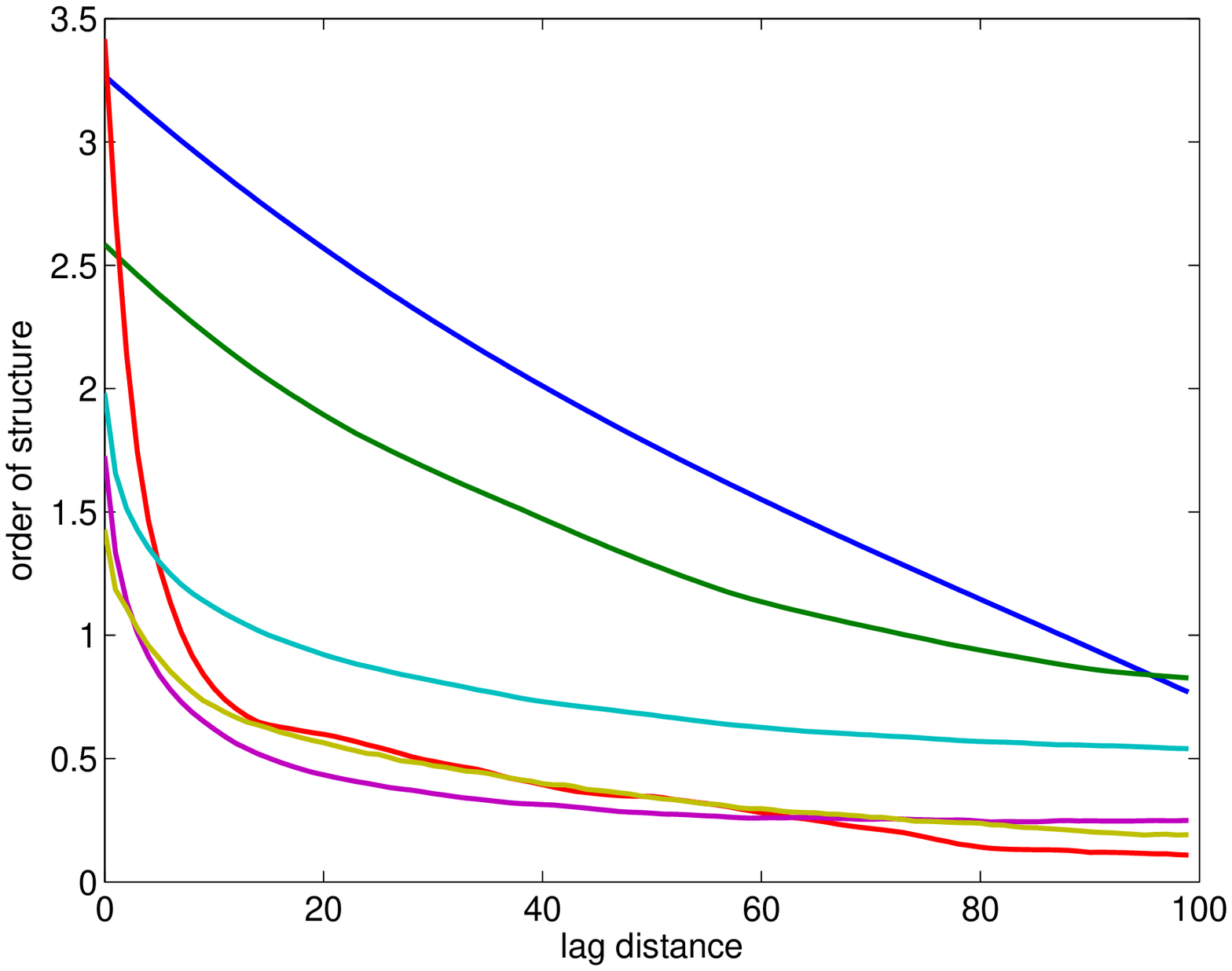}
\includegraphics[width=8.25cm,height=6.2cm]{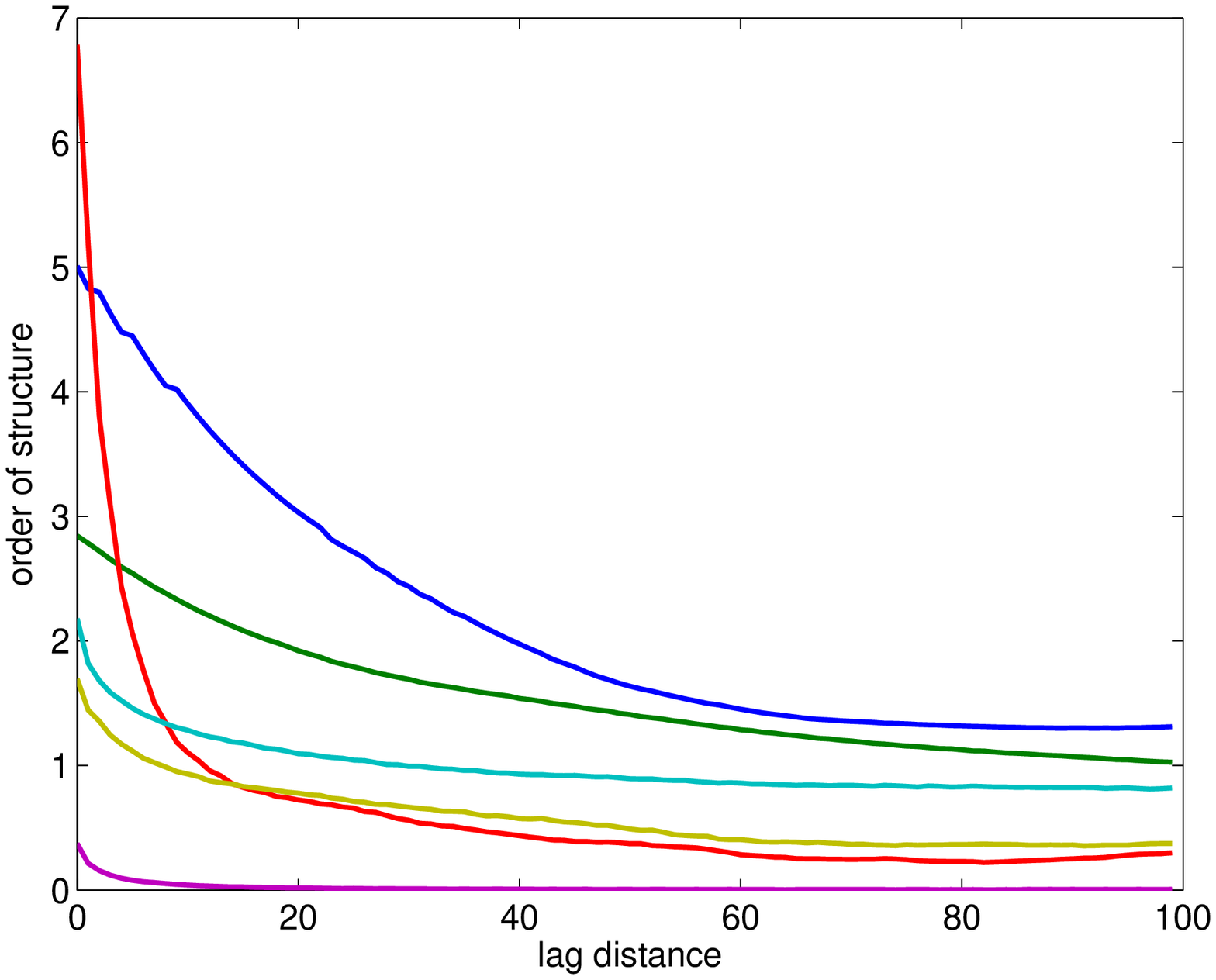}
\includegraphics[width=8.25cm,height=6.2cm]{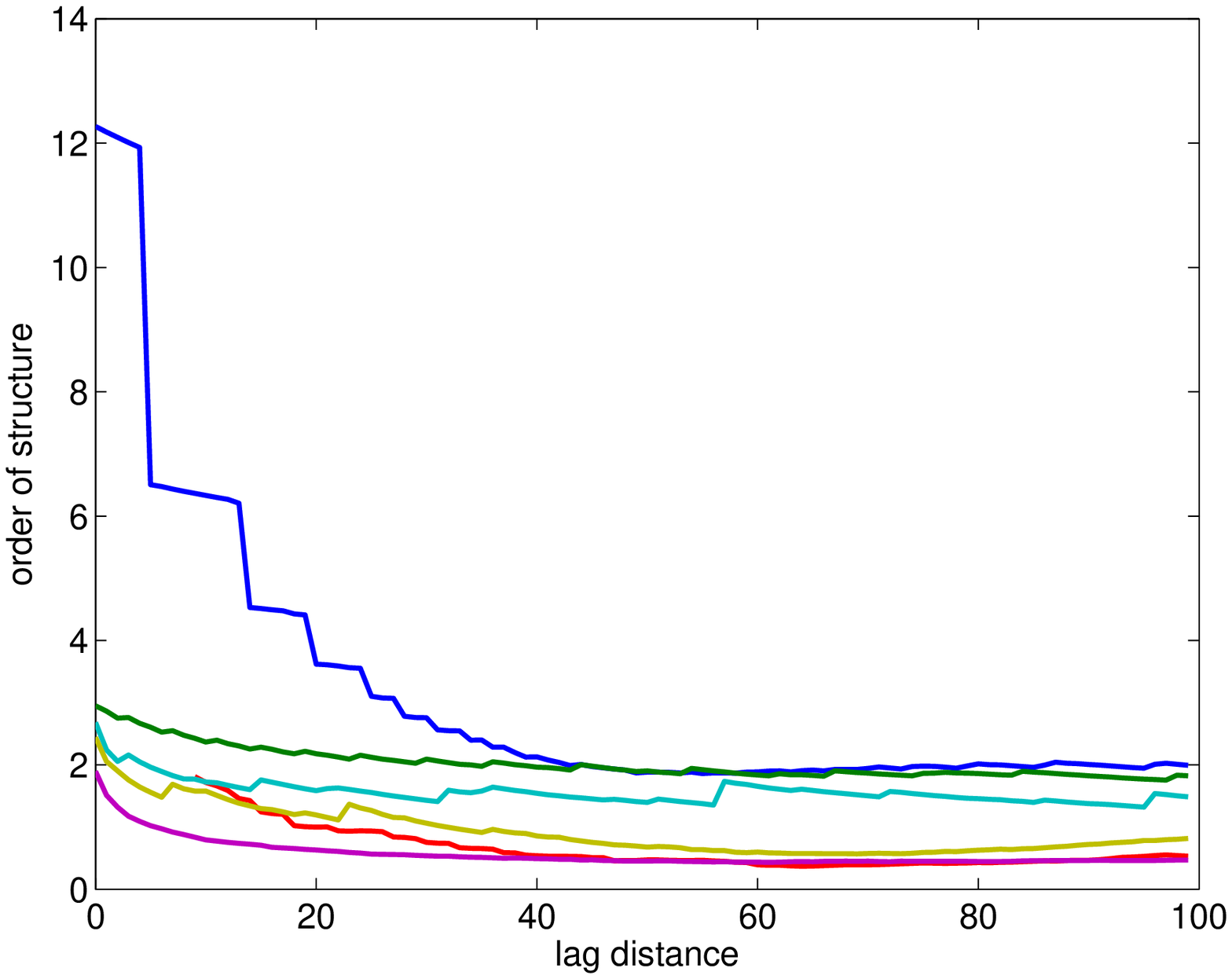}}
\caption[The figure at the top represents the summary of order of structure for templates of 2$\times$2 which includes 16 patterns for the six images]{\small The figure at the top represents the summary of order of structure for templates of 2$\times$2 which includes 16 patterns for the six images. The second figure similarly represents the summary of order of structure for templates of 2$\times$3 that includes 64 patterns. And finally, the last figure represents the summary of order of structure for 512 patterns of 3$\times$3 configurations. All configurations contains only two categories.}
\label{fig:stfC}
\end{figure}


\begin{figure}[t!]\centering
\includegraphics[scale=0.37]{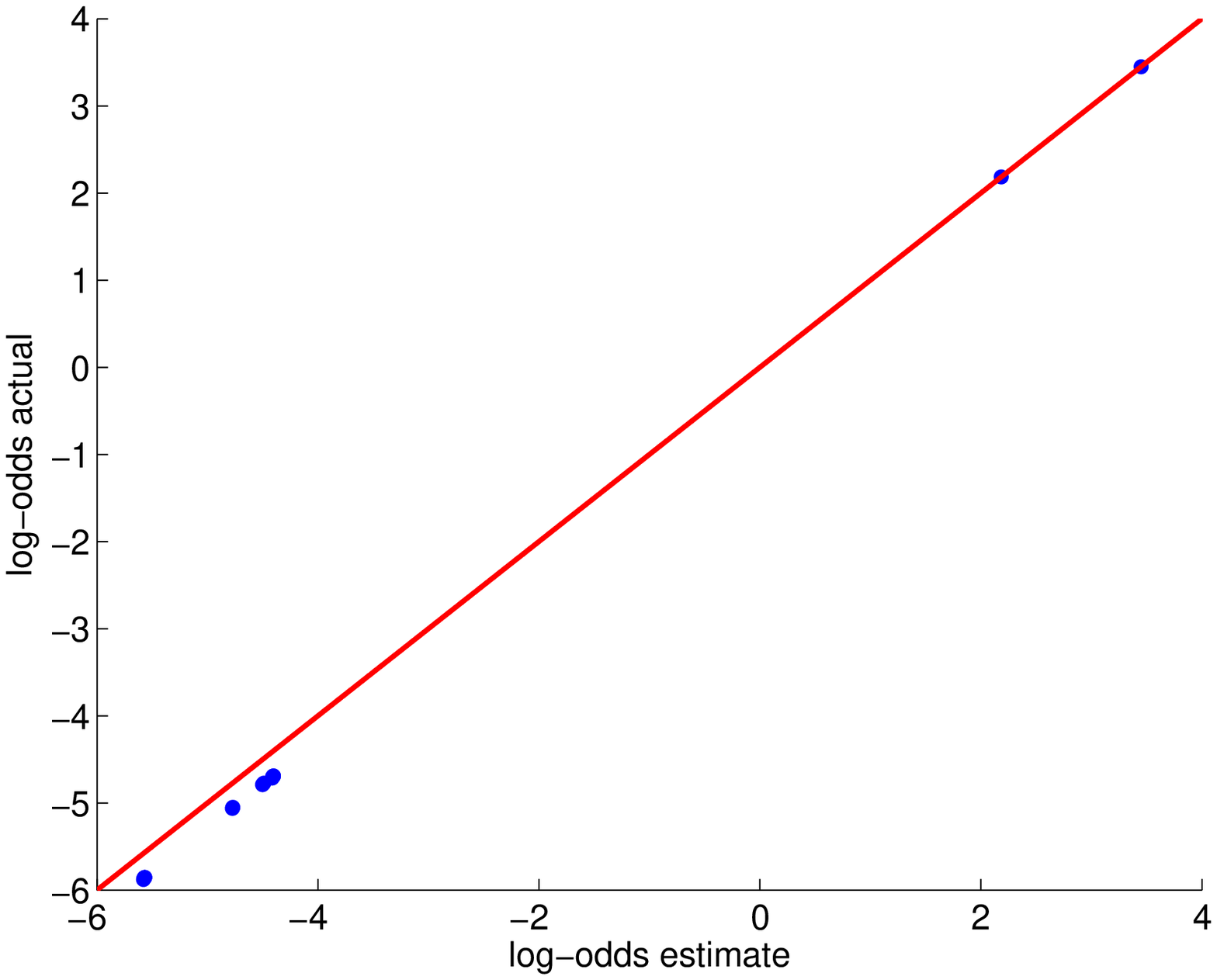}
\includegraphics[scale=0.37]{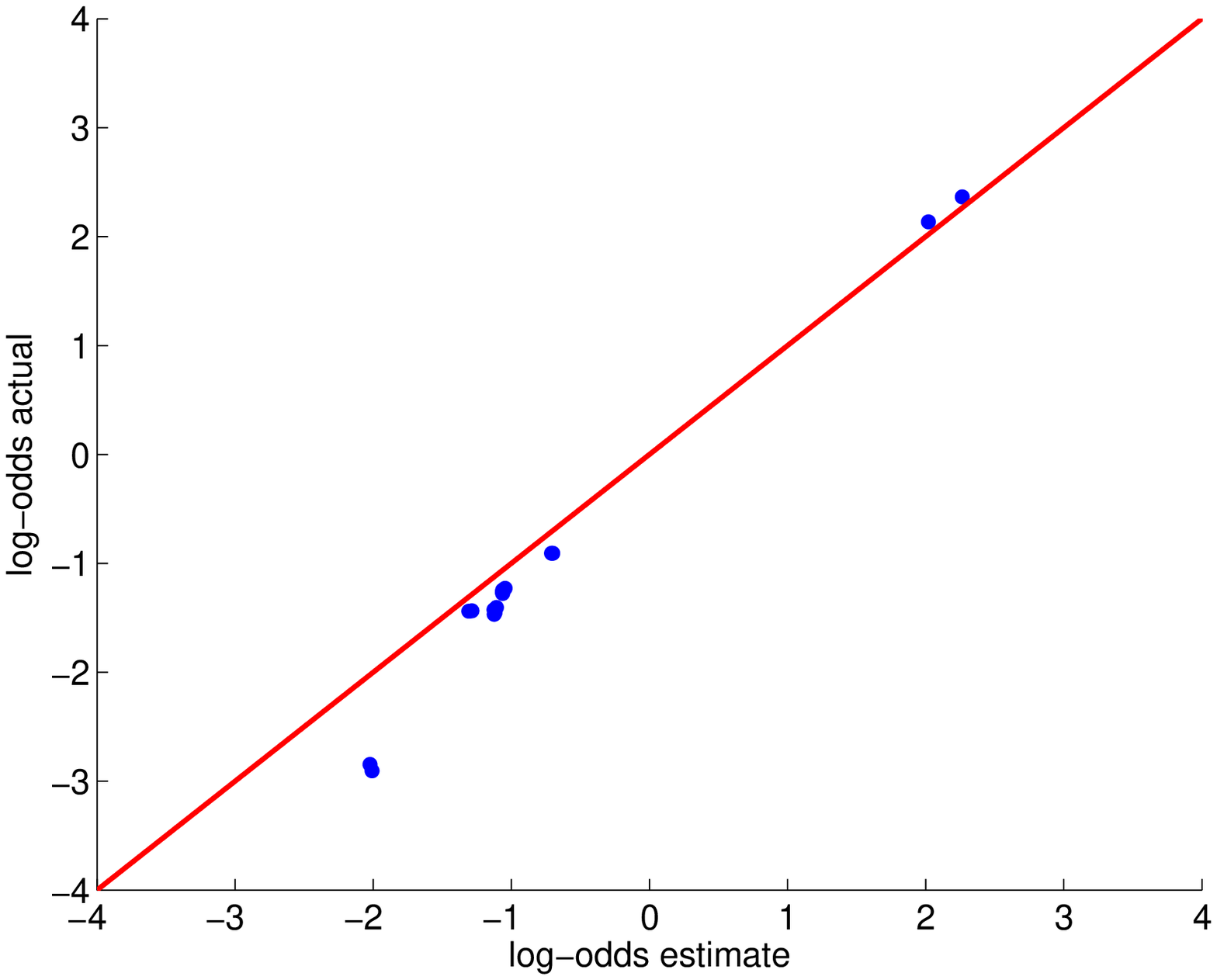}
\includegraphics[scale=0.37]{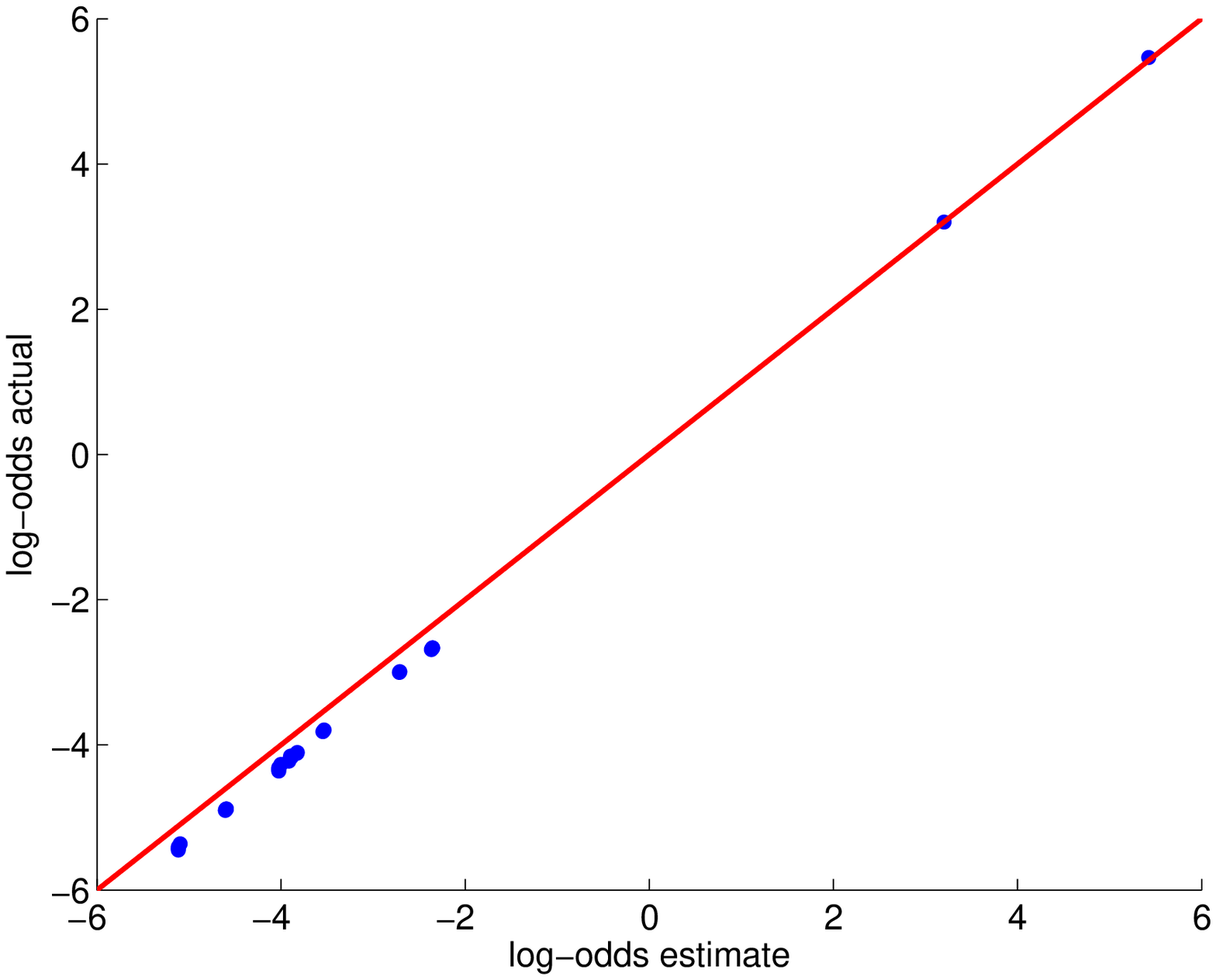}
\includegraphics[scale=0.37]{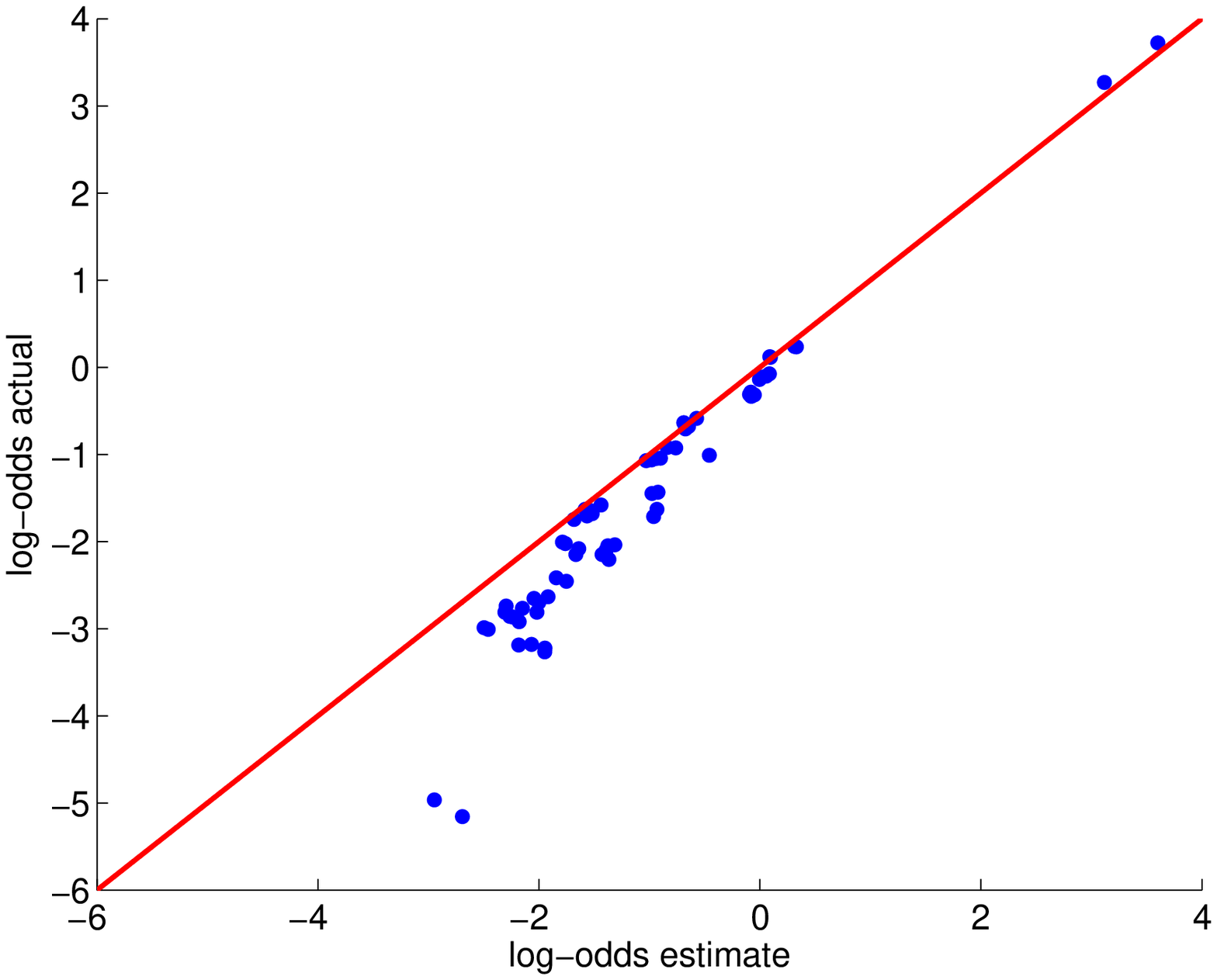}
\includegraphics[scale=0.37]{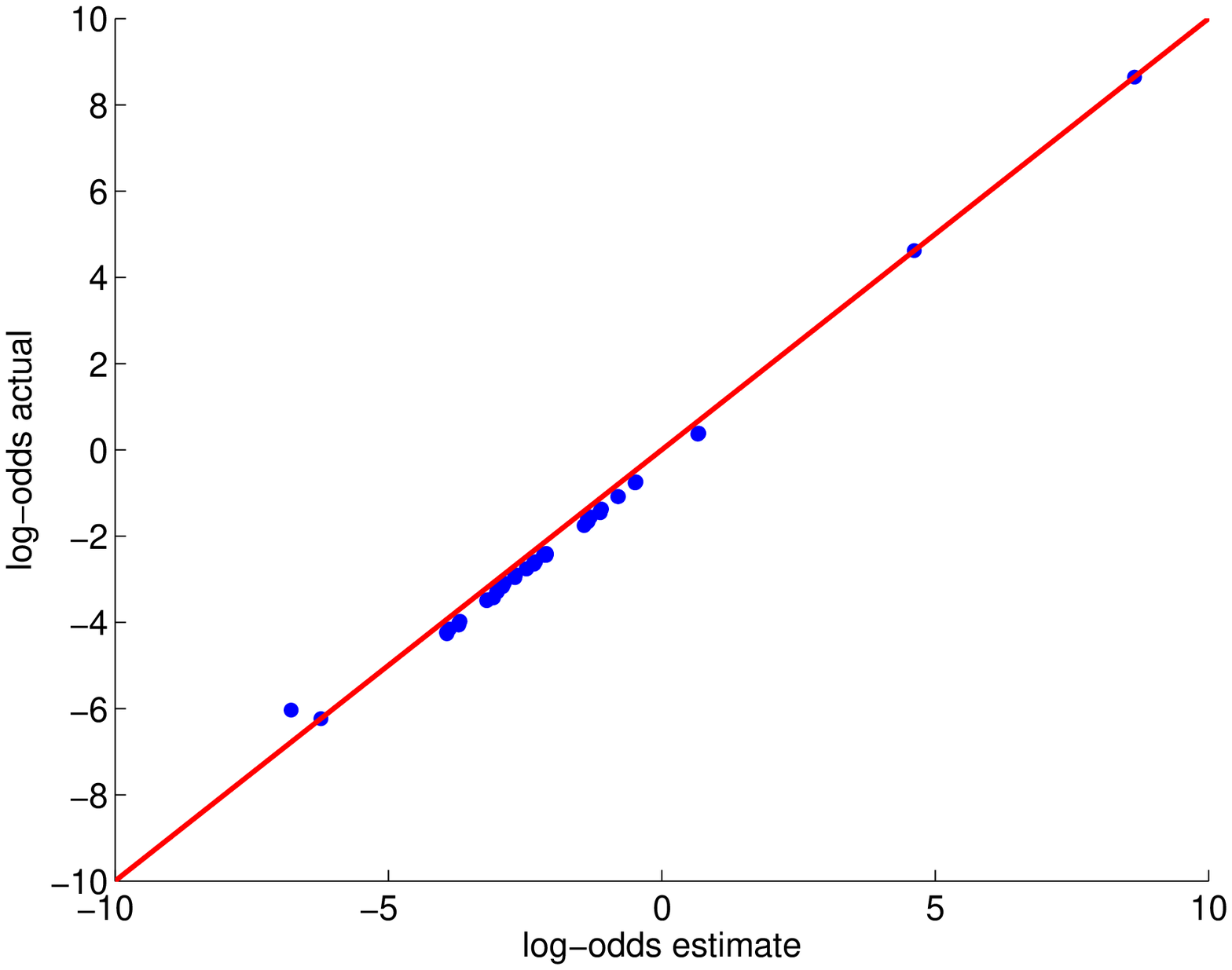}
\includegraphics[scale=0.37]{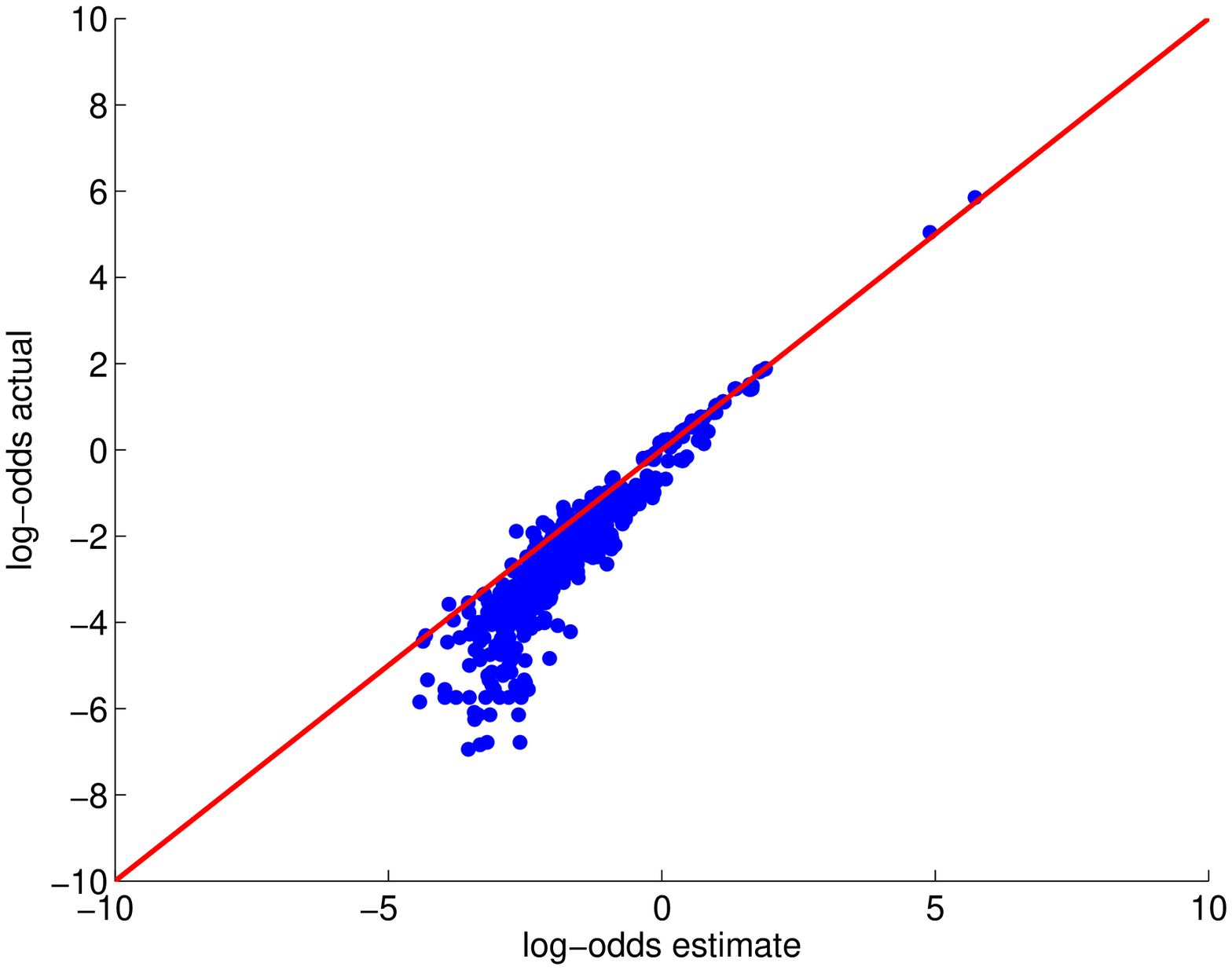}
\caption[These scatter plots represent how the FOP can help with characterizing the short scale variability. The scatter plots on the left belong to the most structured case]{\small These scatter plots represent how the FOP can help with characterizing the short scale variability. The scatter plots on the left belong to the most structured case of the circle shown in Figure~\ref{fig:TI}, and the scatter plots at right belong to the case that shows the tree with lots of branches. The two scatter plots at the top represents the predicting log FOP for 16 patterns of 2$\times$2 configuration. The second row is 3$\times$3 configuration and last row is prediction for 9$\times$9 configuration. It can be observed that the estimation becomes less precise as the configuration scheme becomes more complicated. Similarly, when the map has more complex structure, the estimation becomes more complicated.}
\label{fig:prd}
\end{figure}

\subsection{Examples}
\nin Six binary maps of different structure are considered to examine and understand the FOP introduced in this section (see Figure~\ref{fig:TI}). The maps are roughly ordered from the most structured to the least. The behavior of 16 patterns of 2 by 2 configuration and 2 categories are studied as a function of lag distance. The 2$\times$3 and 3$\times$3 configurations are also considered in this section, but the results are only documented in summary to avoid plotting a very large number of curves on one graph (i.e. $2^6$=64, or $2^9$=512).
Figure~\ref{fig:proportion} represents the FOP of all 16 patterns as a function of lag distance ranging from 0 to 150. Figure~\ref{fig:occur} represents the standardized FOP, as defined in Equation~\eqref{eq:max}.

For the six images, two patterns of 1 and 16 have their odds ratios decreasing as the lag distance increases. These two patterns represent the structuredness in the maps. For larger lag distances, all 16 patterns appear to converge at some point. The logarithm of odds ratio is shown in Figure~\ref{fig:log} since the natural logarithm of odds ratios are more interpretable. The logarithm is often used to exaggerate the values larger than the reference value. As can be seen, the logarithm of patterns is between -4 and 4 where 0 is the reference value. A real zero value of log of odds ratio would be -$\infty$. Log odds greater than 4 and less than -4 are clipped to a maximum of 4 and minimum of -4.

An interesting observation is that the two patterns of 1 and 16 switch position in Figures~\ref{fig:occur} and \ref{fig:log} compared to Figure~\ref{fig:proportion}. The one having the maximum odds for FOP has its maximum occurrence because of the larger proportion of corresponding category.
Also, the first map with the black circle has the most structured feature; some patterns have zero probabilities throughout the range of lag distance. A few patterns have very small occurrences and start at later lags. The same is true for the second image. As can be seen in Figure~\ref{fig:log}, the odds ratio of frequency of patterns 8 and 9 is very small until lag distance of about 40.

A variogram plot represents the nugget effect, variogram range and how the variability increases with the lag distance. Similarly, the order of structure can point out similar concepts in multiple-point data interactions. Figure~\ref{fig:stf} illustrates the order of structure for patterns of 2$\times$2 configuration. At 0 lag distance, the deviation from 1 is at its maximum and the order of structure drops as the lag distance increases. The minimum value for order of structure is 0 at very large lag distances and represents randomness in the model. It starts at a value greater than 0 at zero lag distance and falls as the lag distance increases. The slope of the decay represents how quickly the order of structure is decreasing within the model 
A zero order of structure suggests noise. For example, in Figure~\ref{fig:stf}, the order of structure for the 4th map stabilizes at some point and falls very slowly with lag distance. This indicates that the corresponding map has large scale structure. The steep fall of the order of structure suggests small scale structure. For example, the maximum order of structure at zero lag distance in the third image is at least four times that of the fifth image. This could be the result of short scale features repeating over the map.

Figure~\ref{fig:stfC} illustrates the order of structure for the same templates for all 6 images of Figure\ref{fig:TI} in one plot. The figure at the top is for 2$\times$2 patterns, the one in the middle illustrates the order of structure for 2$\times$3 configurations and finally the one at the bottom corresponds to 3$\times$3 templates. It can be observed that the more structured images start at higher order of structure and decline slowly. Therefore, higher data interactions deal with larger deviation from the baseline.
The maximum also increases as the number of locations in the template increases. However, it seems that the increase is stable for all the images. For example, the red curve in three plots of Figure~\ref{fig:stfC} belongs to the third image representing a tree which has the largest order of structure at zero lag in all three pattern extraction. This is not shown for template configuration of 3$\times$3 as the maximum value is very large. Another observation is that the order of structure for the template configuration of 3$\times$3 is not as smooth. The overall behavior of the order of structure however is very similar in all three pattern configurations; it starts at higher values and falls slowly as the lag distance increases.

Figure~\ref{fig:prd} illustrates the estimation for FOP values at scale 0 based on the information regarding FOP at scales 1 and 2. This is to see how accurate the small scale information could be infered at smaller lags based on the information from larger lags. This is analogous to extrapolation at smaller scales with variograms. This estimation is calculated from two points, that is why an FOP curve with linear behavior at smaller scale estimates more precisely. As can be seen in the corresponding figure, the characterization of occurrence of patterns at smaller scales is more accurate when the selected template is smaller in size and the map is less spatially complex. The bottom right plot represents the FOP estimation for the fifth image using 3$\times$3 patterns which is a relatively poor estimation.

\begin{figure}[t!]\centering
\includegraphics[scale=0.36]{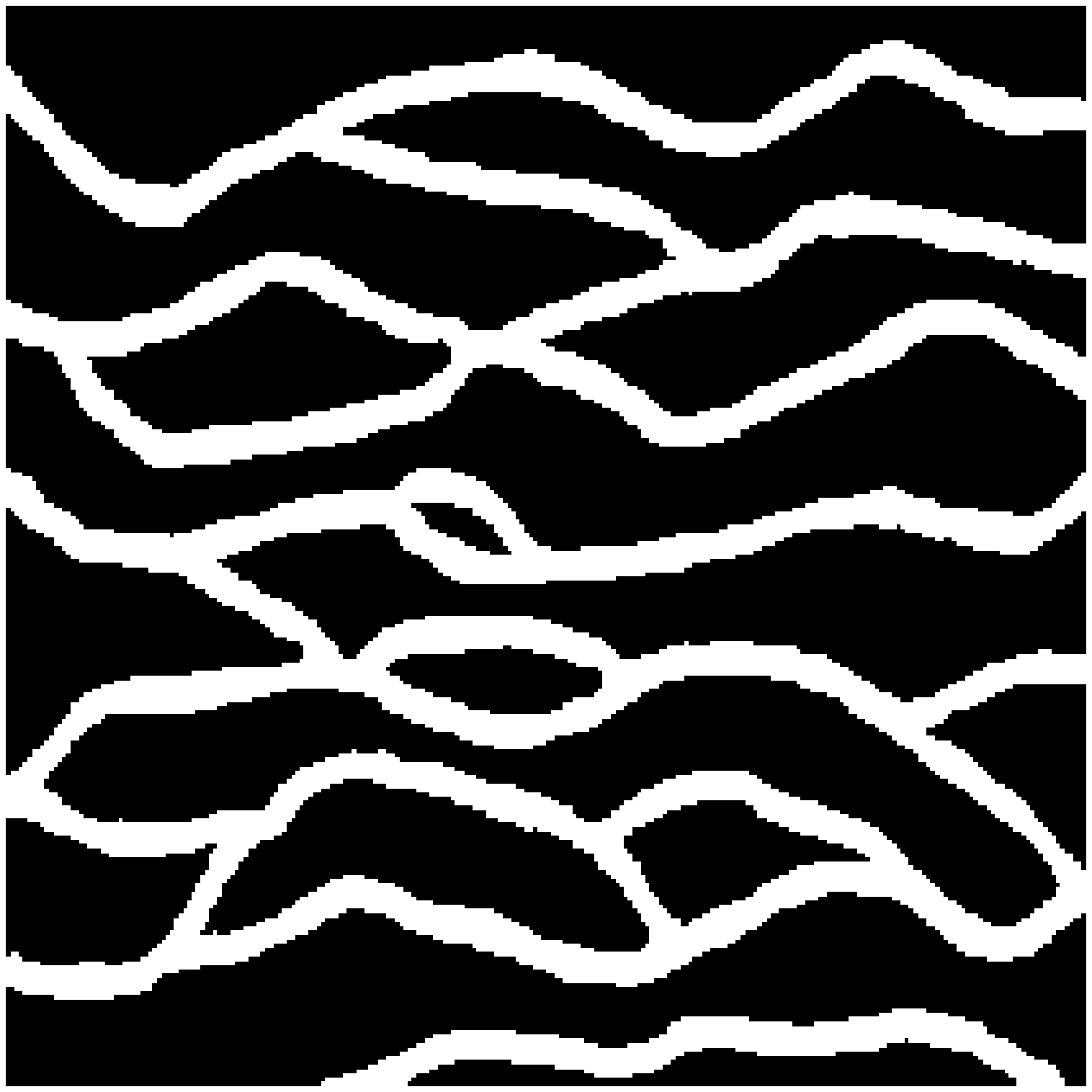}
\includegraphics[scale=0.43]{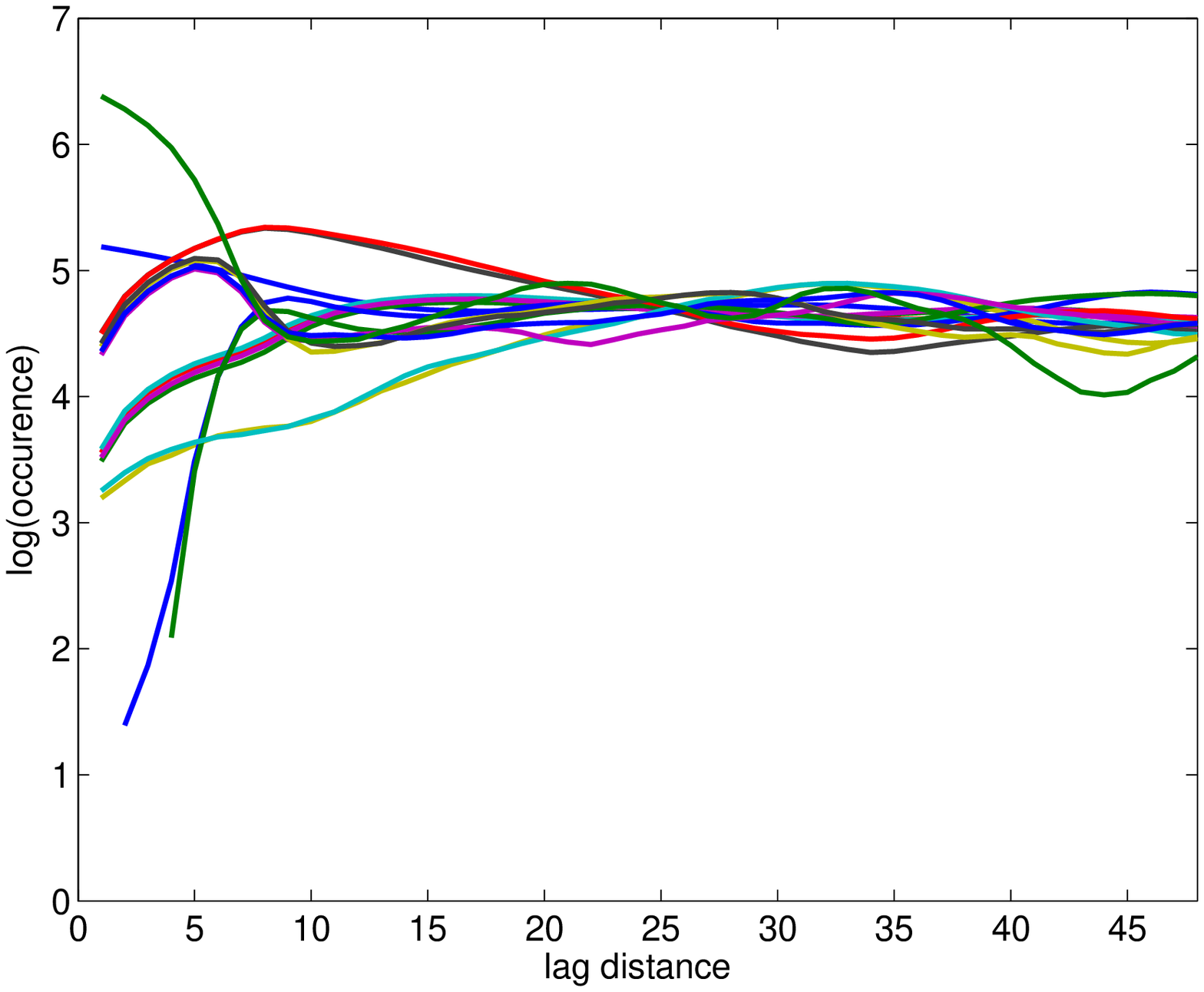}\\
\includegraphics[scale=0.36]{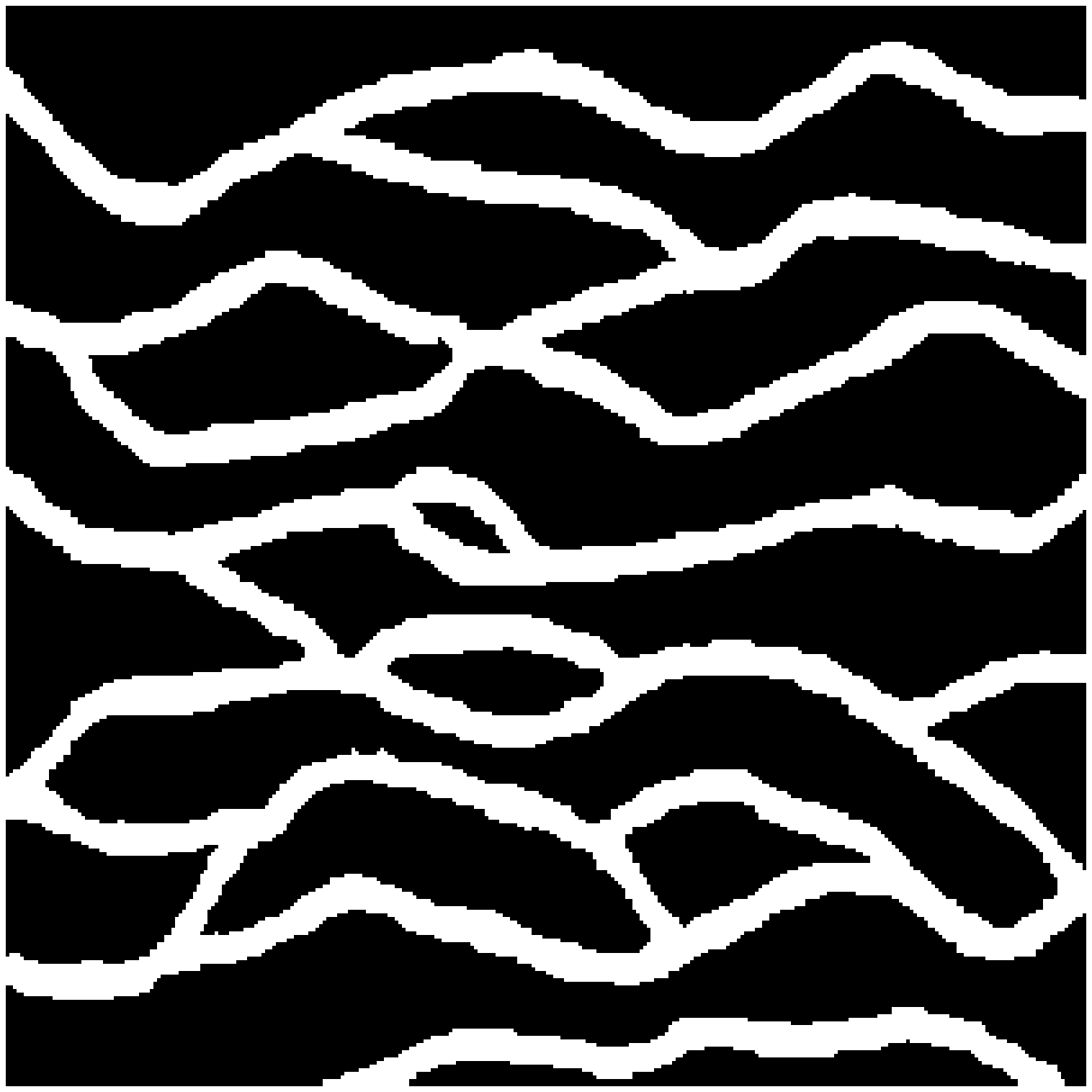}
\includegraphics[scale=0.43]{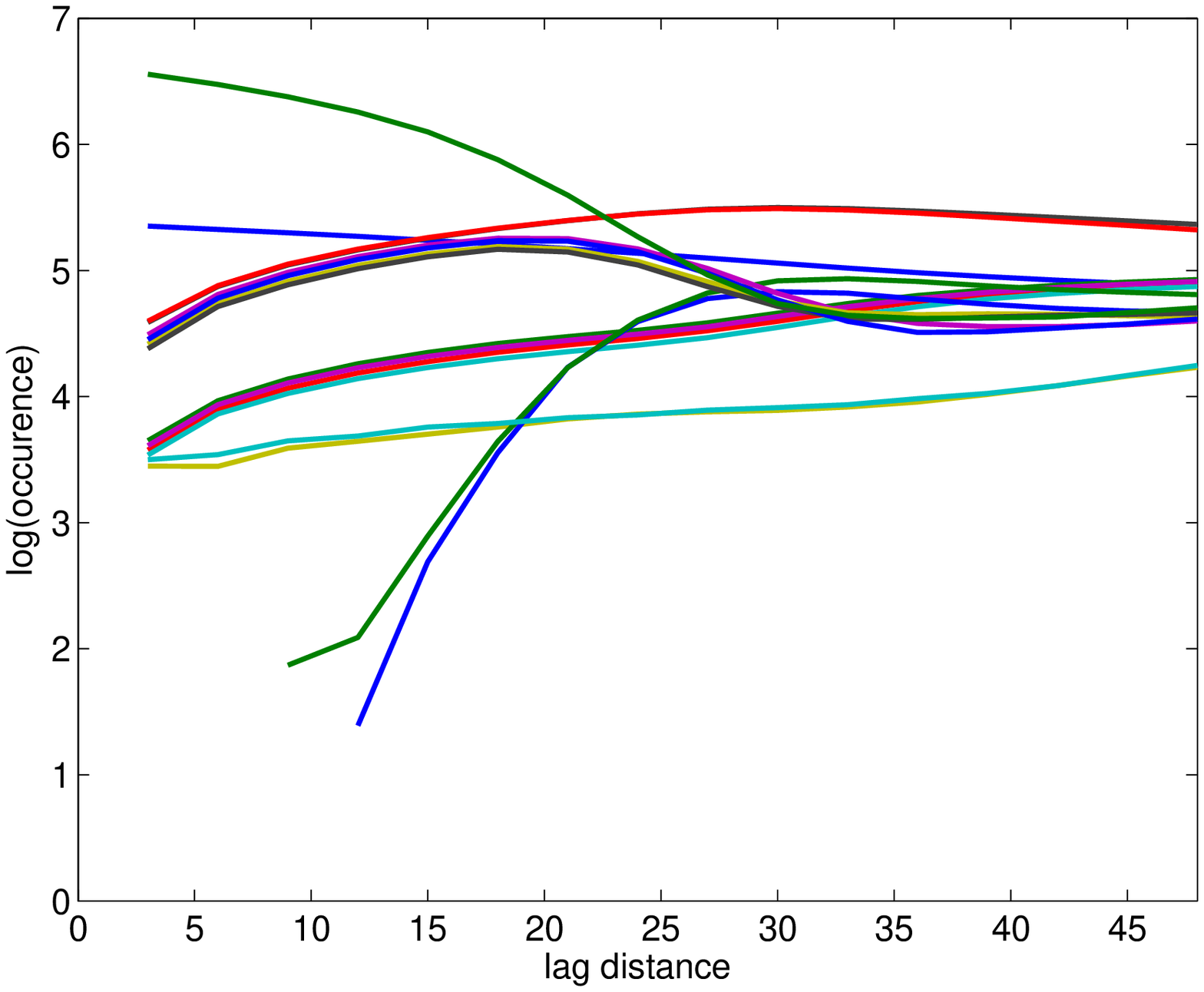}\\
\includegraphics[scale=0.36]{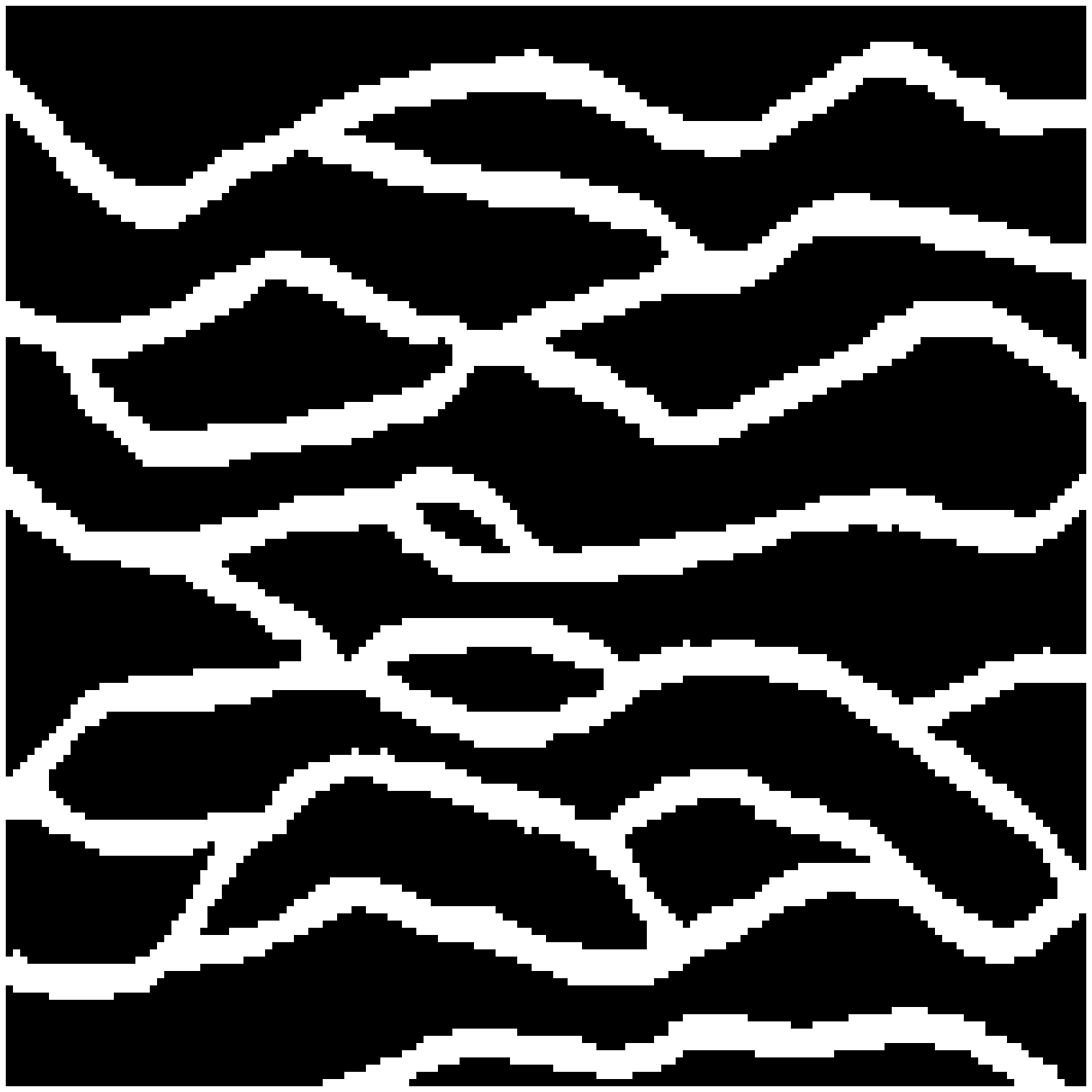}
\includegraphics[scale=0.43]{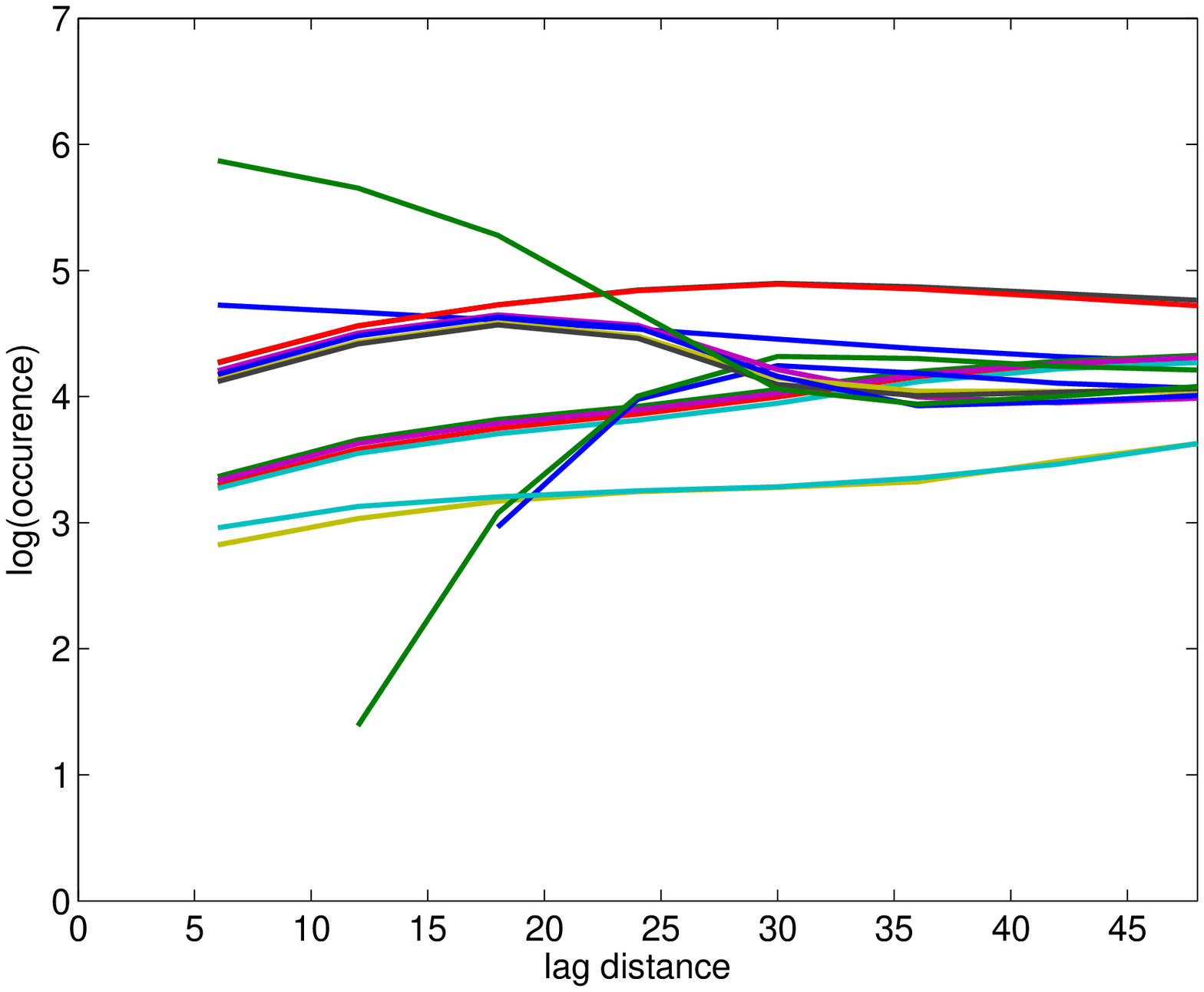}\\
\caption[Illustration of the occurrence of patterns plot and the corresponding training image. The occurrence of patterns are shown for 16 patterns of a binary 2$\times$2 configuration as a function of lag distance]{\small Illustration of the occurrence of patterns plot and the corresponding training image. The occurrence of patterns are shown for 16 patterns of a binary 2$\times$2 configuration as a function of lag distance for interval of 0 to 48. The first row represents the original map and its corresponding occurrence plot, the second row represents the upscaled map (block of 3$\times$3) and its corresponding occurrence plot, and the last row represents the upscaled map (blocks of 6$\times$6) and its corresponding occurrence plot.}
\label{fig:snesim}
\end{figure}

\section{Challenges in Utilizing FOP}
\nin In this section, an exercise is designed to understand the relationship between the training image at different resolutions, and their respective FOP profiles. The 16 patterns of $2\times2$ binary configurations are considered. The preferential patterns are scanned and detected for two different training images depicting petroleum reservoirs. The first training image is the typical channel type petroleum reservoir \citep{ss:02,yl:06} that has been considered in three maps of high, medium and low resolution. The high resolution map is identified by utilizing the original training image. The medium and low resolution maps are the result of performing upscaling on the original training image in two stages. The categorical variables are upscaled based on assigning the most common category \citep{book2}; $3\times3$ grid cells in original training image is translated to $1\times1$ grid cell in medium resolution map, and $6\times6$ grid cells in original training image is translated to $1\times1$ grid cell in low resolution map. The corresponding FOP plots are determined for lag distances changing from 1 to 48. The $y$-axis of FOP represents the occurrence of patterns in logarithmic units. The FOP values for all 16 patterns versus the lag distances are shown in Figure~\ref{fig:snesim}.

The frequency of every pattern in the corresponding training image is determined as follows: (1) the training image is scanned for a specific lag distance and data is extracted for the template configuration (i.e. 4-points template); (2) the probability of binary category is evaluated on the extracted data, i.e., $P_w$ and $P_b$. For example, for the classic training image shown in Figure~\ref{fig:snesim} at the top, $P_w$ is approximately 28\% and $P_b$ is approximately 72\% (these proportions correspond to the lag 0 in original map). Then; (3) the number of occurrence of a pattern is counted in the scanned model; (4) finally, the FOP independent of global proportion is evaluated as in Equation~\eqref{eq:max}.
If the scale of grid cell in original training image is assumed to represent 1 unit, then a 48 units lag distance in FOP covers about 1/5 of the domain in the corresponding 2-D map ($250\times250$). As can be seen in Figure~\ref{fig:snesim}, the occurrences of 16 patterns somewhat converges after relatively large lag distances. This confirms an earlier claim that the patterns with higher orderness are more probable at smaller lag distances as opposed to the more random patterns. This suggests the models tend to get more random at larger lag distances and therefore the probability of occurrences for all patterns becomes similar (FOP converges for 16 patterns when lag distance is relatively large). 

It can also be concluded that inferring information about high resolution maps from low resolution (upscaled) maps is not trivial; the information regarding small lag distance variability is not preserved in the upscaled process. Thus, to enhance the resolution of a training image, the information regarding frequency of patterns of a coarse resolution training image would not be sufficient to characterize smaller scale variability. Another observation is that some patterns do not appear until certain lag distances at higher resolution. Extrapolating such patterns from large scale information would be impossible.

\begin{figure}[t!]\centering
\includegraphics[scale=0.33]{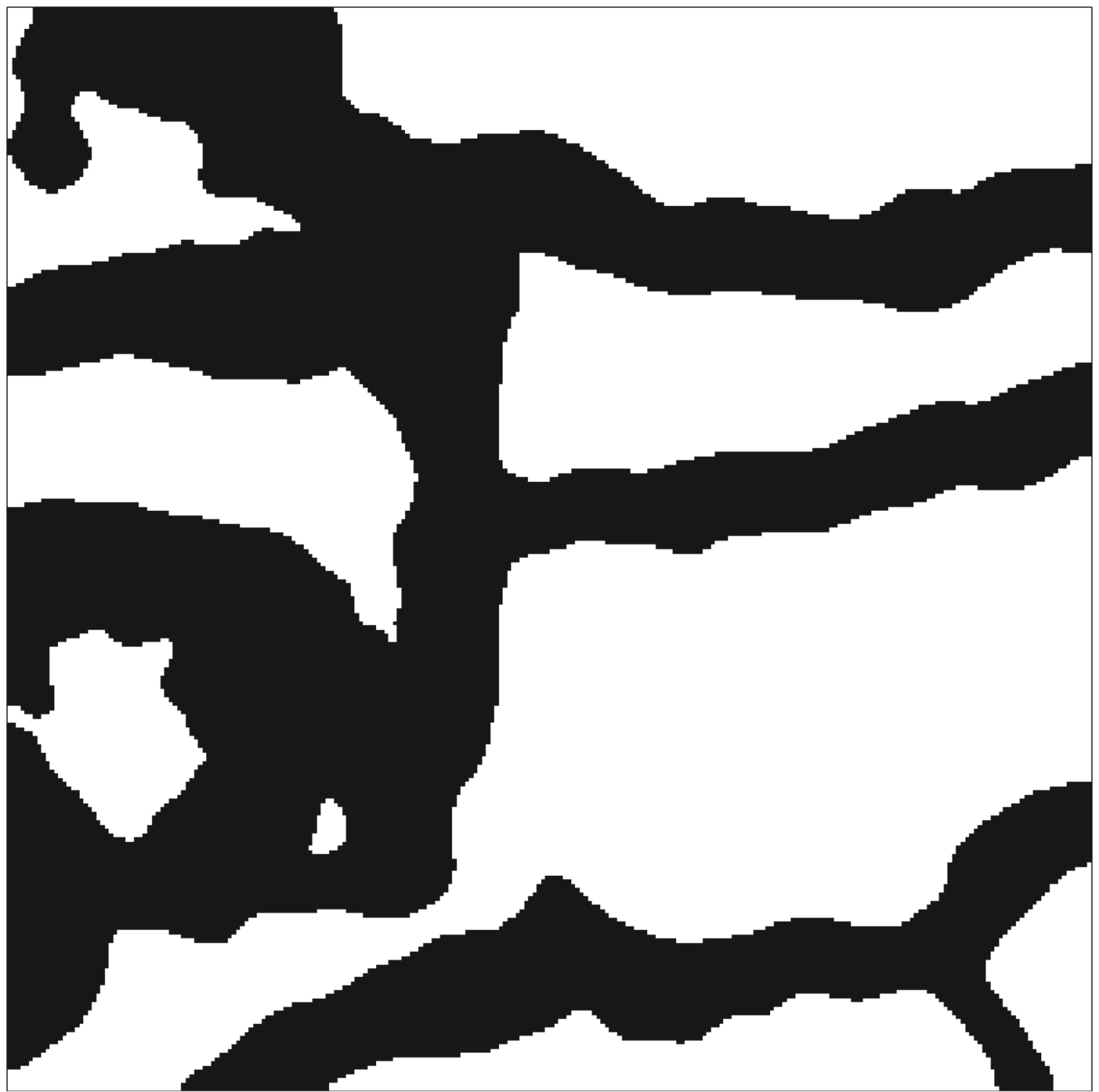}
\includegraphics[scale=0.4]{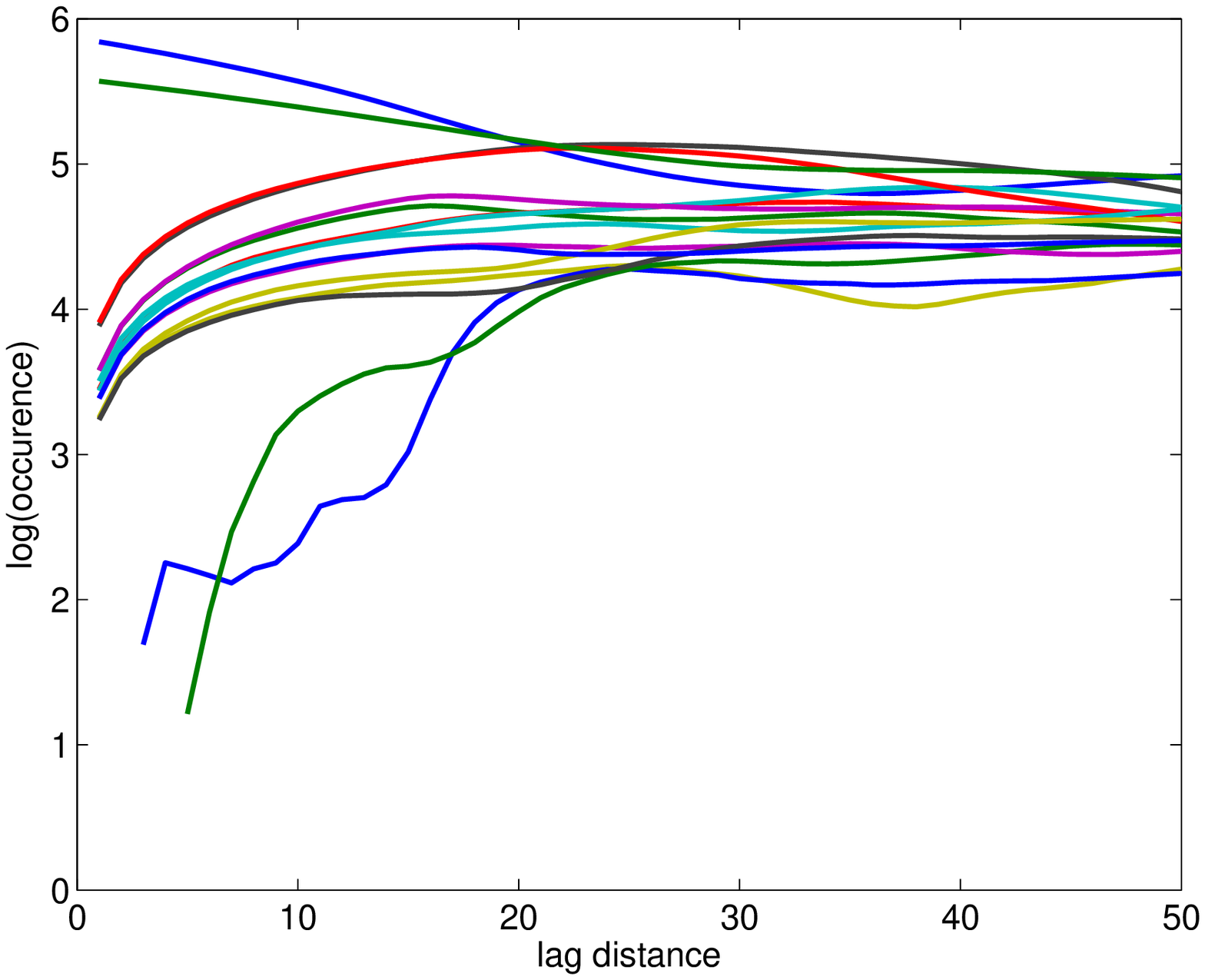}\\
\includegraphics[scale=0.33]{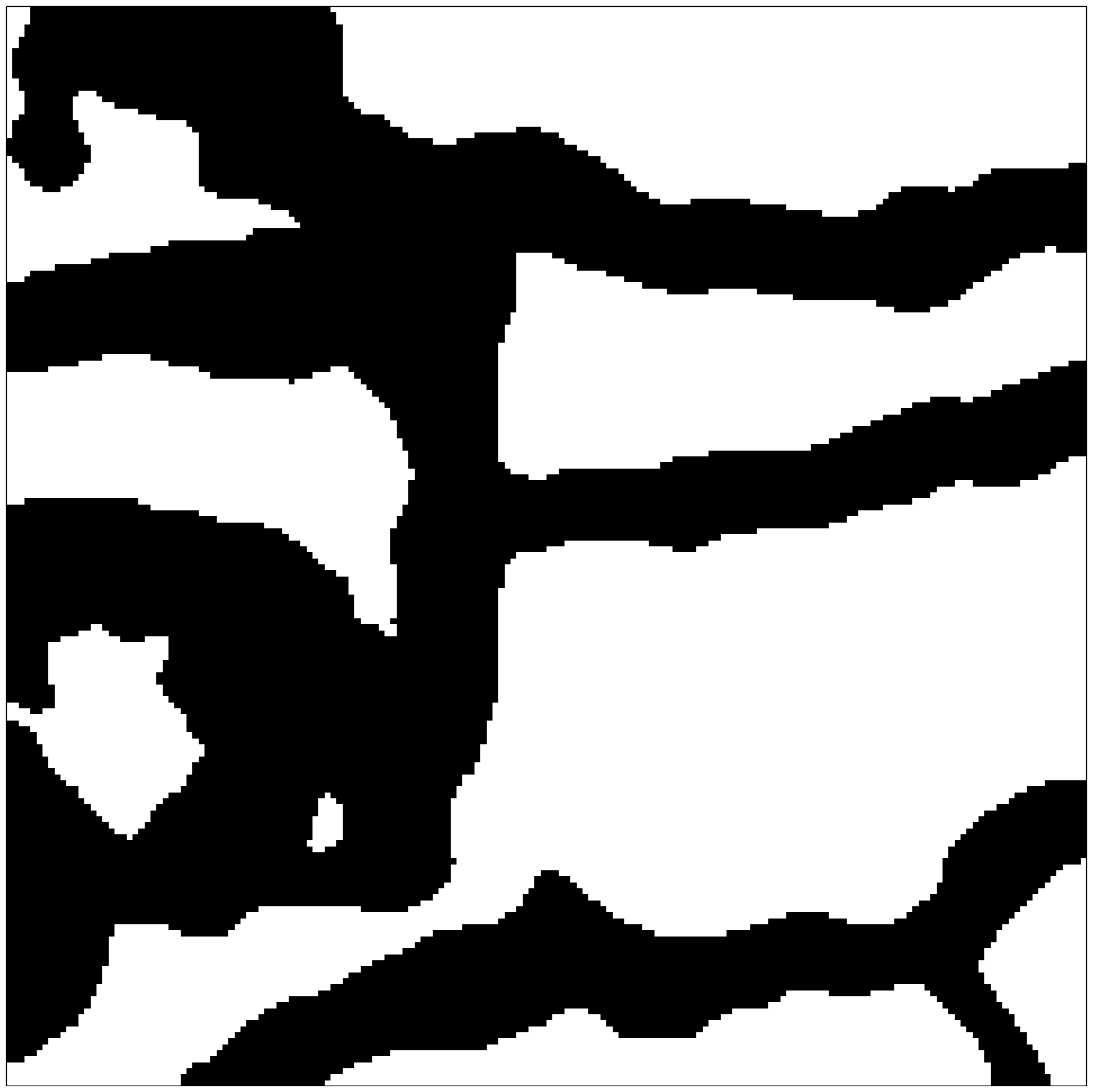}
\includegraphics[scale=0.4]{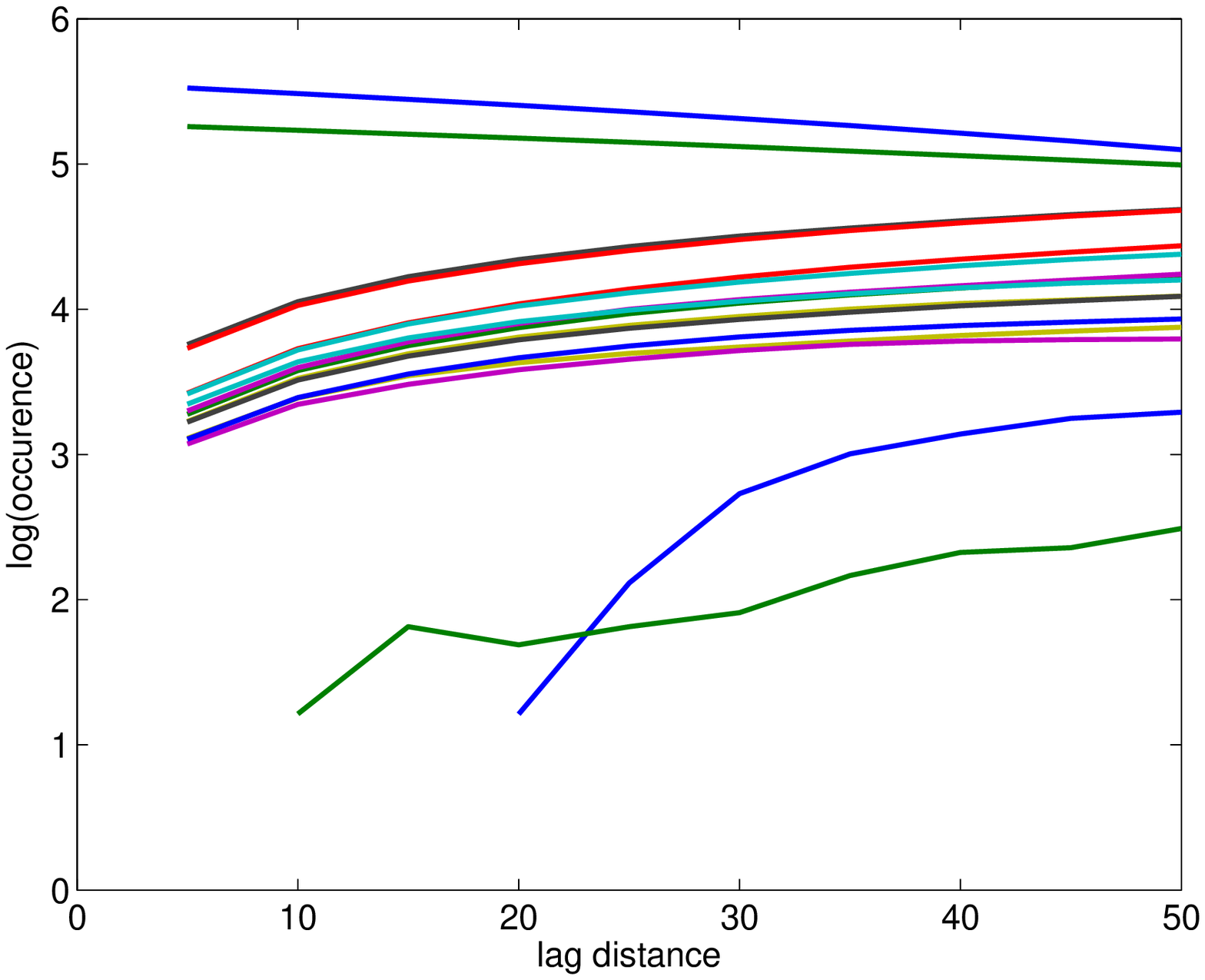}\\
\includegraphics[scale=0.33]{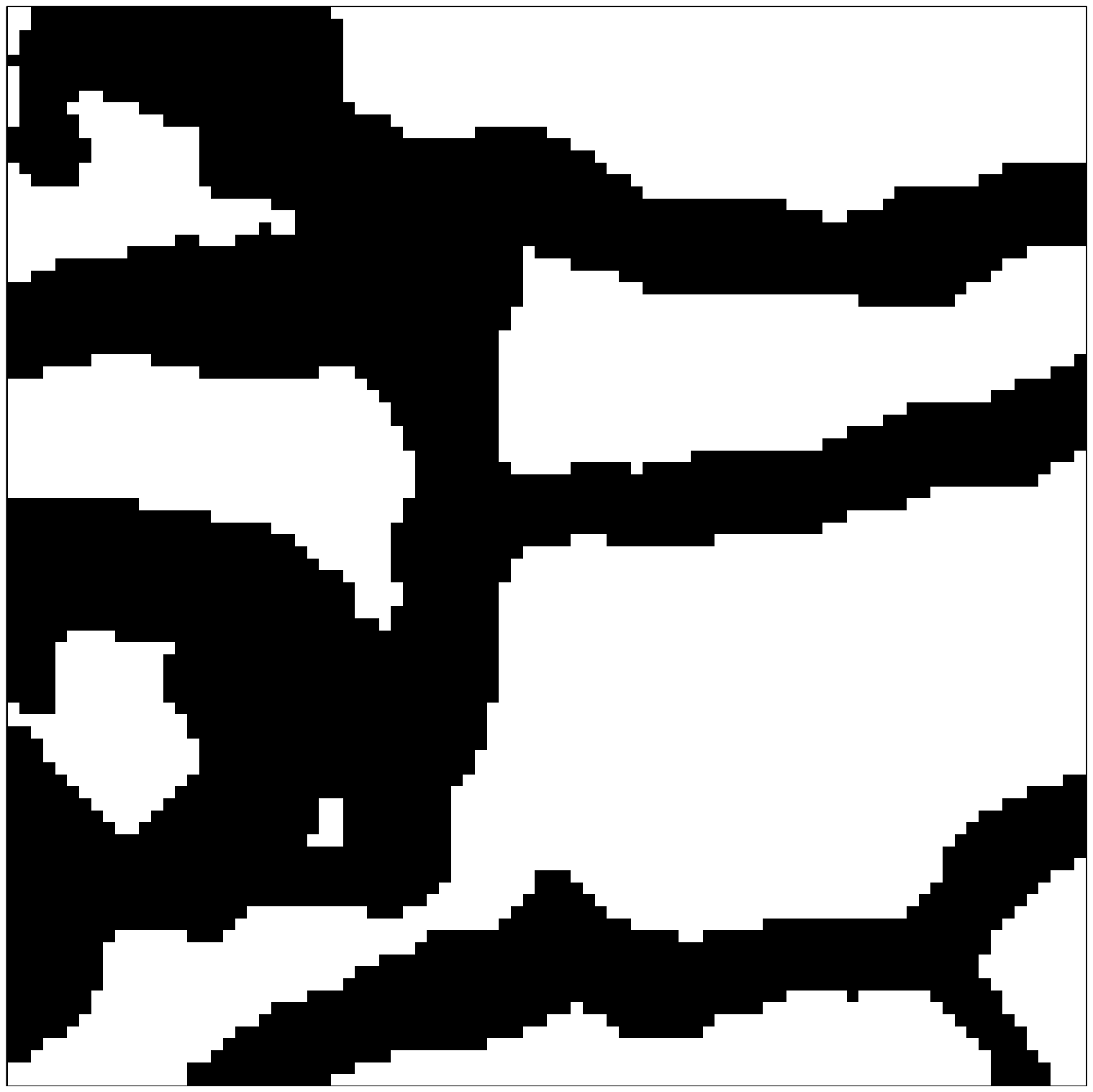}
\includegraphics[scale=0.4]{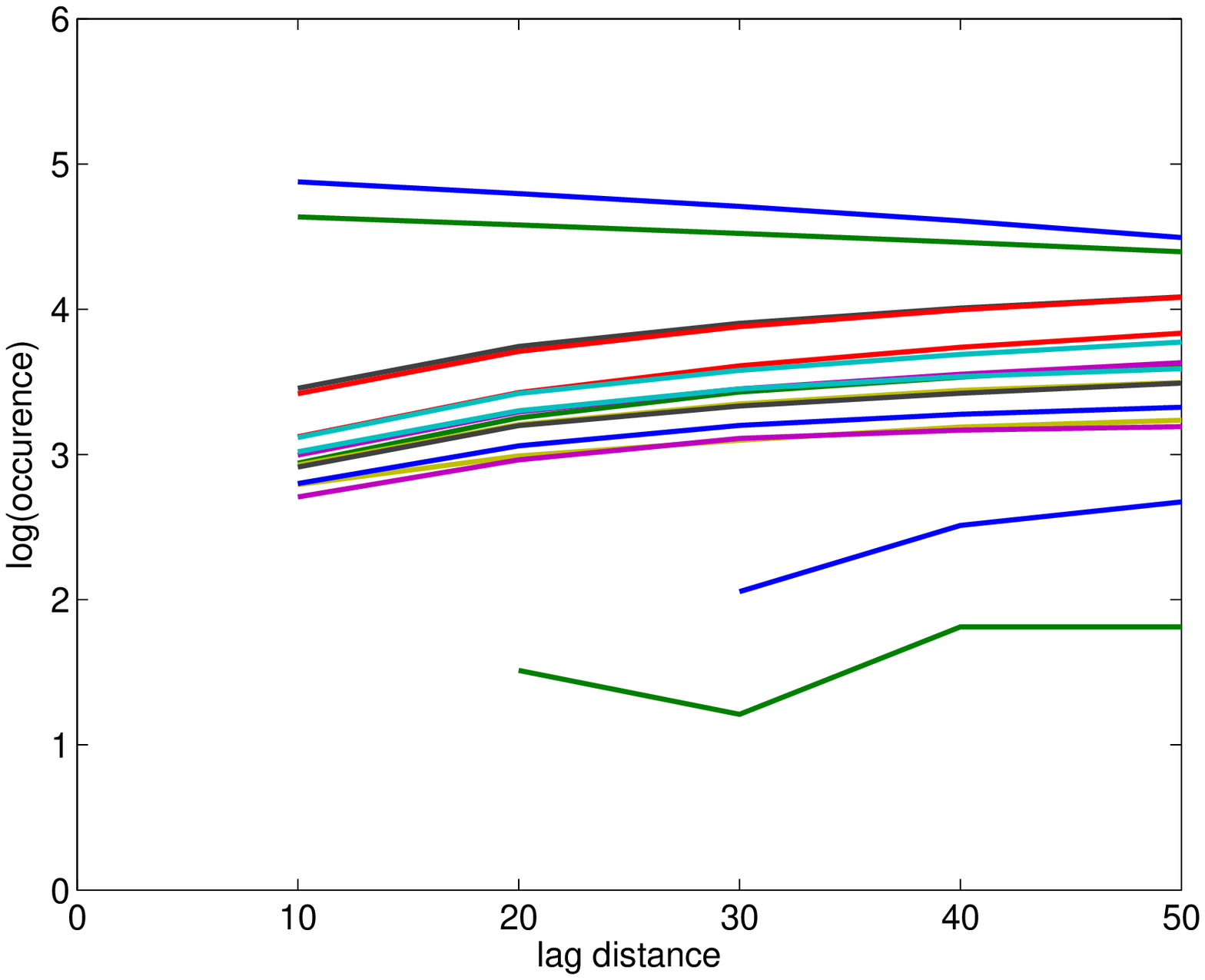}
\caption[Illustration of the occurrence of patterns plot and the corresponding training images]{\small Illustration of the occurrence of patterns plot and the corresponding training images. The occurrence of patterns are shown for 16 patterns of a binary 2$\times$2 configuration as a function of lag distance for interval of 0 to 50. The first row represents the original map and its corresponding occurrence plot, the second row represents the upscaled map (block of 5$\times$5) and its corresponding occurrence plot, and the last row represents the upscaled map (blocks of 10$\times$10) and its corresponding occurrence plot.}
\label{fig:daniel}
\end{figure}


For the sake of illustration, the same process is applied to another training image that is shown in Figure~\ref{fig:daniel}. The first map (top) is the original training image of size 256$\times$256 and its corresponding FOP is determined for up to 50 units lag distance (almost 1/5 of the domain). The second map (middle) is upscaled 5 times so that 5 units lag distance in its corresponding FOP translates to 1 unit in high resolution map. The last map (bottom) has 10 times less resolution than the original training image; less continuity could be detected at the edges.
The same conclusions are drawn for this training image. Almost after 25 units lag distance, FOP plot for the high resolution map appears to stabilize for all 16 patterns.
The next section discusses the proposed solution where the resolution of the training image is increased directly.

\section{High Resolution Training Image}
\nin The generation of high resolution training images is studied for specific types of deposits with smooth features like channels. 
Image processing techniques could be implemented on a training image in order to enhance its resolution.
Most image resolution enhancement techniques suffer from undesired smoothness at high frequency regions of the image. The high frequency regions are often the edges that are essential to people's visual perception. There is a different aspect to edges in geological modeling; features of natural phenomena more often have rounded and smooth edges. For example, specific types of fluvial channels, including straight, meandering, and anastomosis present smooth features due to erosion during deposition processes \citep{gn:09}.
The smoothing aspect of interpolation techniques could therefore be an advantage in the process of enhancing the resolution of the training image. 

Well-known and simple interpolation techniques include nearest neighbor, bilinear, bicubic and sinc interpolations (bi refers to 2-D image).
The nearest neighbor interpolation simply replicates the pixels to estimate the new grid points which results in blocky edges. Bilinear method employs a 2$\times$2 (4 pixels) neighborhood pixels and average them to define new grid cells. Bicubic interpolation employs 4$\times$4 (16 pixels) surrounding pixels and averages for the new location. The results of averaging methods are typically improved when some form of weighting is considered.
The interpolation techniques are discussed considering 2 dimensional maps. The sinc method applies interpolation in the frequency domain. Utilizing more surrounding pixels at the stage of averaging could lead to a better quality high resolution image.
The spline interpolation techniques tend to preserve the low frequency content of image and lose the high frequency content \citep{hh:78}. This could result in artifacts in regions where the map contains more detailed information. Recently, research in image resolution enhancement is directed towards the preservation of high frequency content. These methods include nonlinear techniques, interpolation techniques in frequency domain, and techniques that utilize the image geometry.

\begin{figure}[t!]\centering
\includegraphics[scale=0.45]{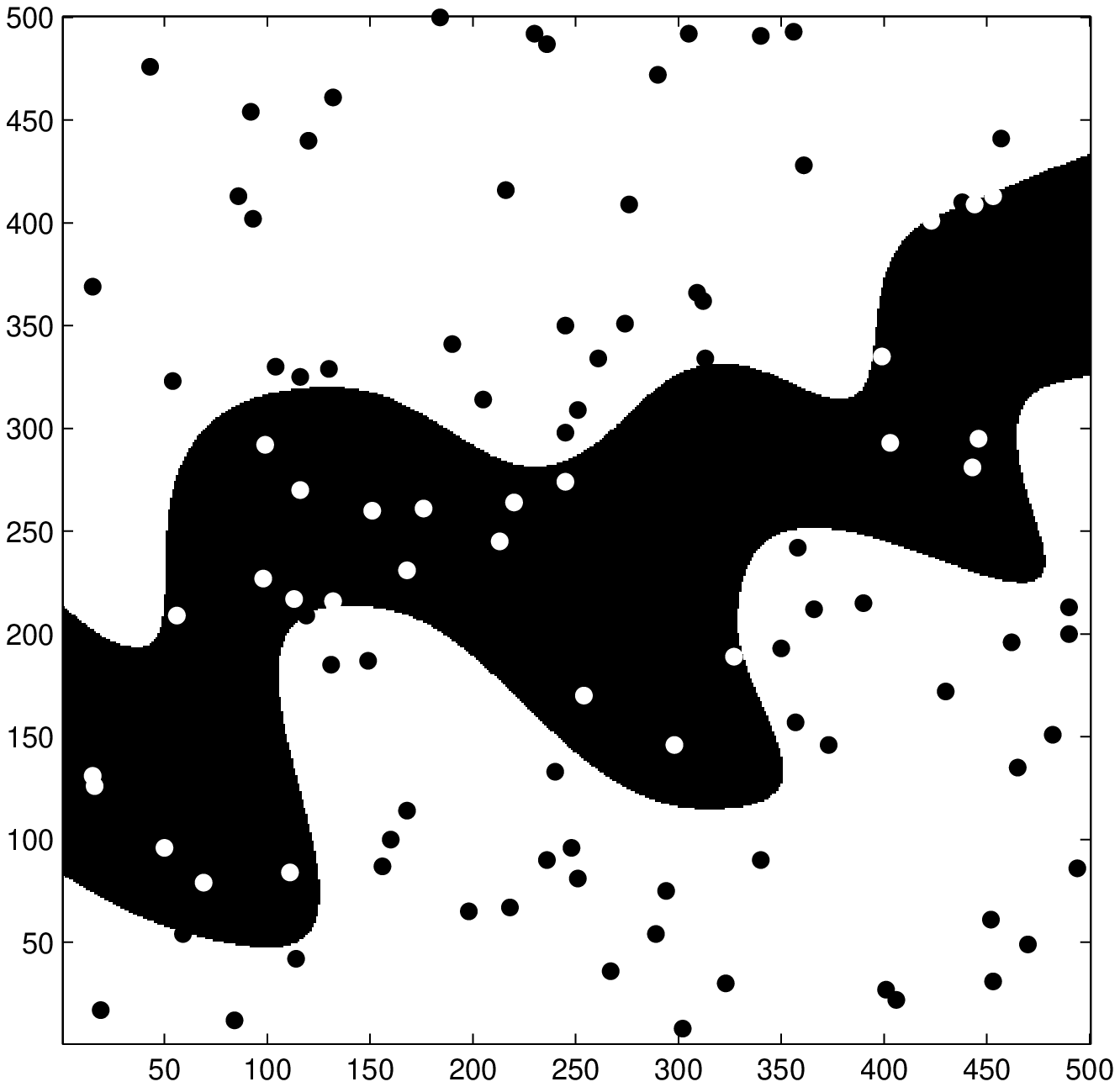}
\includegraphics[scale=0.45]{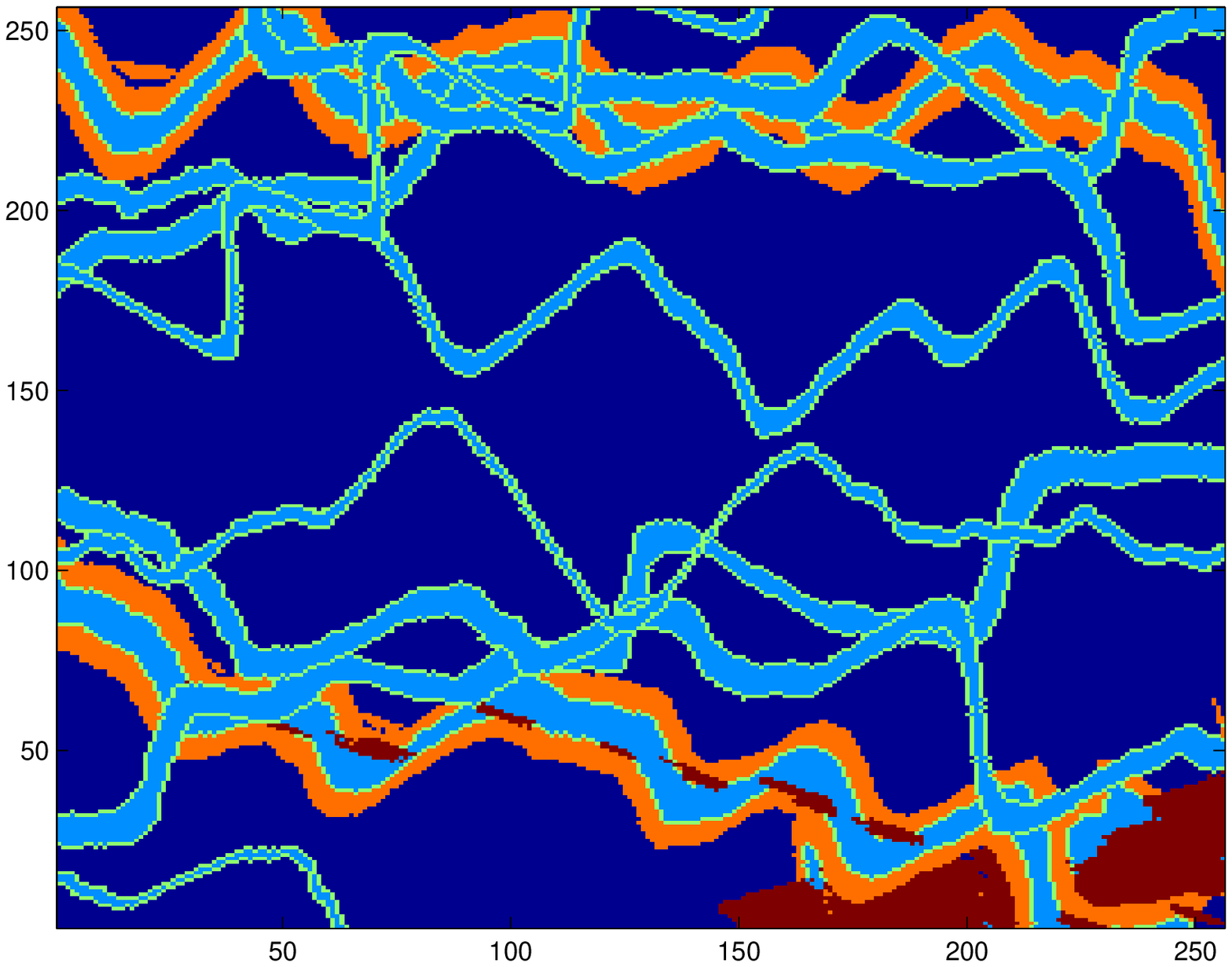}
\caption{\small The left image is a binary channel depiction of size 500$\times$500 with 100 sampled for conditioning realizations. The map on the right represents a training image of 256$\times$256 and cell size of 8 containing 5 categories of Crevasse like reservoir.}
\label{fig:orig}
\end{figure}

These interpolation methods along with the signed distance function (DF) kriging \citep{mer:13} are applied to generate high resolution 2-D maps. The signed distance function is to assign every cell by its shortest Euclidean distance to another rock type.
The distance of a given point $\mb{x}$ from the boundary of a set $\Omega$ is defined by

\begin{align*}
f(\mb{x}) = \left\{
        \begin{array}{ll}
          d(\mb{x},\Omega^c)   \;\;\;\;\;  &\tx{if}~~\mb{x} \in \Omega \\
        -d(\mb{x},\Omega)   \;\;\;\;\;  &\tx{if}~~\mb{x} \in \Omega^c
        \end{array} \right.
\end{align*} where $d(\mb{x}, \Omega)=\inf_{y\in \Omega}d(\mb{x}, y)$ which infimum indicates the greatest lower bound, that is, shortest distance to the boundary. In the case of binary map, the DF map would simply include the positive (inside), negative (outside) and zero (boundary) values. In terms of non-binary maps, the signed DF map is evaluated sequentially by considering every rock type as the main object and outside the object as the other rock type. Assigning the positive value to data within the object, the maximum distance for every cell over all DF maps is considered as the final value of signed distance in the final DF map of the non-binary map. The distance values are then considered as the available data to kriging process in order to generate data at higher resolution.

Two examples are designed to demonstrate the performance of interpolation tools generating high resolution training images. The first one is a binary depiction of a channel which is sampled and simulated at coarse resolution. Interpolation techniques are then utilized to construct the high channel at the original scale from the low resolution. In the second example, a more complex training image containing 5 categories is considered (see Figure~\ref{fig:orig}). The coarse resolution maps---in both examples---are constructed by extracting every forth pixel in both directions uniformly. 

\subsection{Example: Binary Training Image}
\begin{figure}[t!]\centering
\includegraphics[scale=0.37]{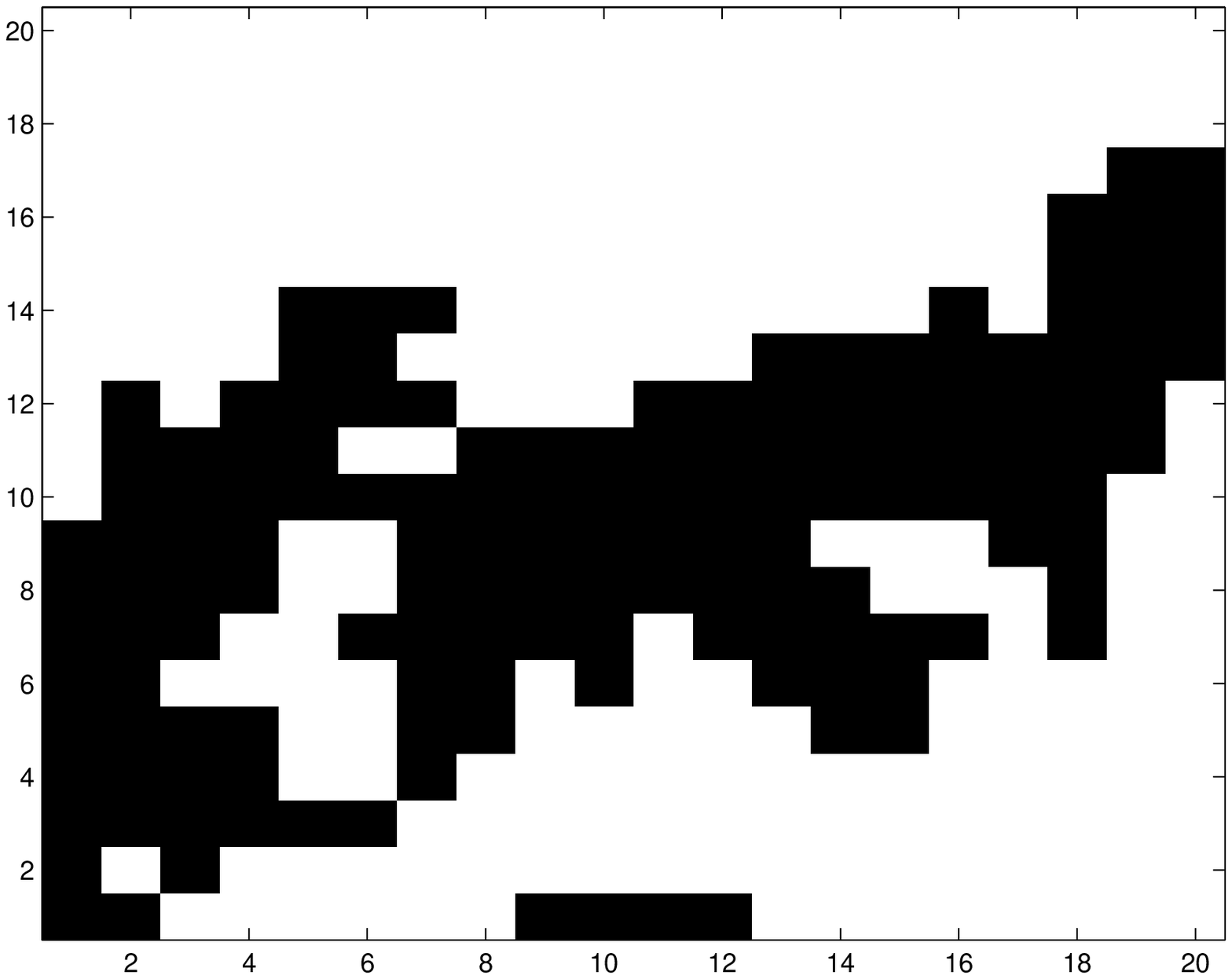}
{\includegraphics[scale=0.37]{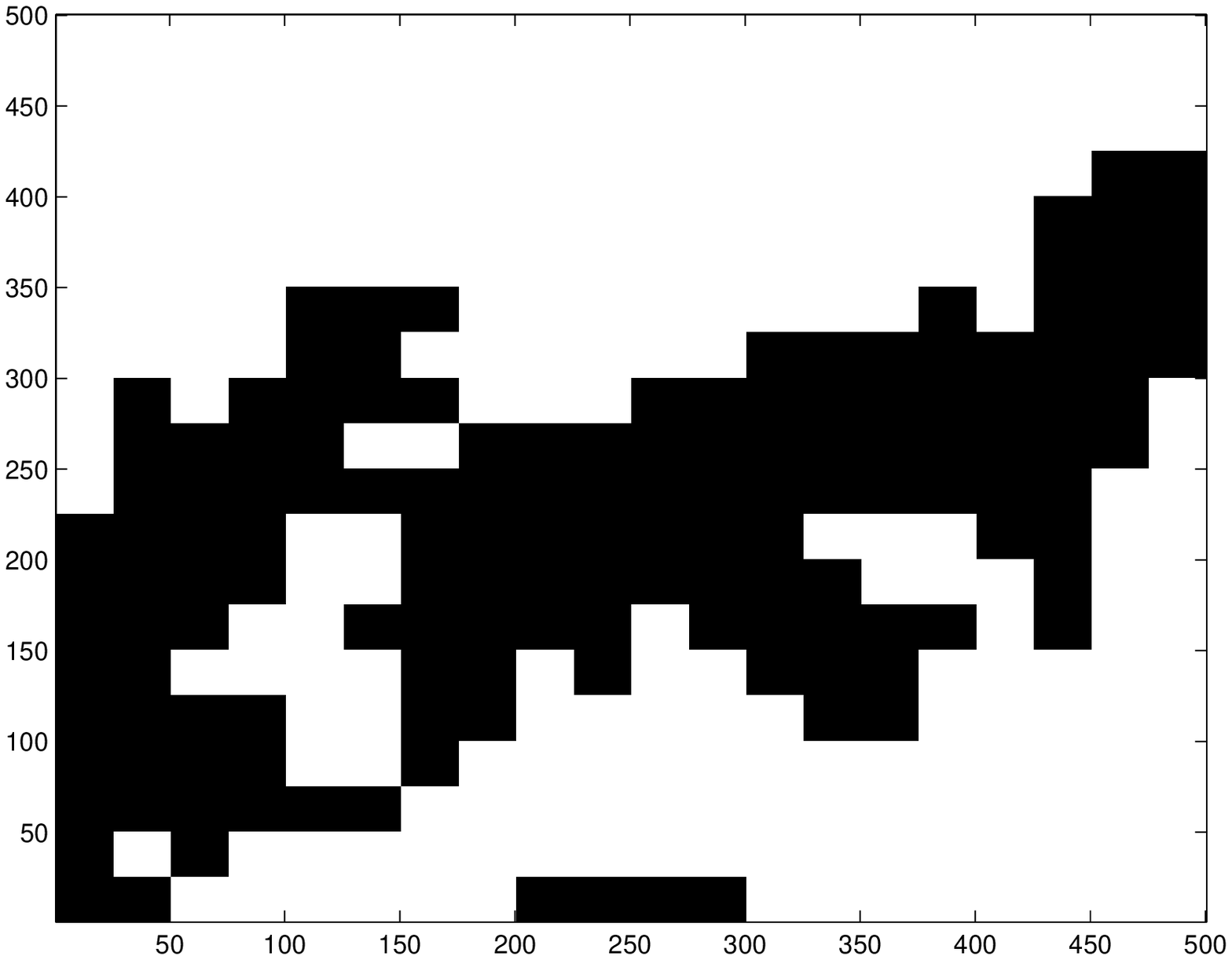}}\\
\includegraphics[scale=0.37]{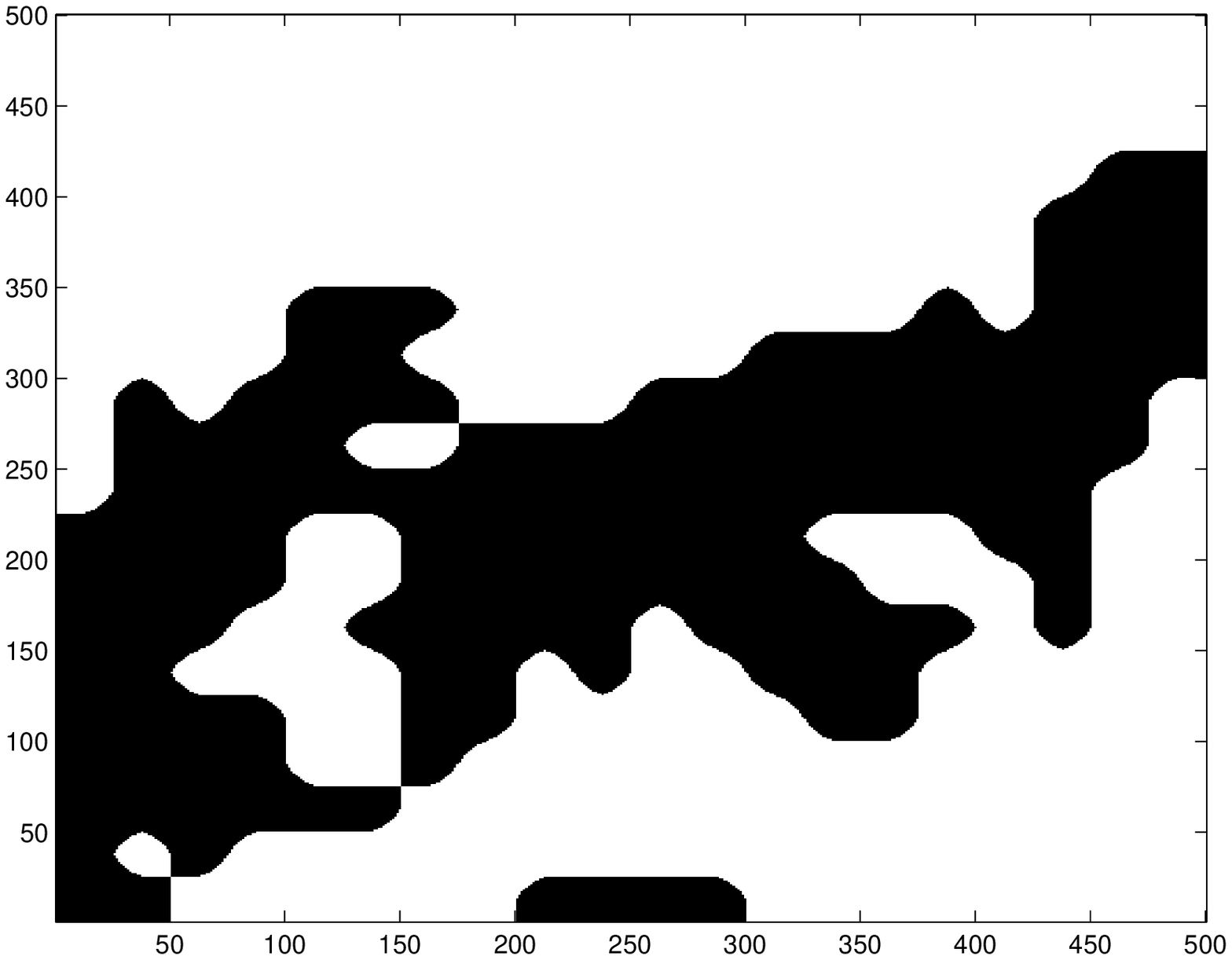}
{\includegraphics[scale=0.37]{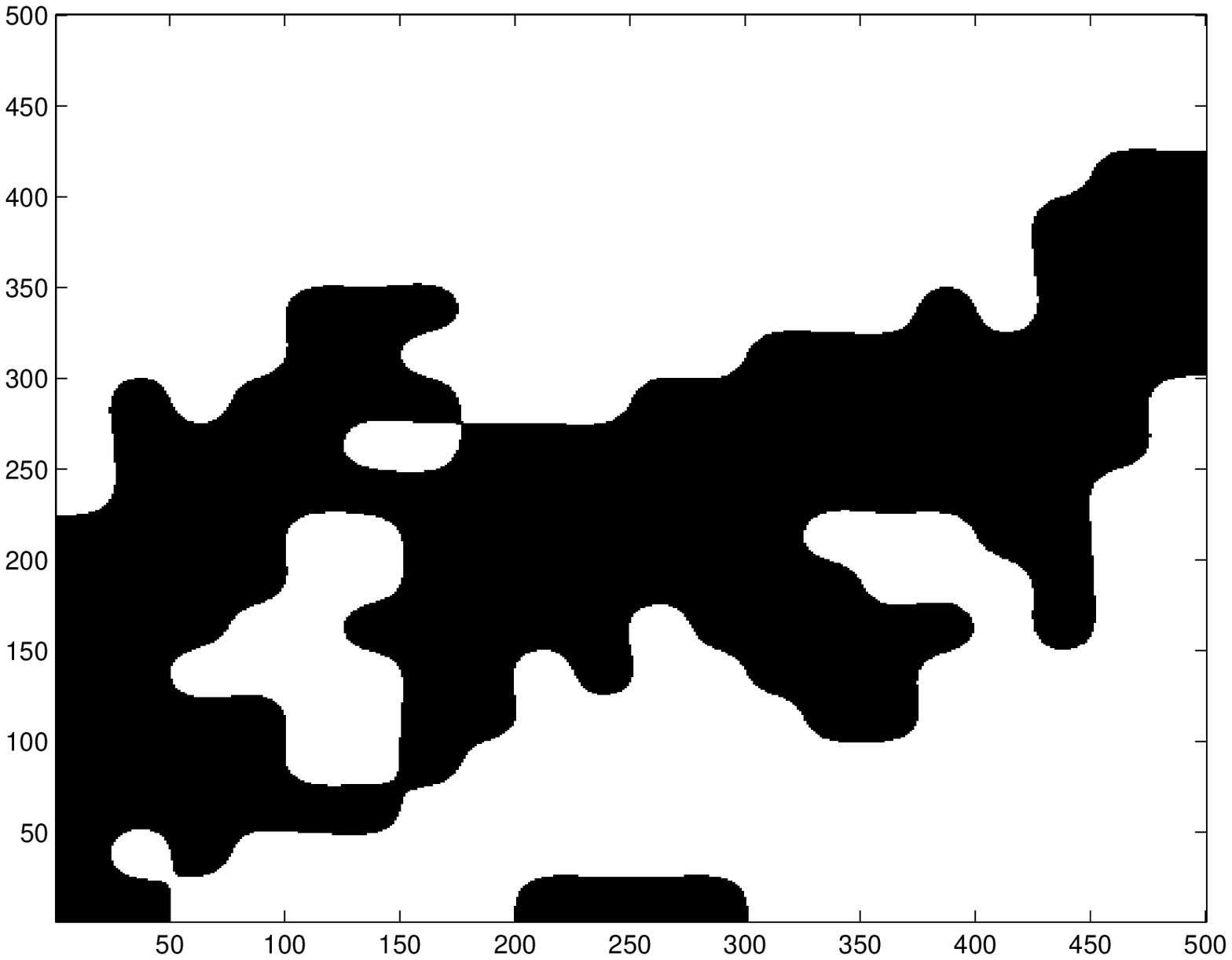}}\\
\includegraphics[scale=0.37]{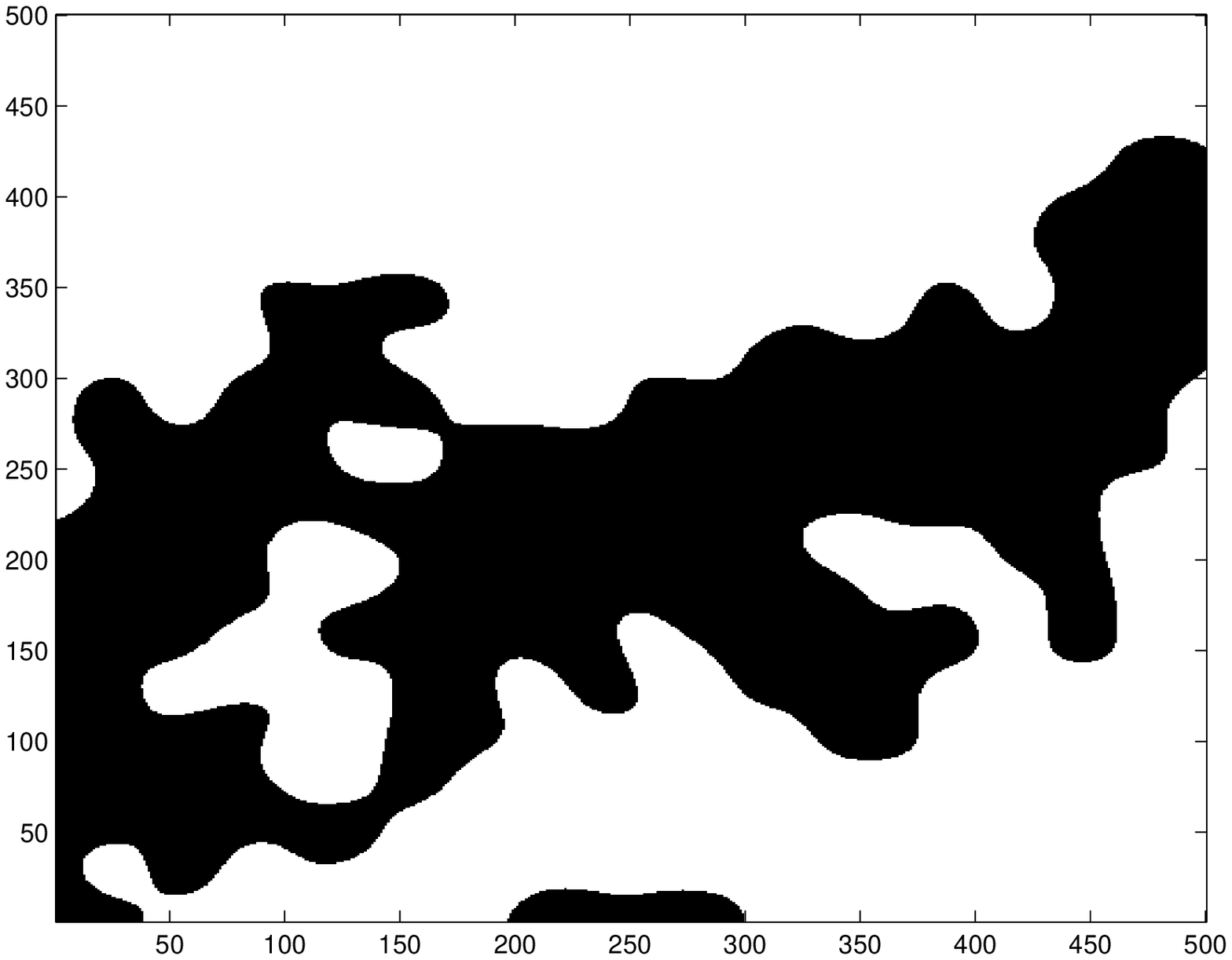}
{\includegraphics[scale=0.37]{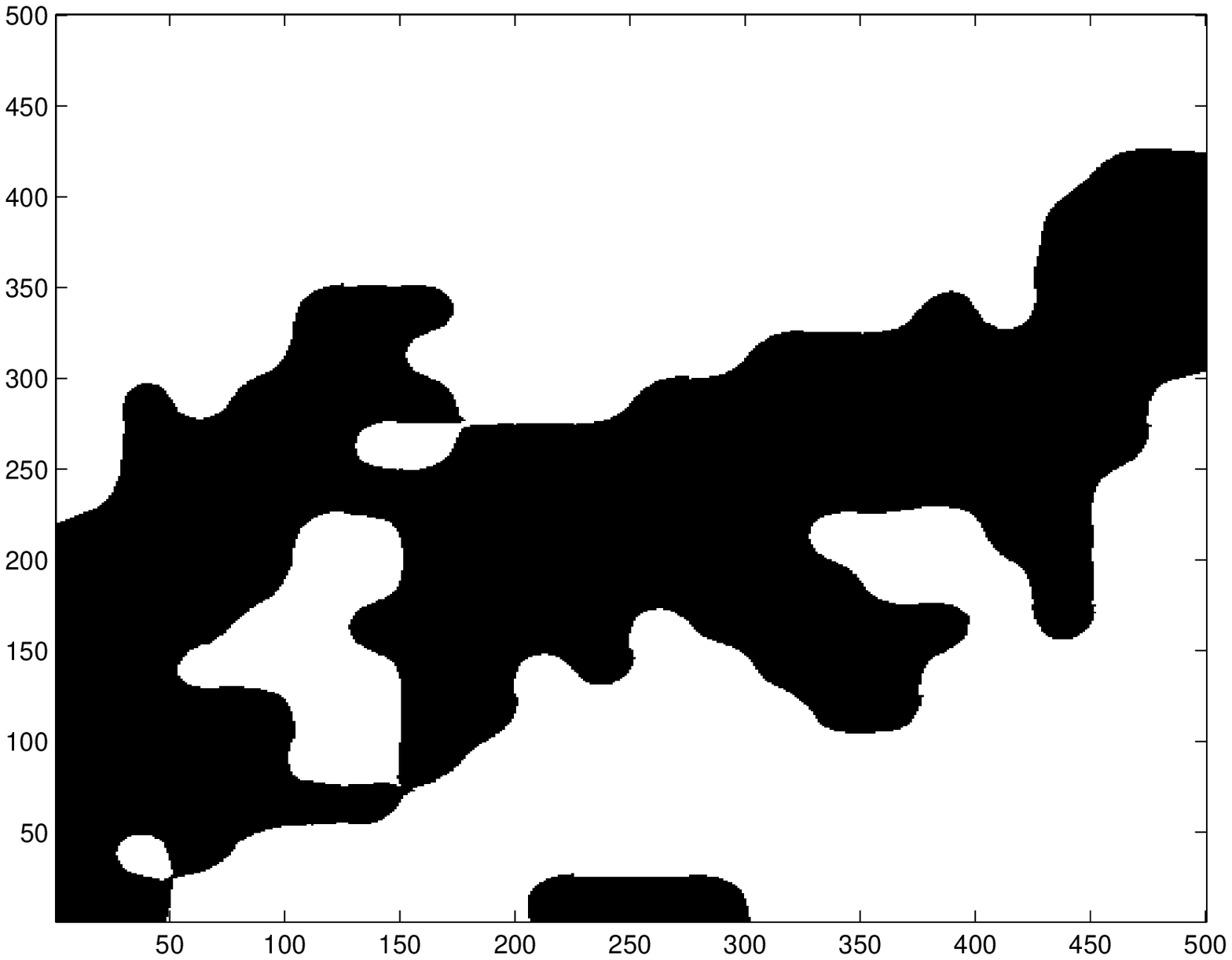}}\\
\caption[The figures should be read from left to right and top to bottom: The coarse map realization that is simulated at the coarse resolution conditioning 100 samples data of binary channel]{\small The figures should be read from left to right and top to bottom: The coarse map realization that is simulated at the coarse resolution conditioning 100 samples data of binary channel in Figure~\ref{fig:orig}, its resolution enhanced once using nearest neighbor interpolation (map in top right), using bilinear interpolation (map in middle left), using bicubic interpolation (map in middle right), using sinc function interpolation (map in bottom right), and using distance function kriging (map in bottom left).}
\label{fig:RLZ1}
\end{figure}

\begin{figure}[t!]\centering
\includegraphics[scale=0.27]{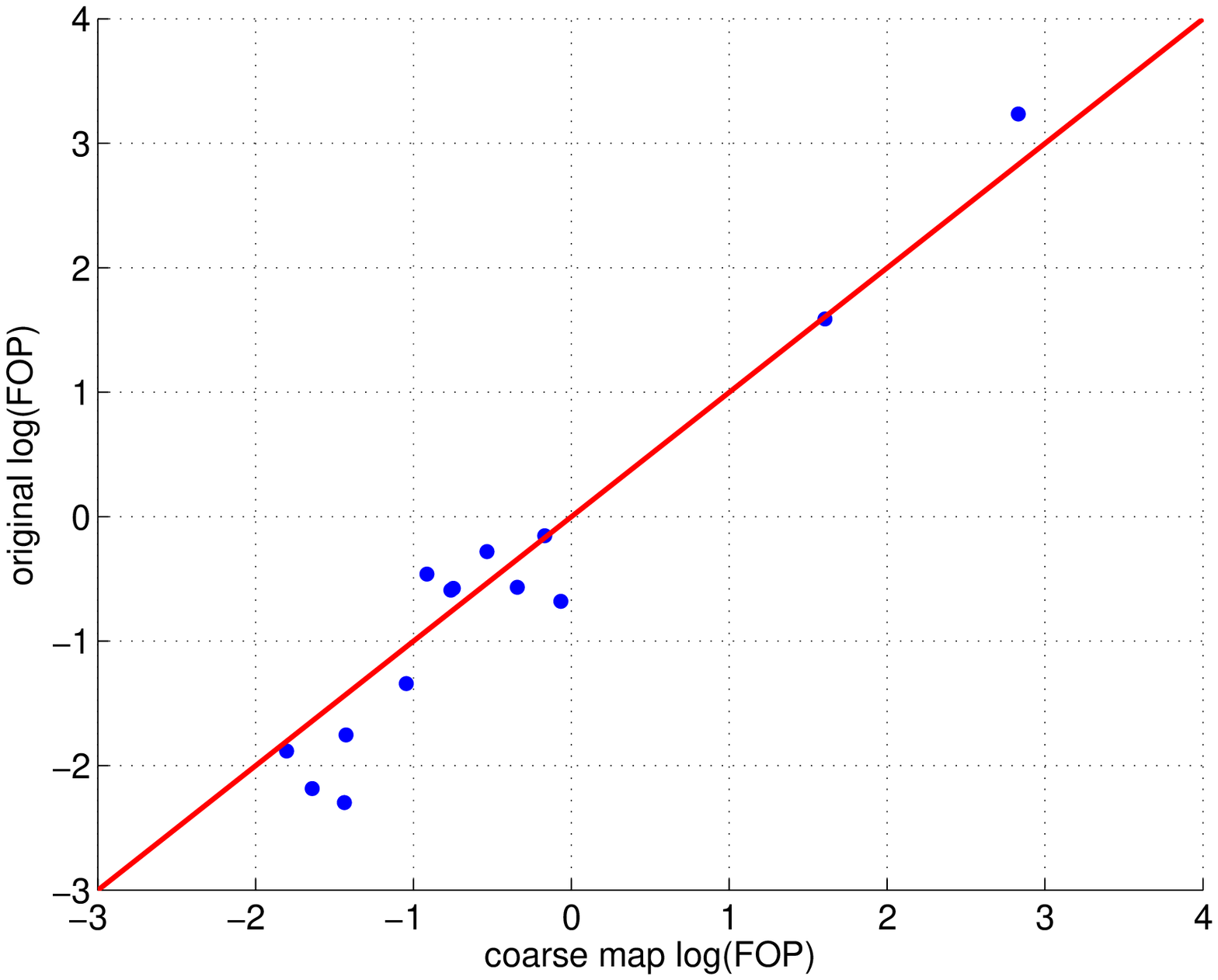}
\includegraphics[scale=0.27]{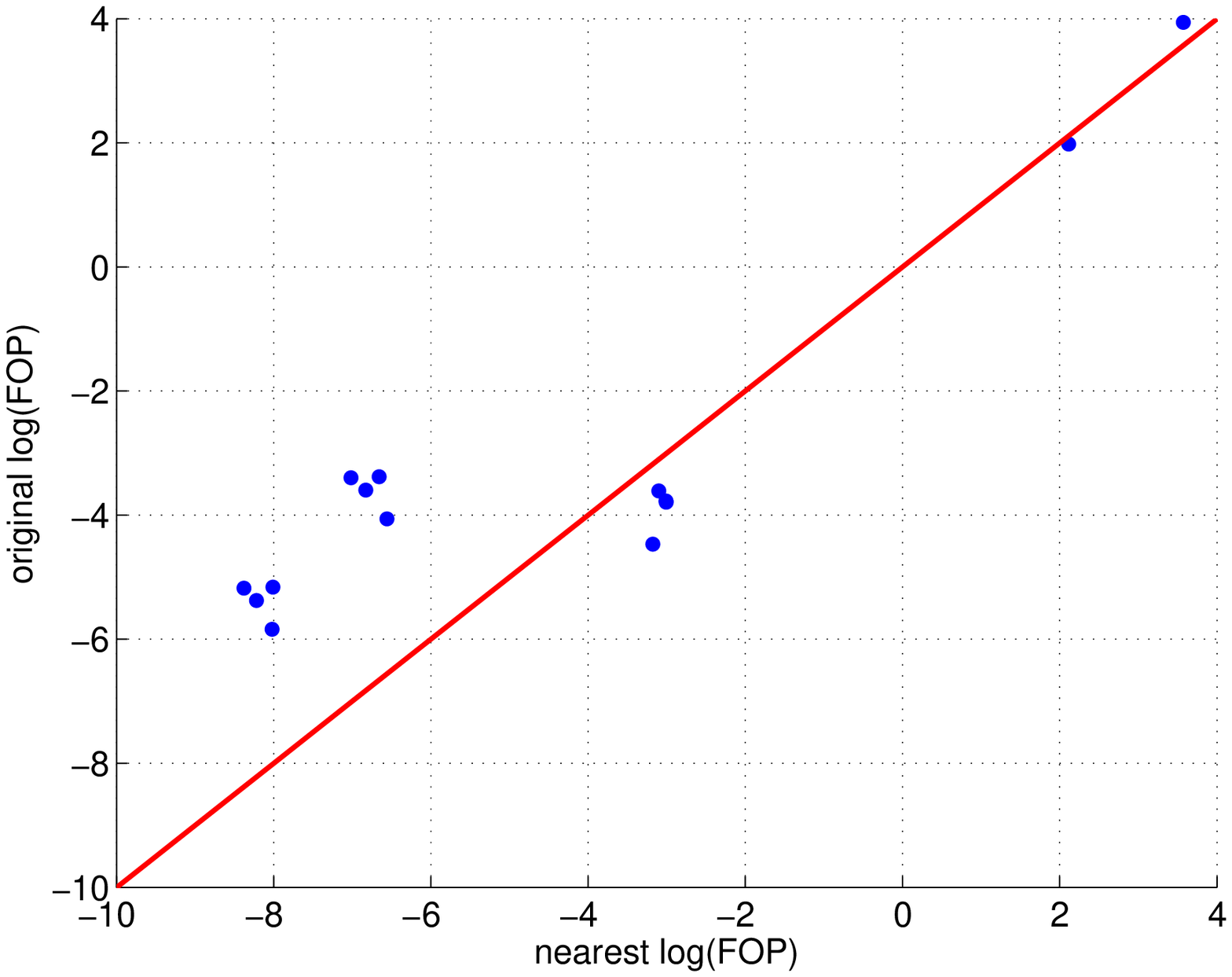}
\includegraphics[scale=0.27]{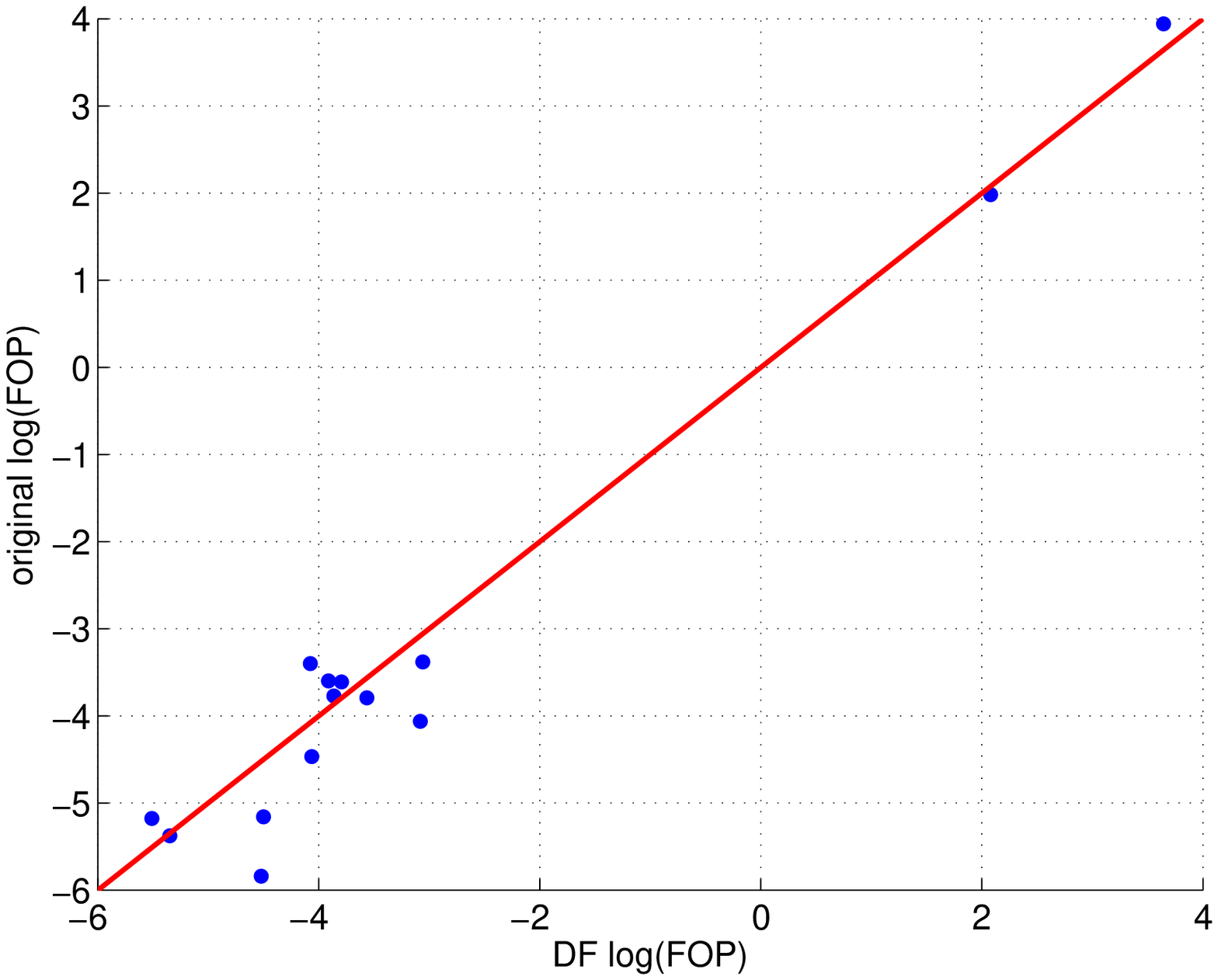}
\includegraphics[scale=0.27]{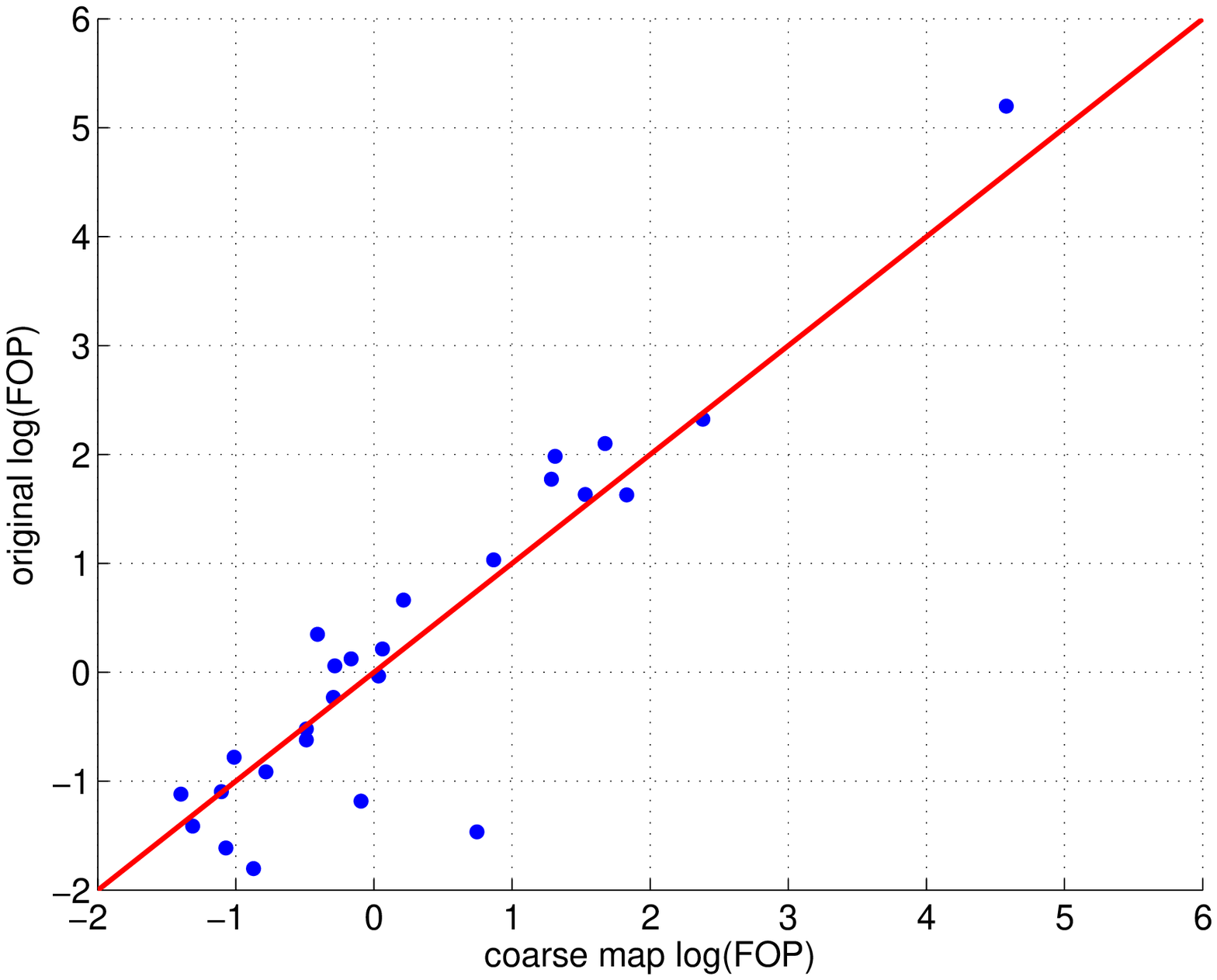}
\includegraphics[scale=0.27]{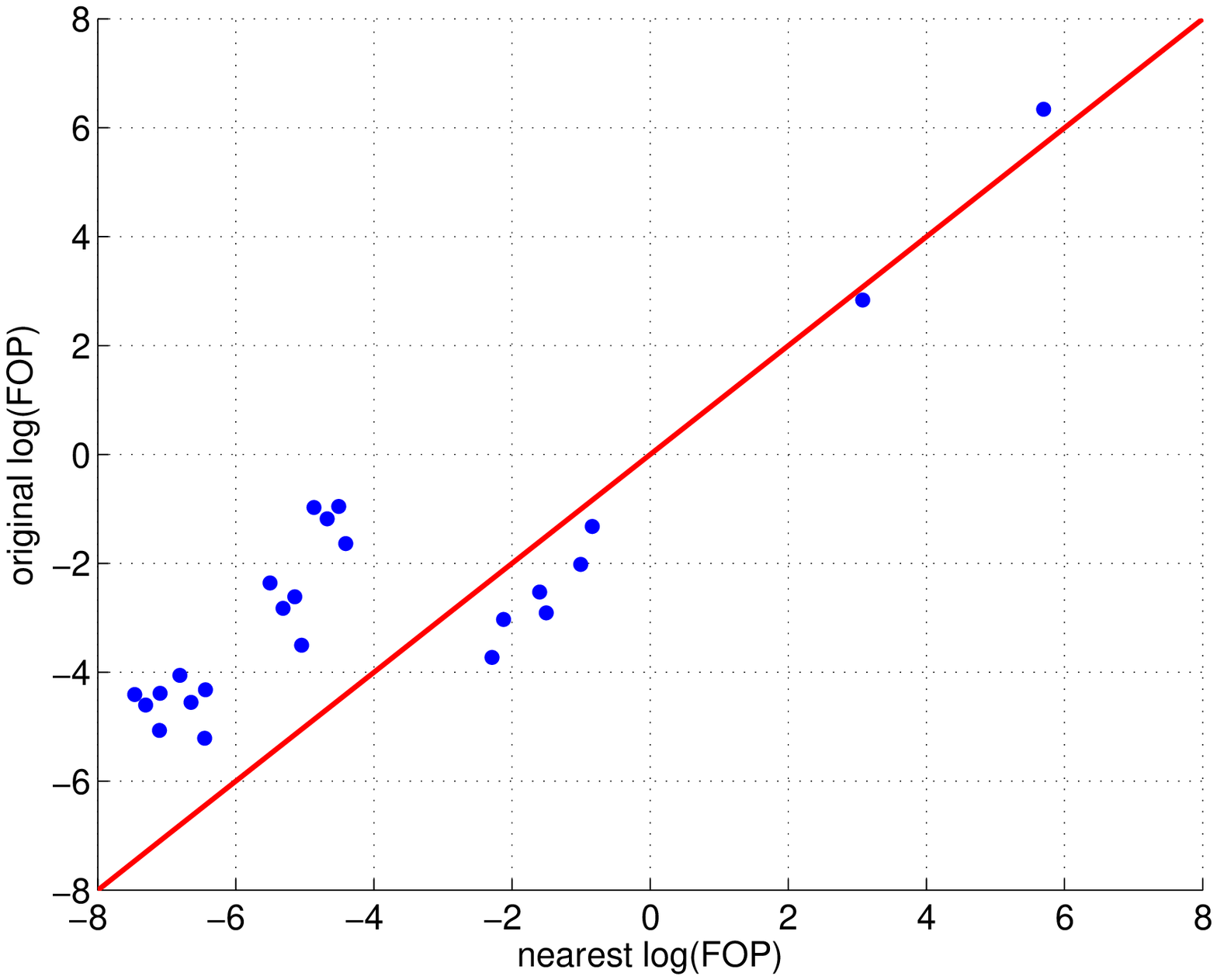}
\includegraphics[scale=0.27]{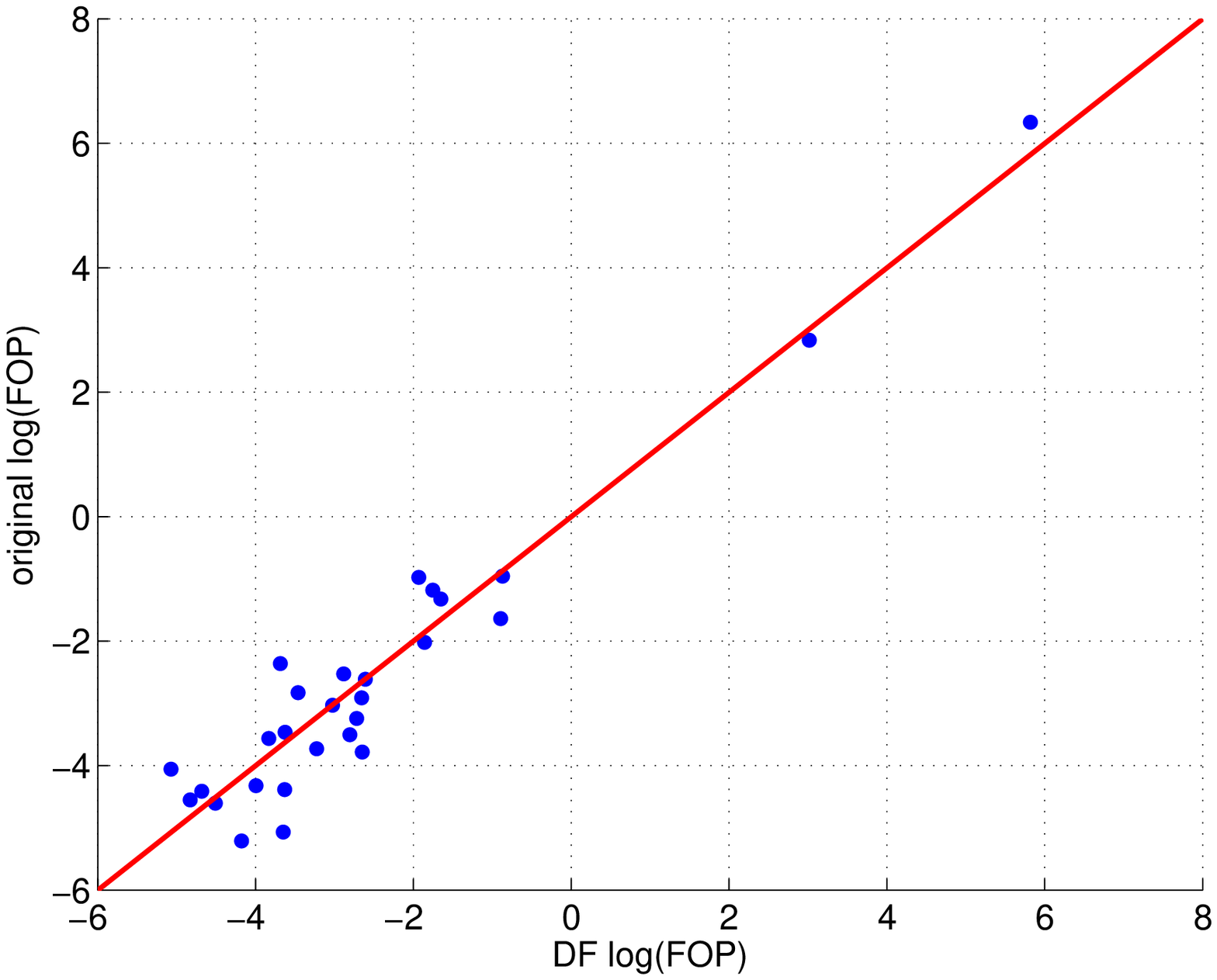}
\includegraphics[scale=0.27]{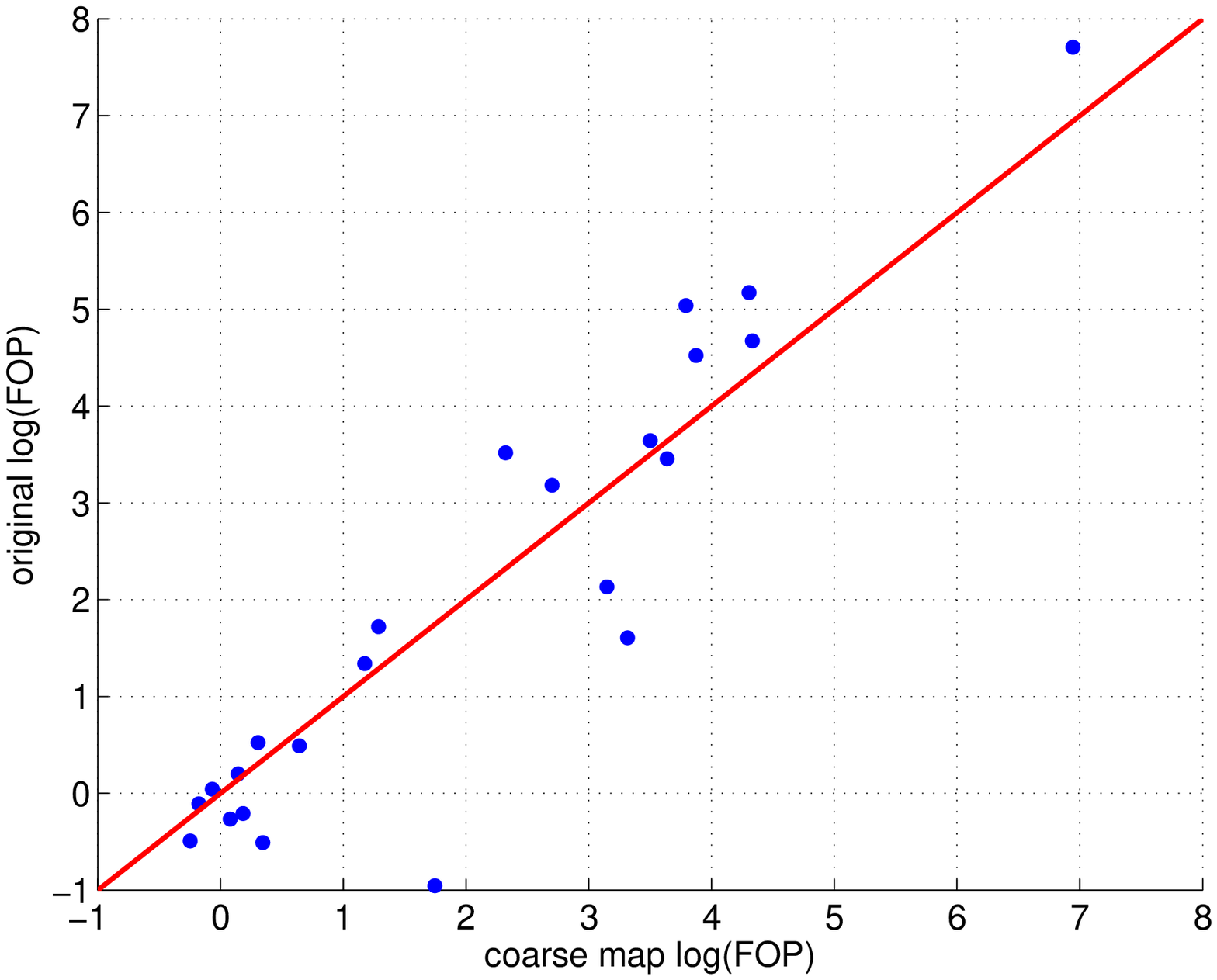}
\includegraphics[scale=0.27]{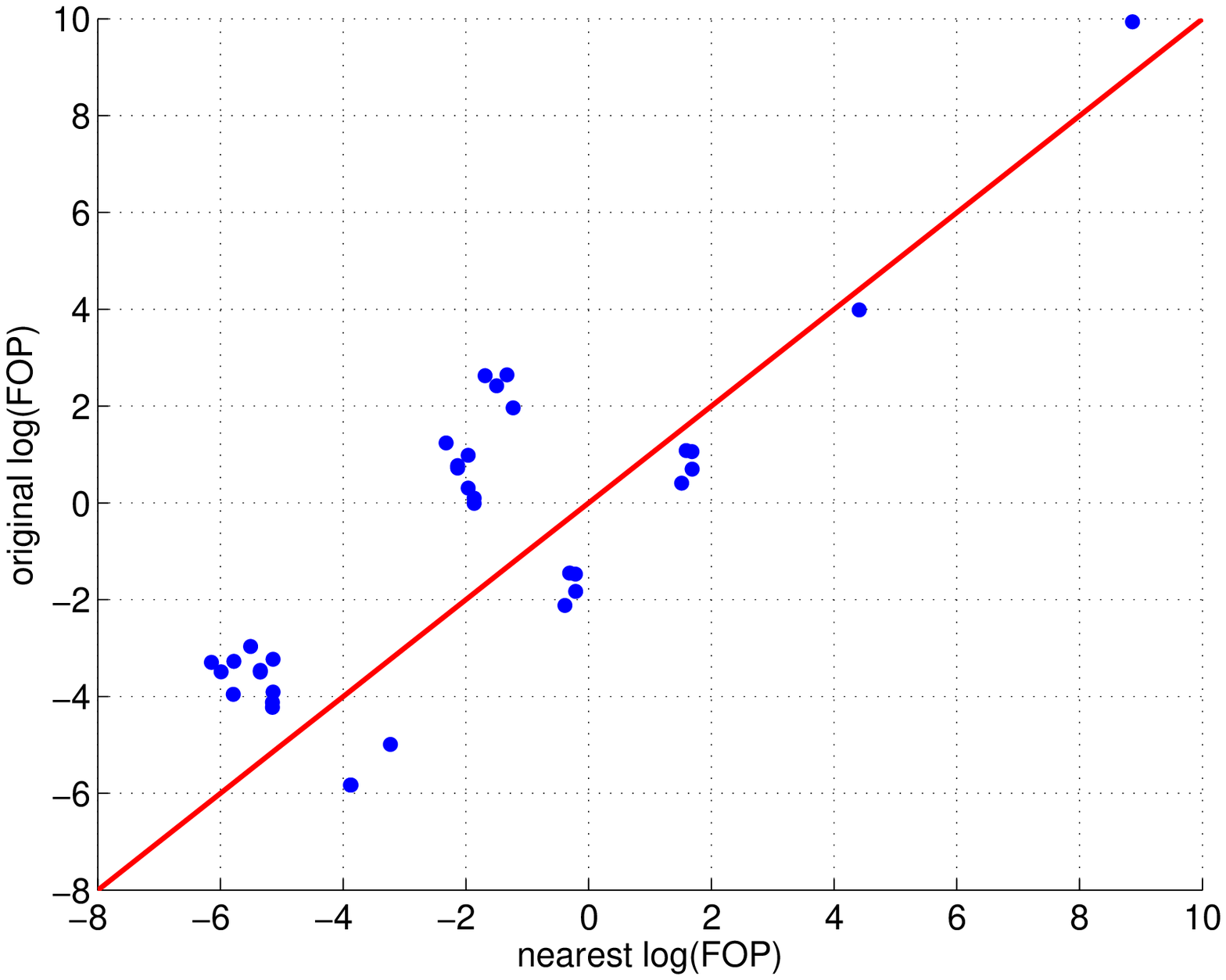}
\includegraphics[scale=0.27]{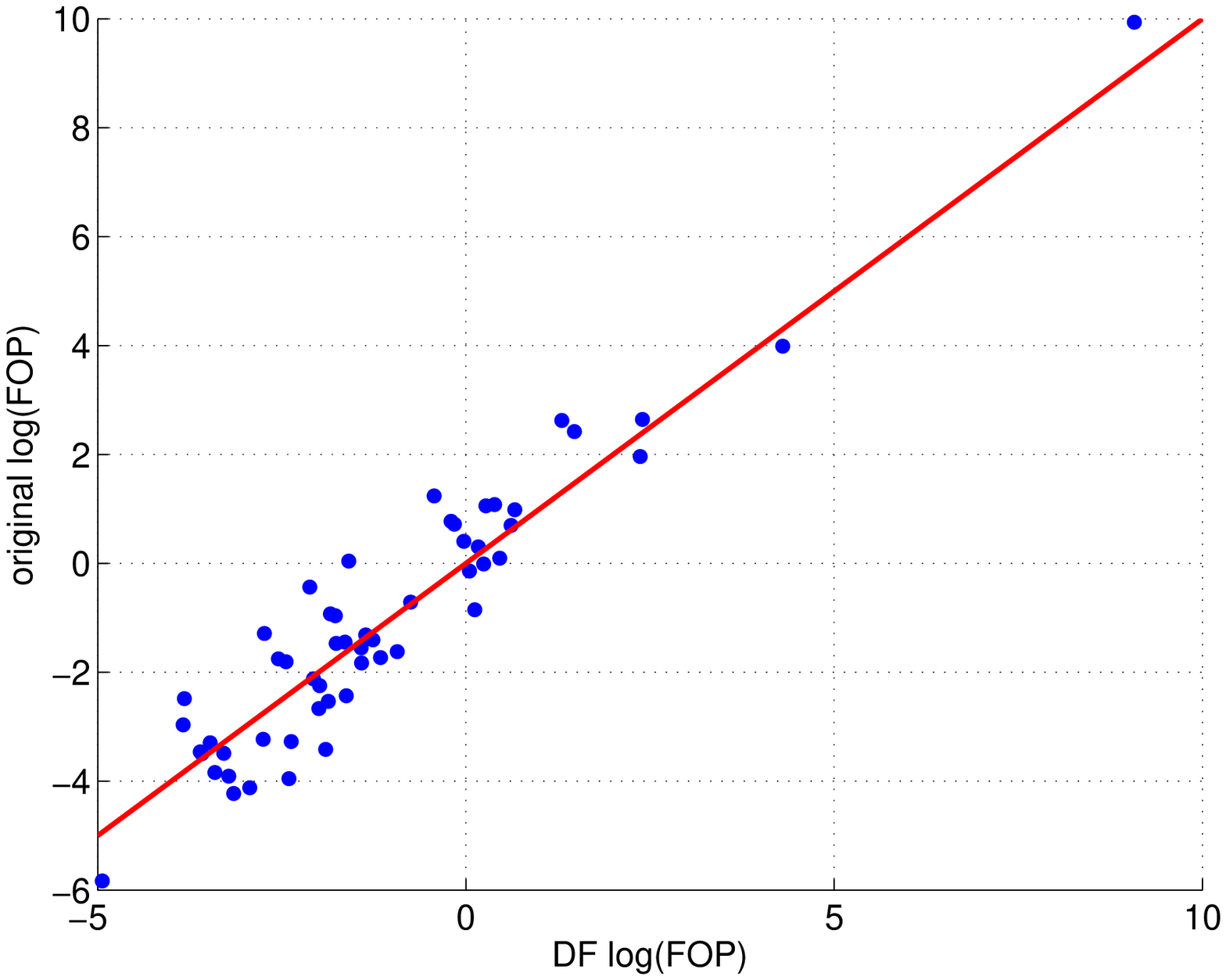}
\caption[These scatterplots illustrate the precision in data statistics at smaller scale based on interpolation of two known statistics at larger scale in logarithmic form of FOP]{\small These scatterplots illustrate the precision in data statistics at smaller scale based on interpolation of two known statistics at larger scale in logarithmic form of FOP. The first row represents the scatter plot for the 2$\times$2 configuration, second row for 2$\times$3 configuration and last row is 3$\times$3 configuration. The first column represents the relationship of FOP for zero lag of coarse map compared to its associated lag in the original map. The second column represents the relationship of FOP for lag zero of nearest neighbor map compared to its associated lag in the original map. And finally, the third column represents the relationship of FOP for lag zero of distance function kriging map compared to its associated lag in the original map. Extrapolation of FOP values to smaller scale for generated high resolution map using nearest neighbor has less accuracy compared to the generated high resolution map using distance function kriging.}
\label{fig:fop1}
\end{figure}

\nin A binary example illustrates the interpolation process to enhance the resolution of the available training image. An original channel depiction of 500$\times$500 shown on the left side of Figure~\ref{fig:orig} is considered as the initial high resolution training image. 100 samples are selected randomly from the original 2-D map to be considered as the only available data to the simulation process. The realizations are generated utilizing stochastic simulation at 20$\times$20 grid nodes and cell sizes of 25 m $\times$ 25 m. This coarse scale realization is considered as the available training image at this point. Different approaches could be taken at this stage to enhance the coarse resolution model to that of the original map. One is to follow geostatistical methods and determine the distance function to be considered in the kriging process and further in the generation of high resolution realizations. The other approaches relate to known interpolation techniques in image processing discussed earlier as shown in Figure~\ref{fig:RLZ1}). The resulting estimation at higher resolution for different techniques could be compared visually and further through the comparison of high-order statistics (i.e. FOP) with that of the original training image (see Figure~\ref{fig:fop1}).


\subsection{Example: Non-binary Training Image}
\nin This example is conducted to investigate complications that arise with more categories. Dealing with more than two categories can make the process complex. For the purpose of this exercise a training image with 5 categories is considered. An X-Y slice of the training image, shown on the right side of Figure~\ref{fig:orig}, is extracted and considered as the training image for this exercise. It can be visually examined that presence of more than two categories in the reservoir results in more abrupt changes from one category to another which makes it harder for the features to be captured.

\begin{figure}[t!]\centering
\includegraphics[scale=0.4]{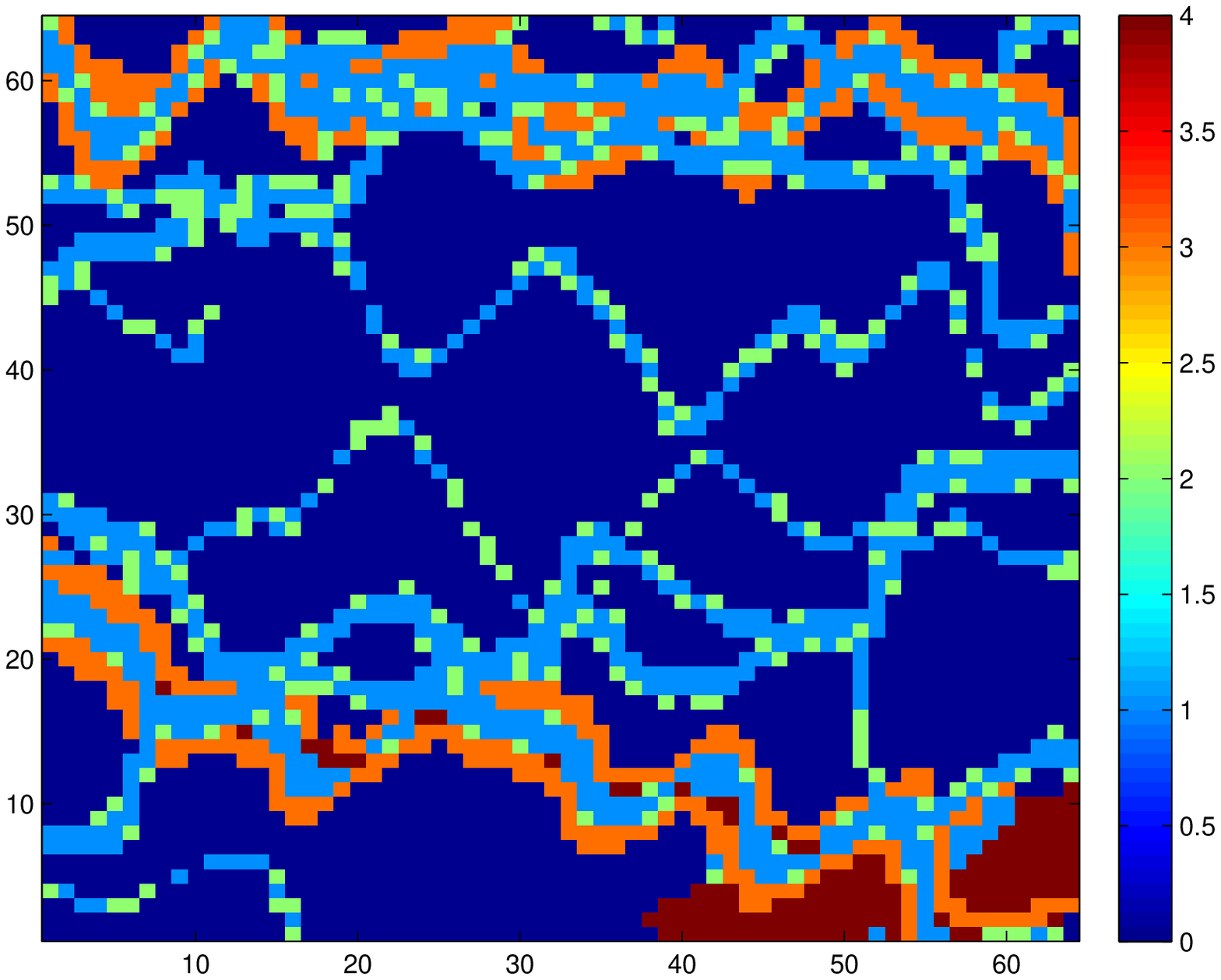}
\includegraphics[scale=0.4]{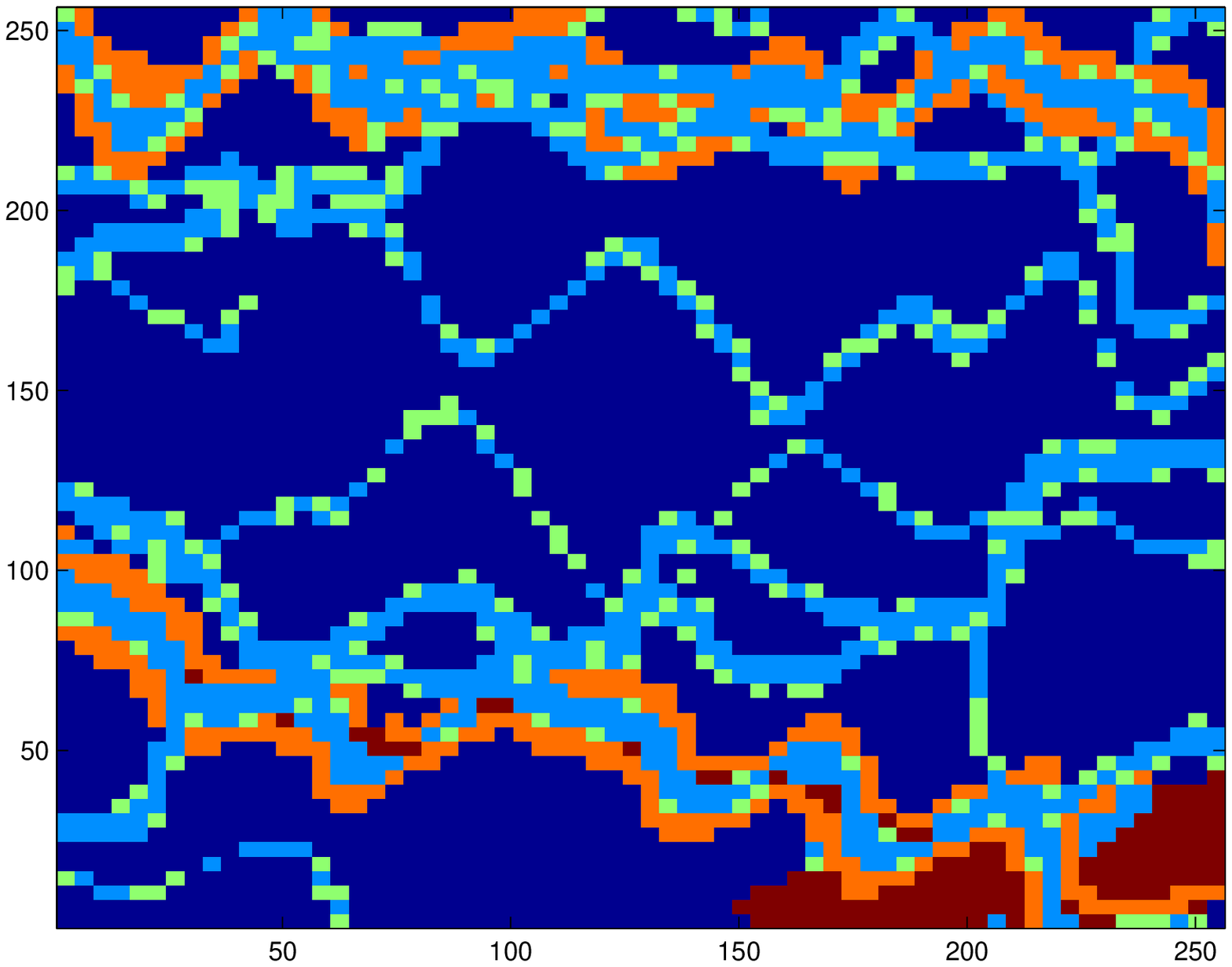}\\
\includegraphics[scale=0.4]{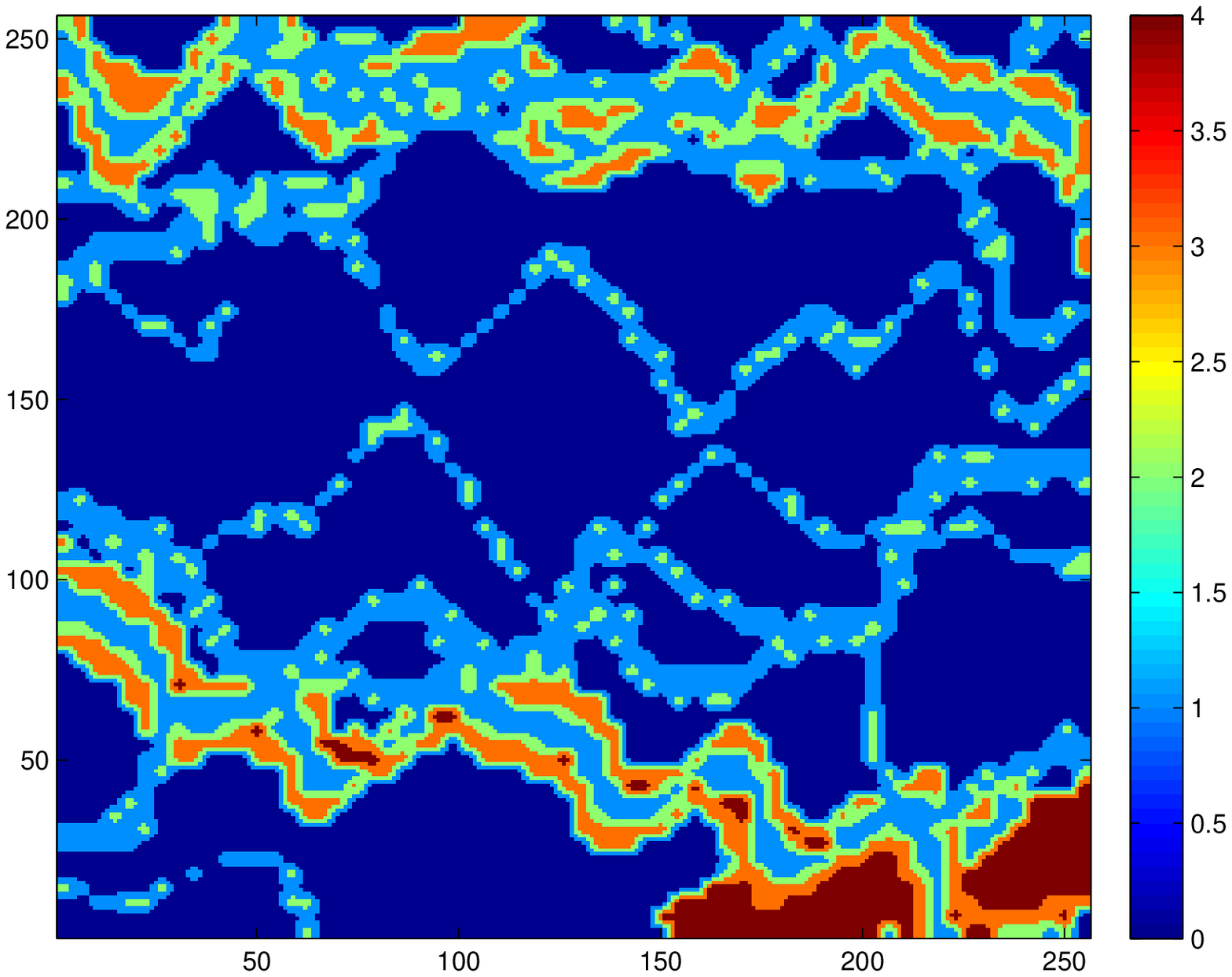}
\includegraphics[scale=0.4]{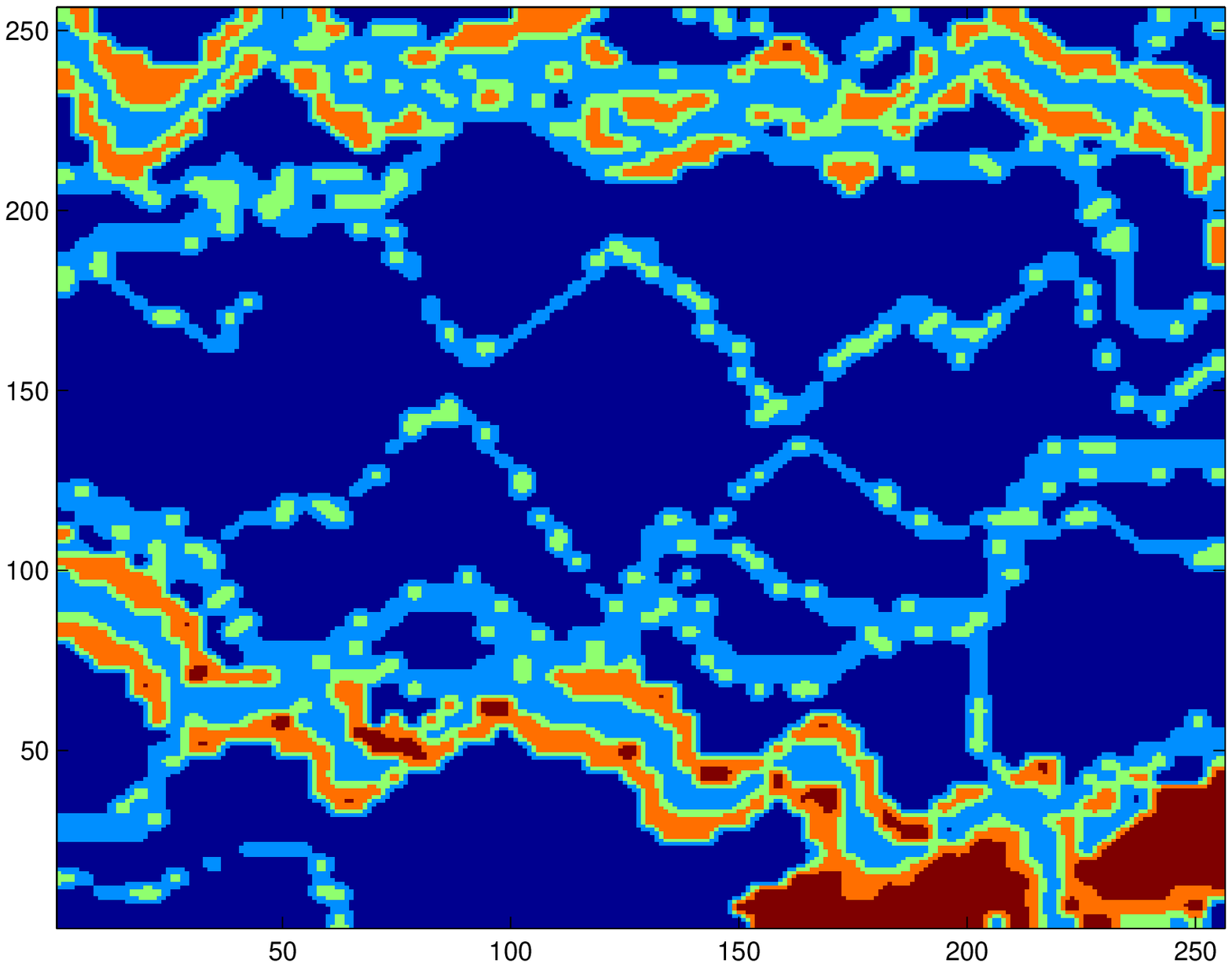}\\
\includegraphics[scale=0.4]{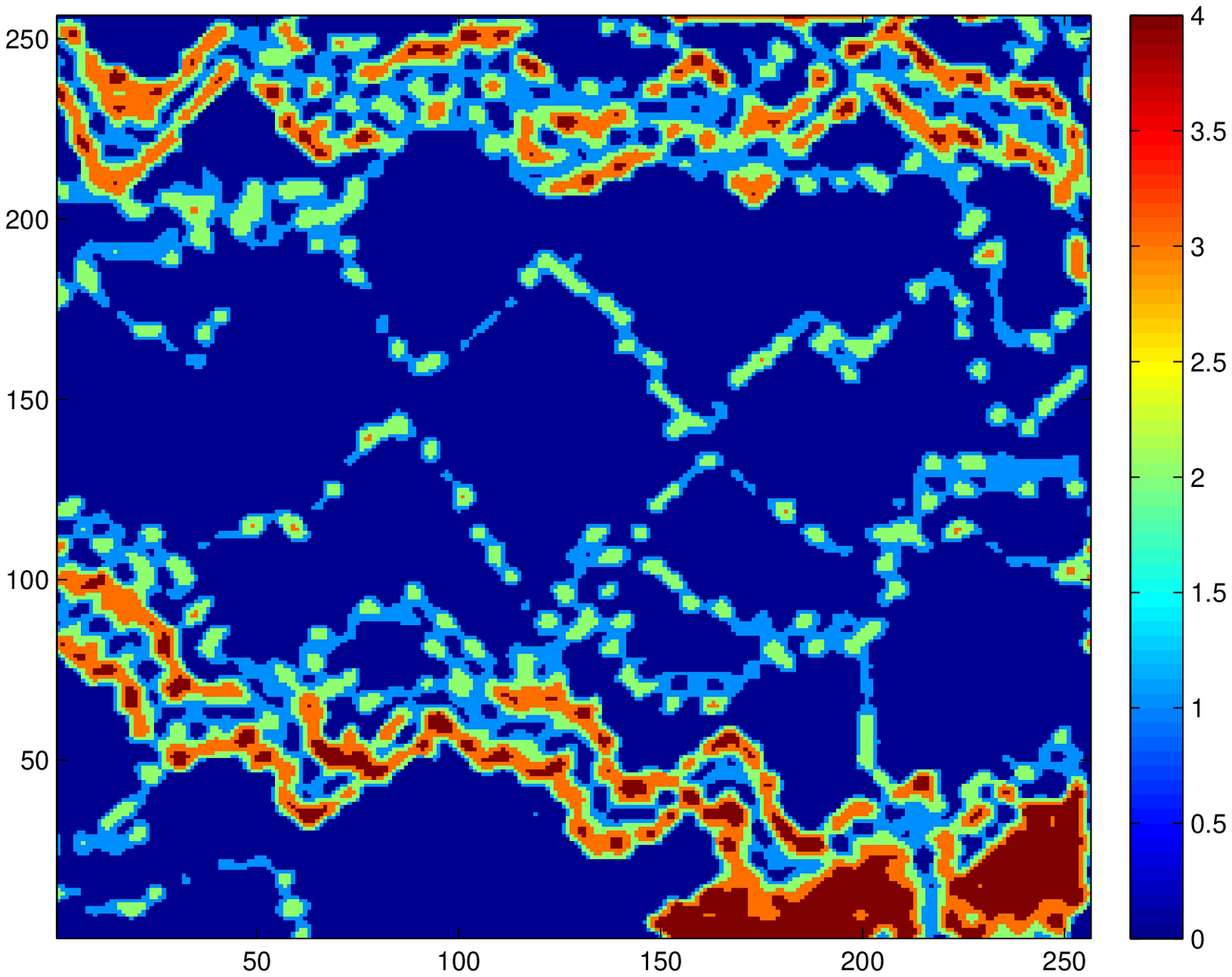}
\includegraphics[scale=0.4]{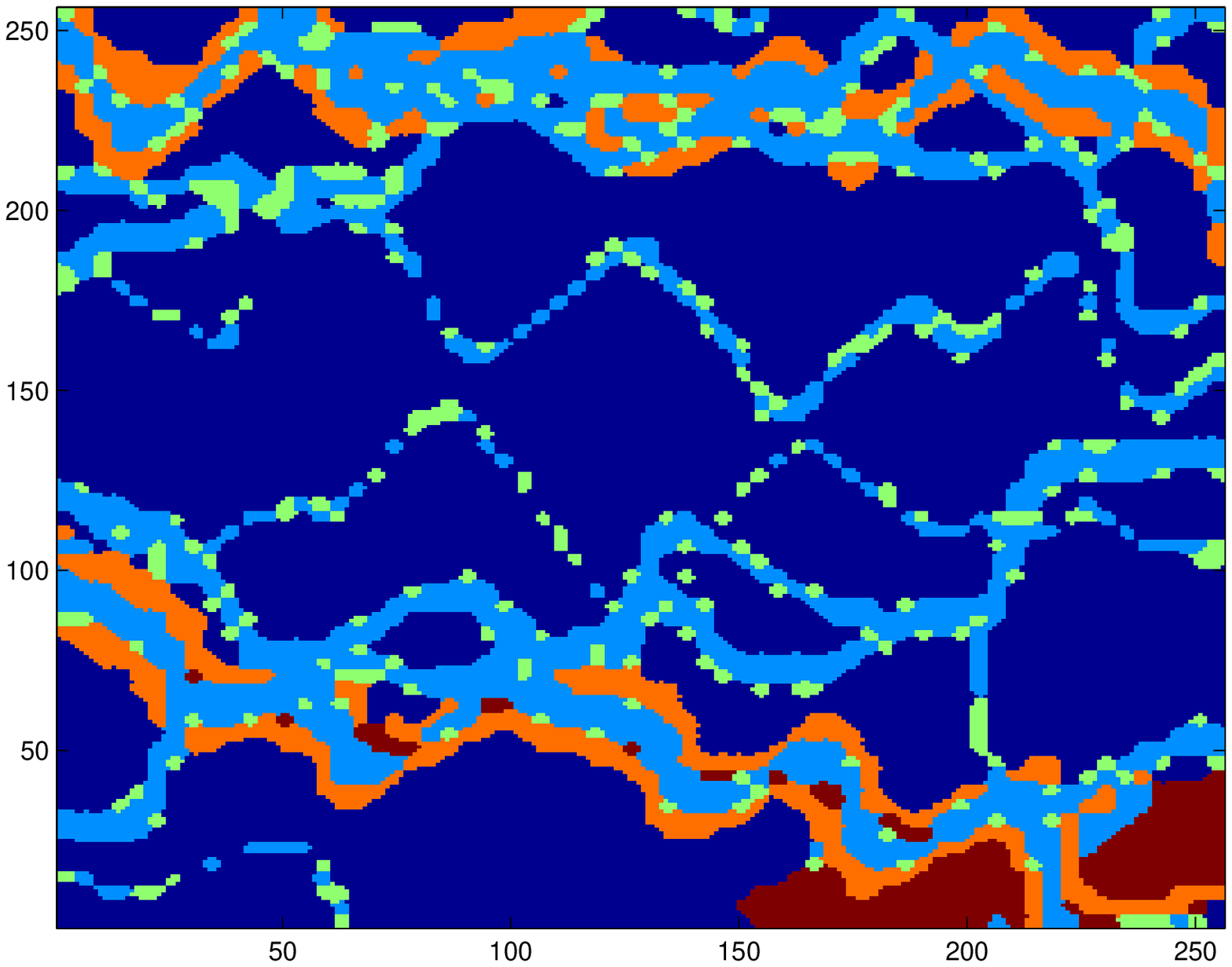}\\
\caption[Resolution enhancement is applied through interpolation techniques on the Crevasse training image]{\small Resolution enhancement is applied through interpolation techniques on the Crevasse training image. The plots represent the coarsen map of, nearest neighbor interpolation, bilinear interpolation, bicubic interpolation, sinc function, and distance function kriging from left to right and top to bottom. For three cases of bilinear, bicubic and sinc function, the transition between categories should happen slowly. The green lines in between two categories of orange and dark blue is evident to such an observation. This is not the case for nearest neighbor interpolation and distance function kriging. Yet, nearest neighbor is largely undesired because of generating blocky artifacts.}
\label{fig:5cat}
\end{figure}

The training image in this example has 256$\times$256 grid cells with size of 8 m $\times$ 8 m. The map is coarsen by extracting every 4 grid cells in both X and Y directions. Notice that, conventional simulation of the sample data in this case does not result in realistic modeling. 
The interpolation methods are applied to a coarser map of 64$\times$64 with cell size of 32 m $\times$ 32 m.

\subsection{Discussion}
\nin The resulting high resolution maps are shown for nearest neighbor, bilinear, bicubic, sinc and distance function kriging in Figure~\ref{fig:5cat}. Most interpolation algorithms lead to similar results for the binary image. Whereas, in the second example, the results are more different. The visual appearance, and absence of artifacts are important criteria to select a technique. One other criterion is robustness; a preferred algorithm should be widely applicable to different images. As can be observed from Figure~\ref{fig:5cat}, the spline like interpolations are blurry, meaning that transitions happen slowly. For example, the green lines in between the dark blue and orange at the bottom of the map is some sort of the ease for transition between the facies. This is not the case for the nearest neighbor and distance function kriging. Nearest neighbor, however, leads to blocky edges. Among these techniques, distance function kriging interpolation seems to be a good candidate because of its simplicity, stability and visually appealing results. Another advantage of this technique is its flexibility. For instance, one can specify the variogram for interpolation.

\begin{figure}[t!]\centering
\includegraphics[scale=0.37]{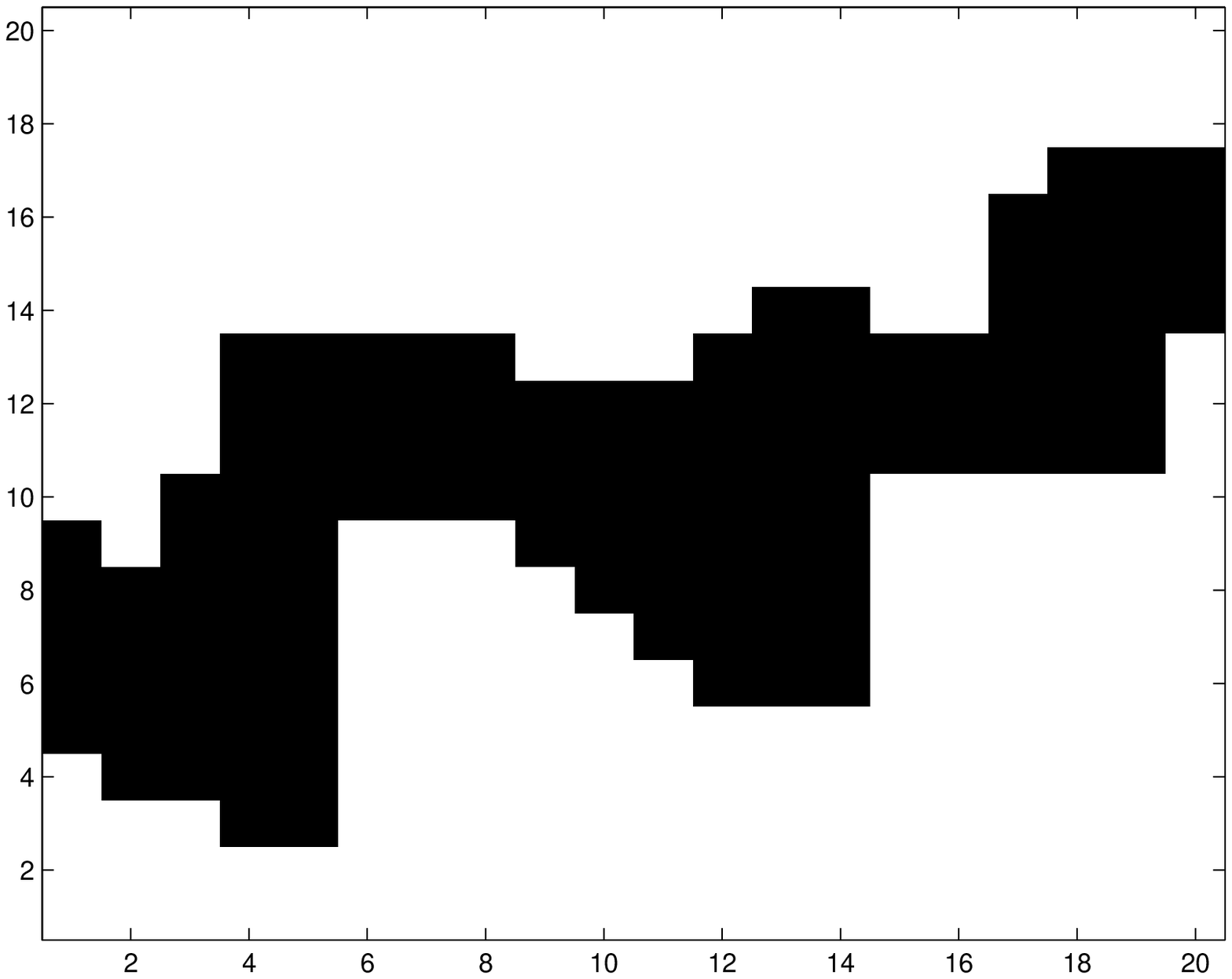}
\includegraphics[scale=0.37]{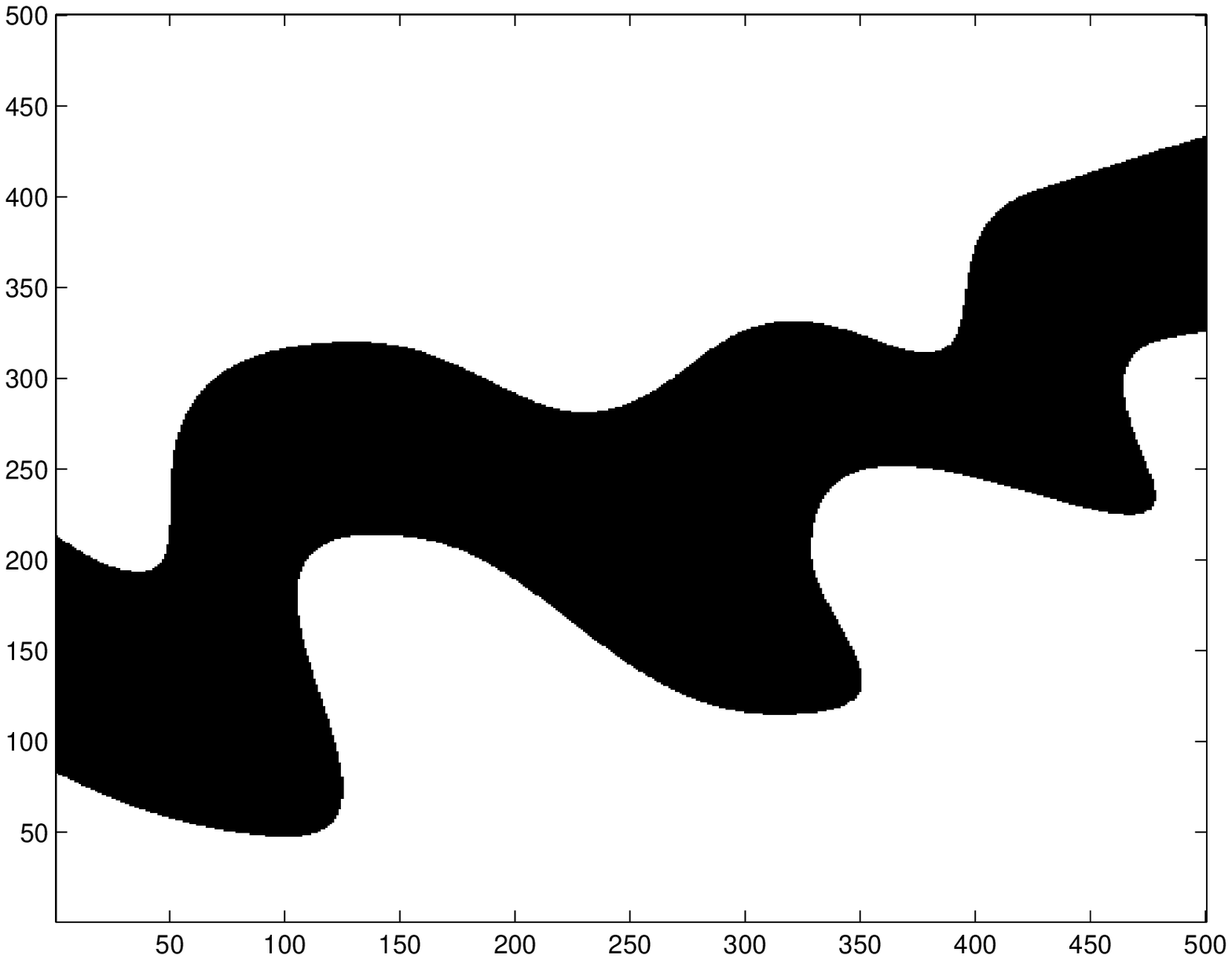}
\includegraphics[scale=0.37]{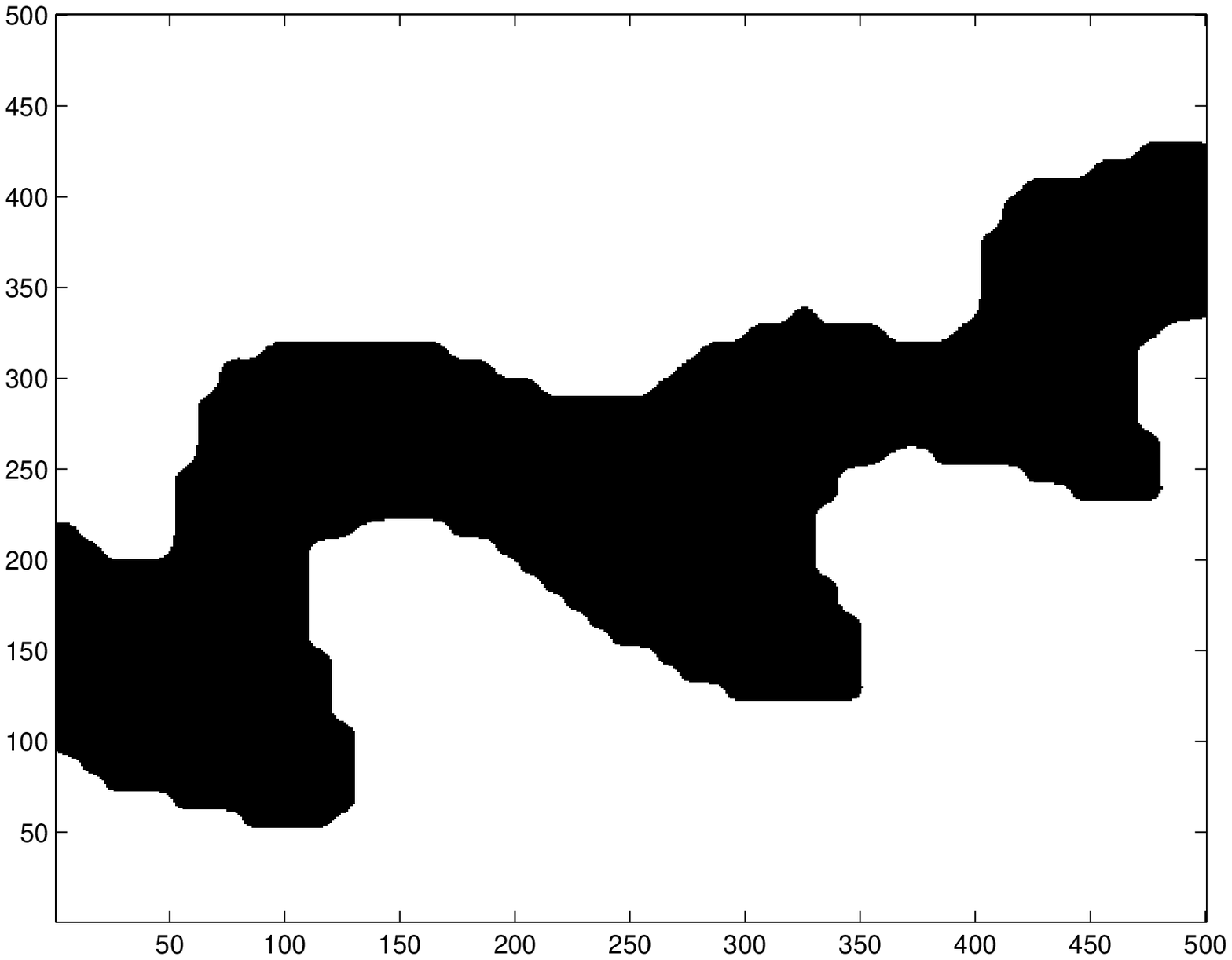}
\includegraphics[scale=0.37]{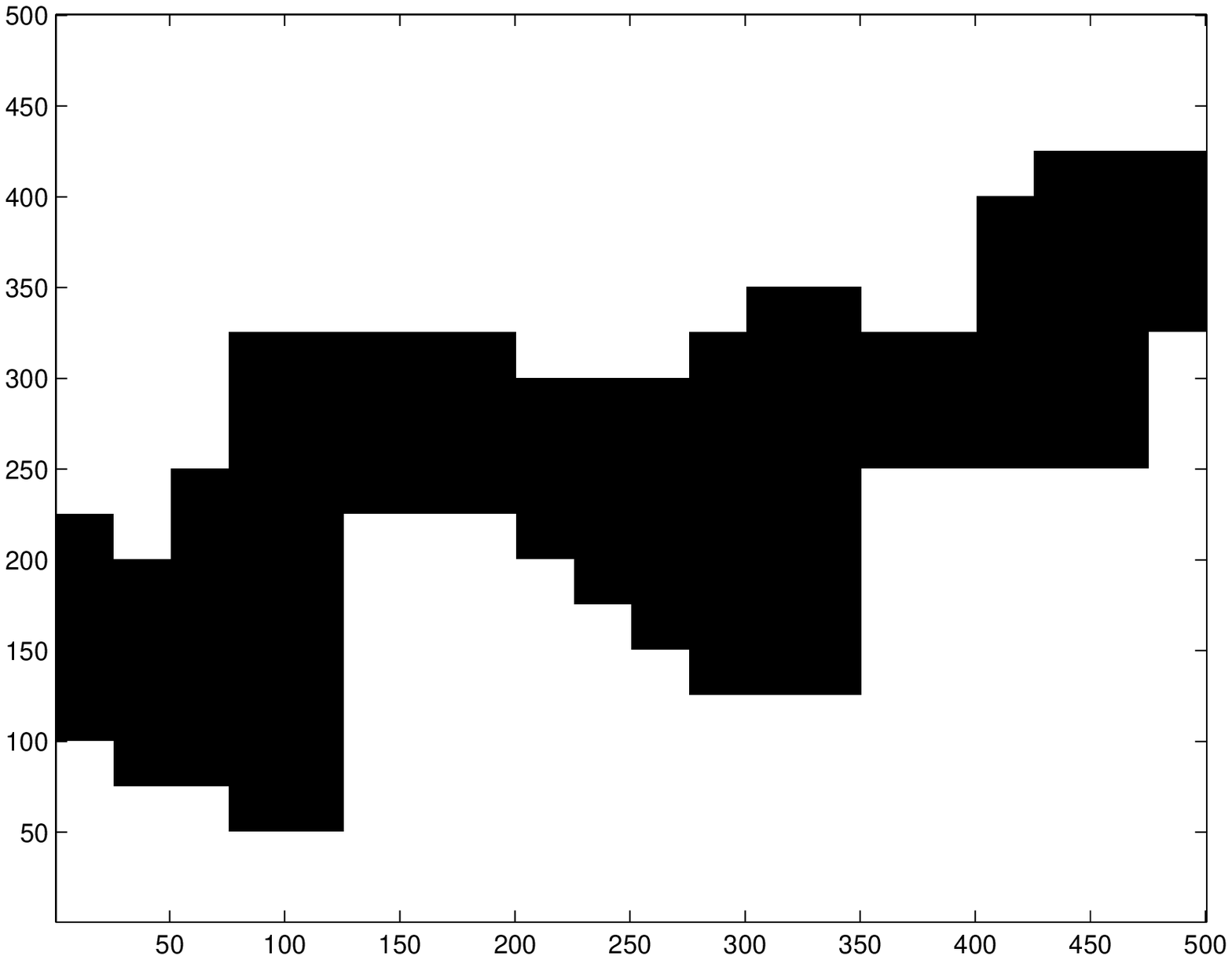}
\caption[The coarse resolution map at the top left represents the training image which is considered to be accessible to this example]{\small The coarse resolution map at the top left represents the training image which is considered to be accessible to this example. The resolution of training image is enhanced to the resolution of interest shown in maps (1) original (right top), (2) using DF kriging interpolation (bottom left), and (3) using nearest neighbor interpolation (bottom right).}
\label{fig:3case}
\end{figure}

\begin{figure}[t!]\centering
\includegraphics[scale=0.37]{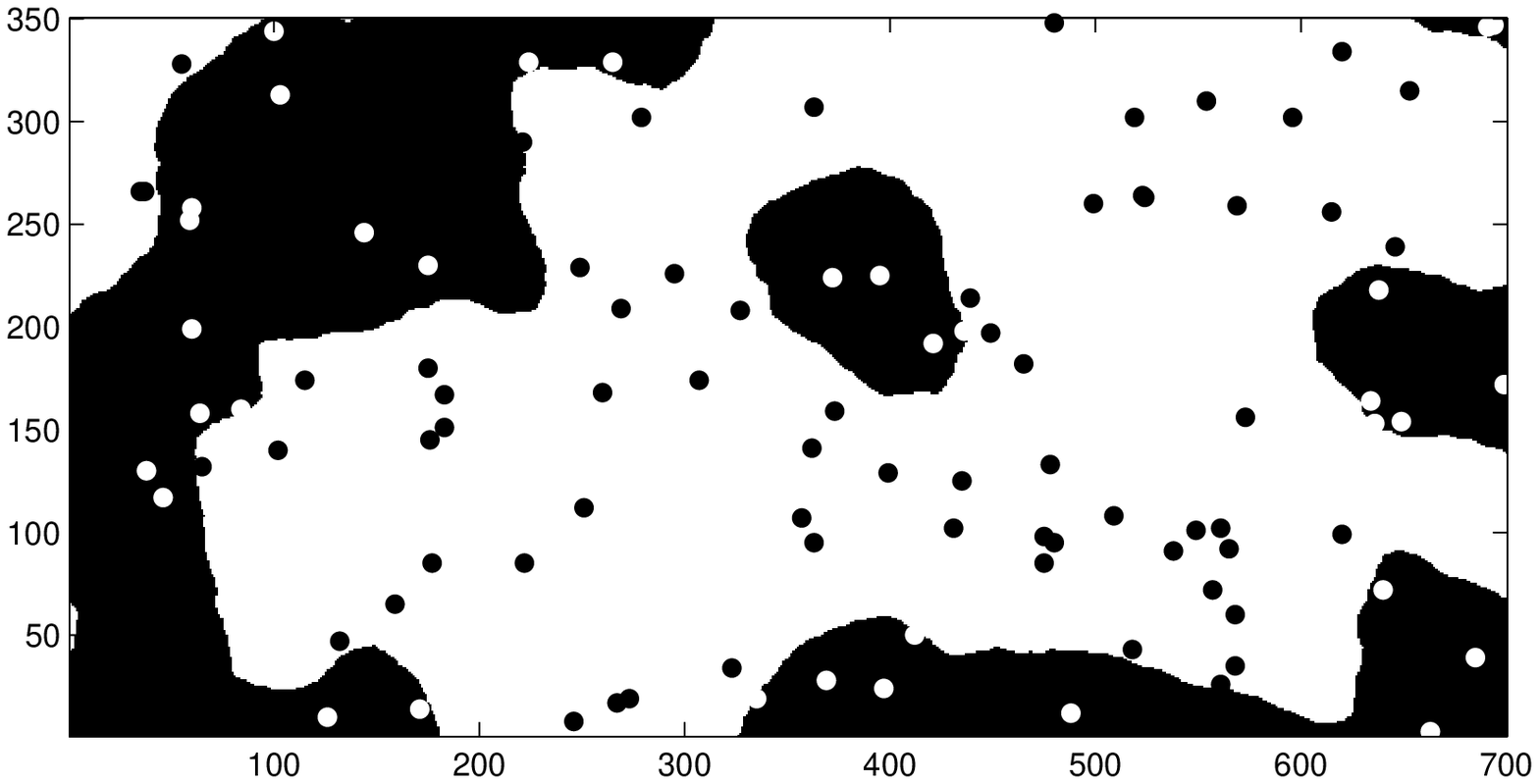}
\includegraphics[scale=0.37]{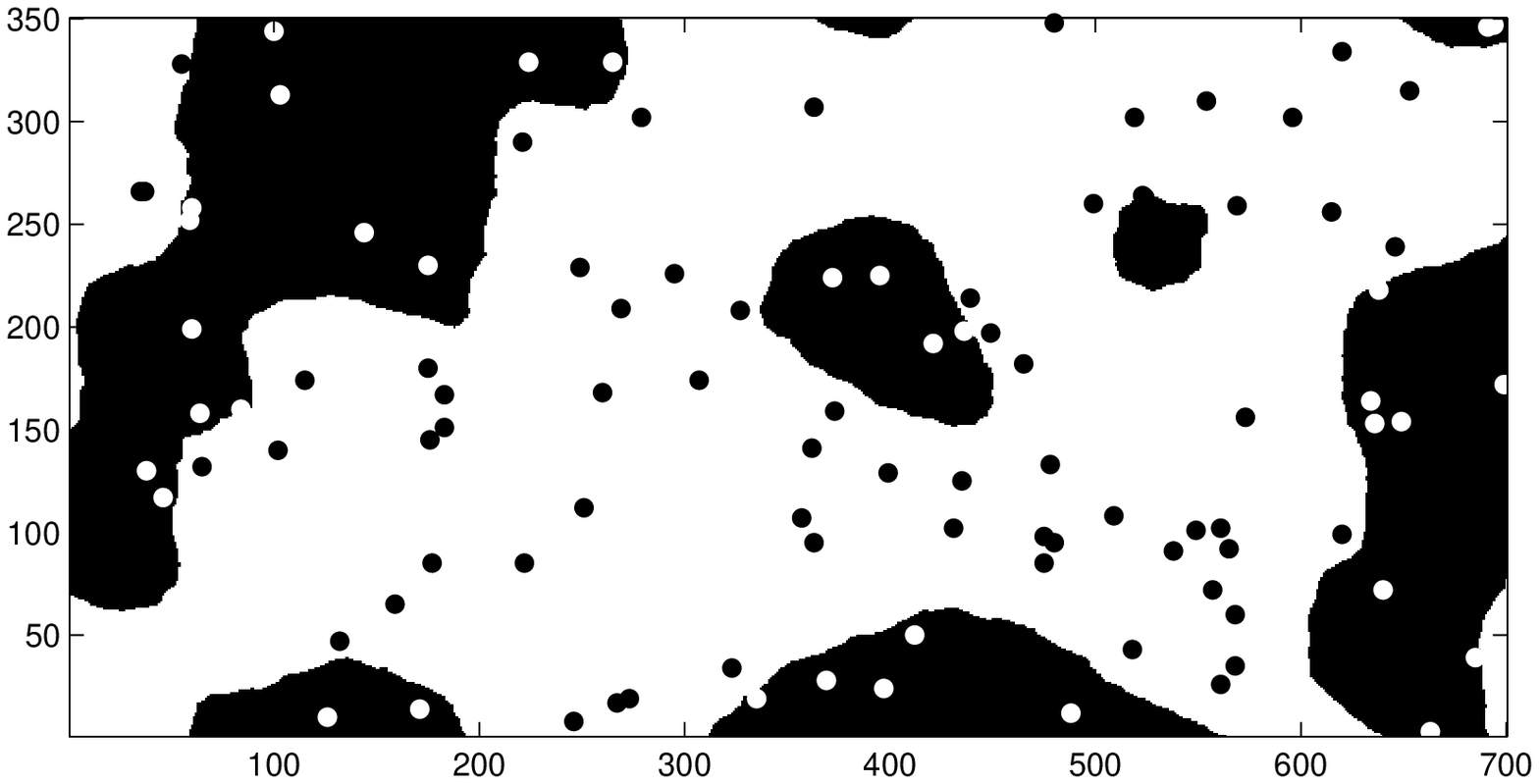}
\includegraphics[scale=0.37]{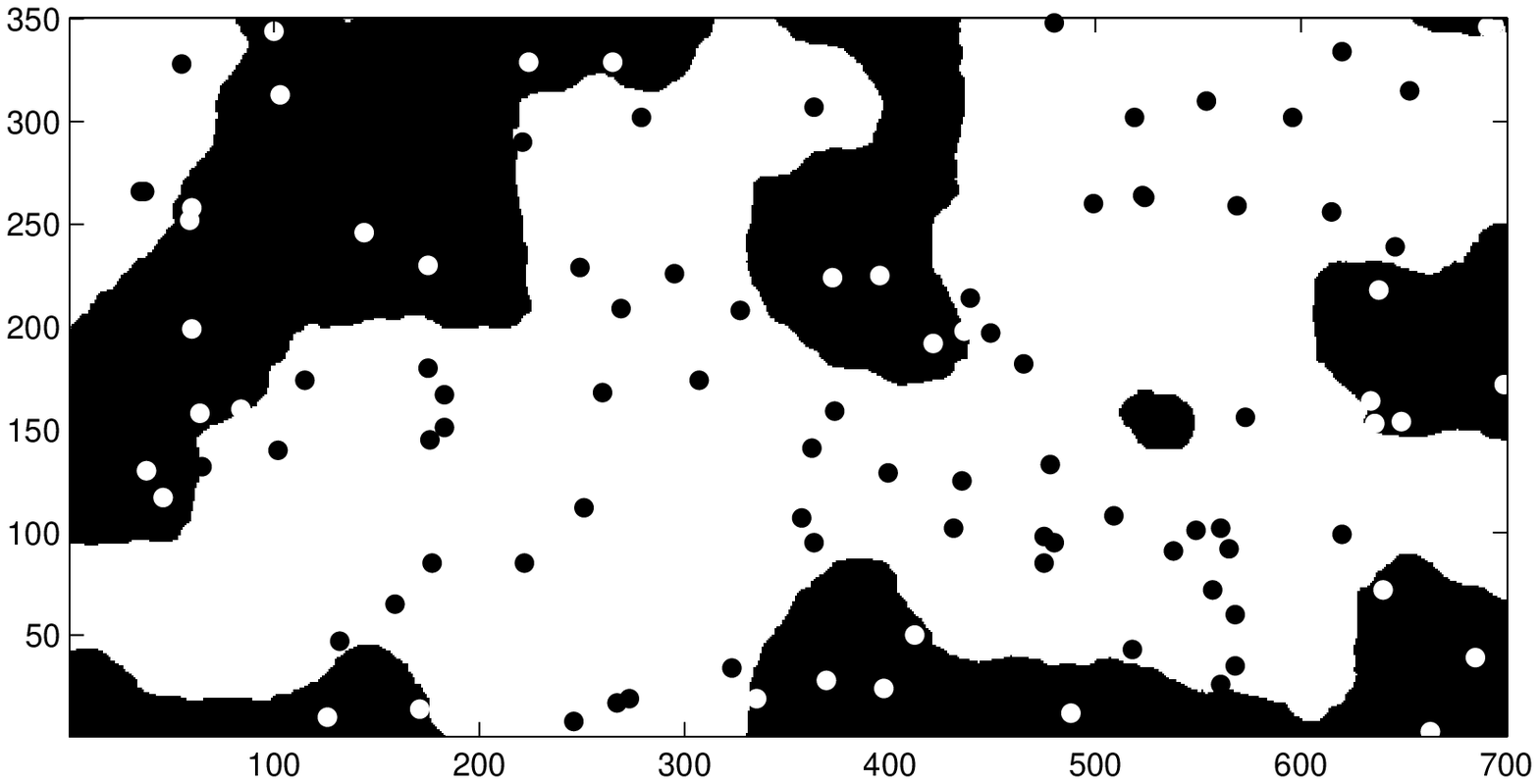}
\includegraphics[scale=0.37]{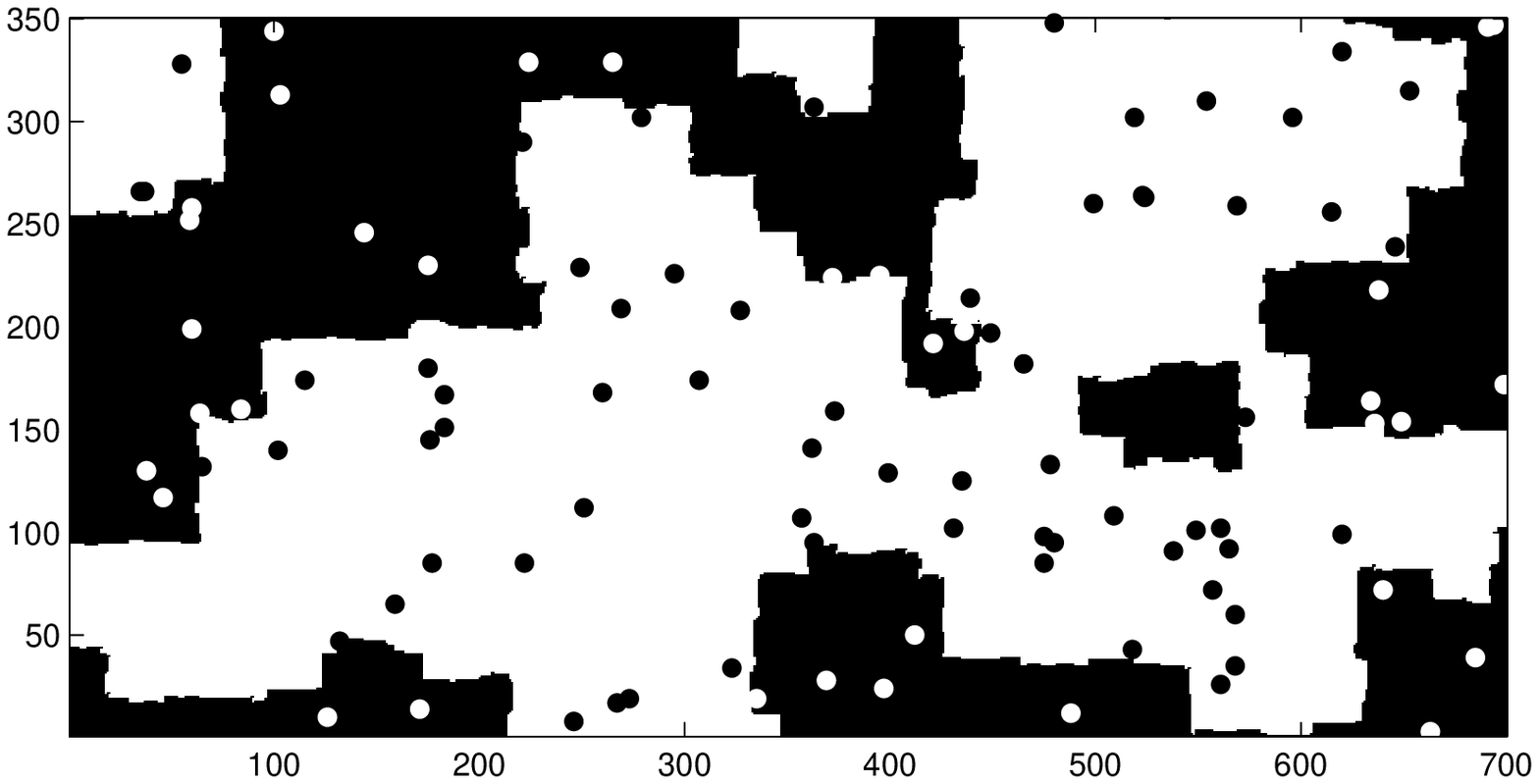}
\caption[The map at the top left is the unconditional realization that has been simulated using MPS and the original training image]{\small The map at the top left is the unconditional realization that has been simulated using MPS and the original training image. The 100 samples are selected randomly (as shown) to condition the further simulation using high resolution training image. The map at the top right is the generated realization using MPS simulation, the conditioning data and the enhanced-resolution training image (original one). The realization at the bottom left is generated similarly but using DF training image, and the one at the bottom right results from the smaller simulation but enhanced-resolution training image of nearest neighbor interpolation.}
\label{fig:locan}
\end{figure}

\begin{figure}[t!]\centering
\includegraphics[scale=0.49]{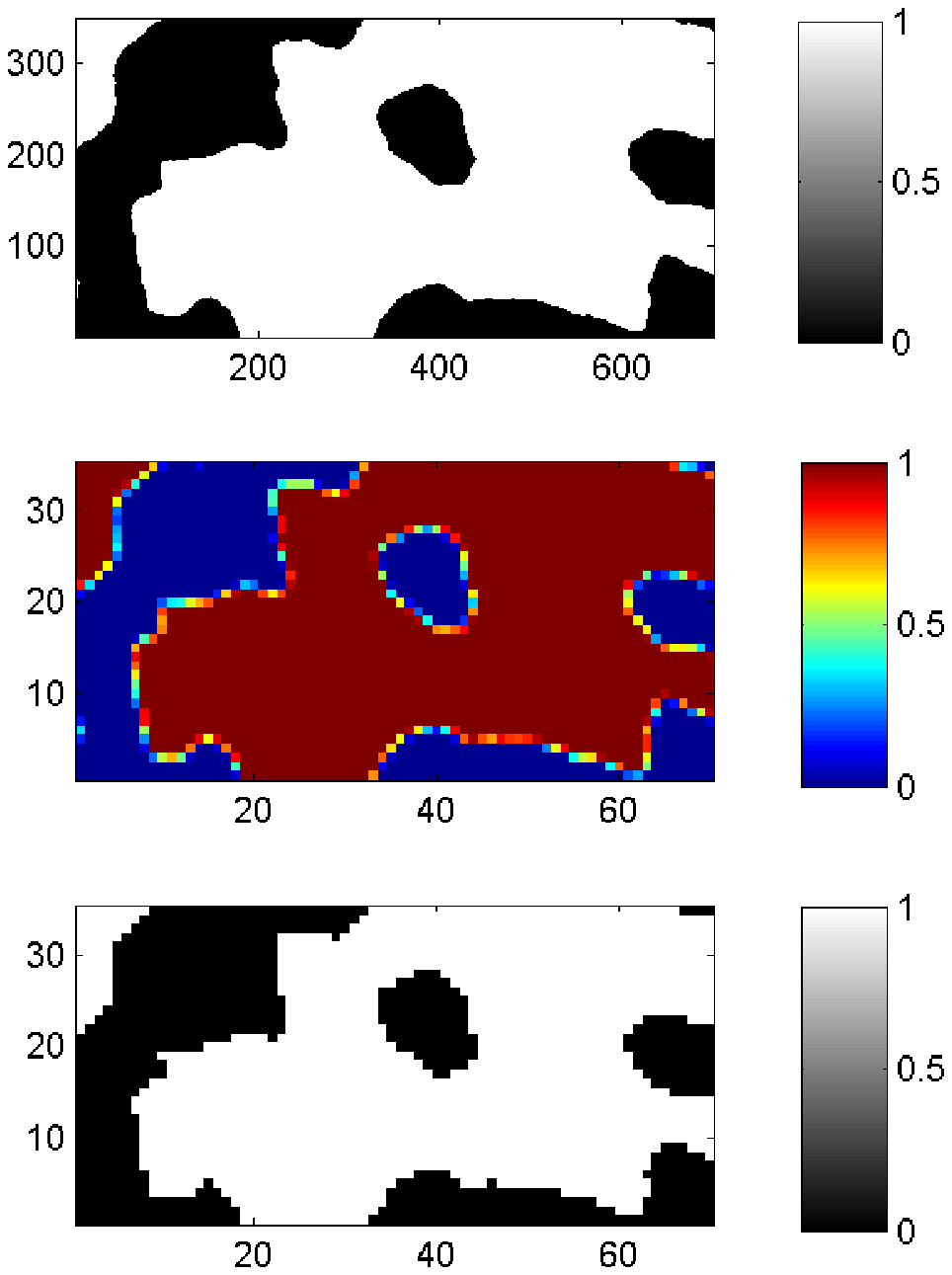}
\includegraphics[scale=0.49]{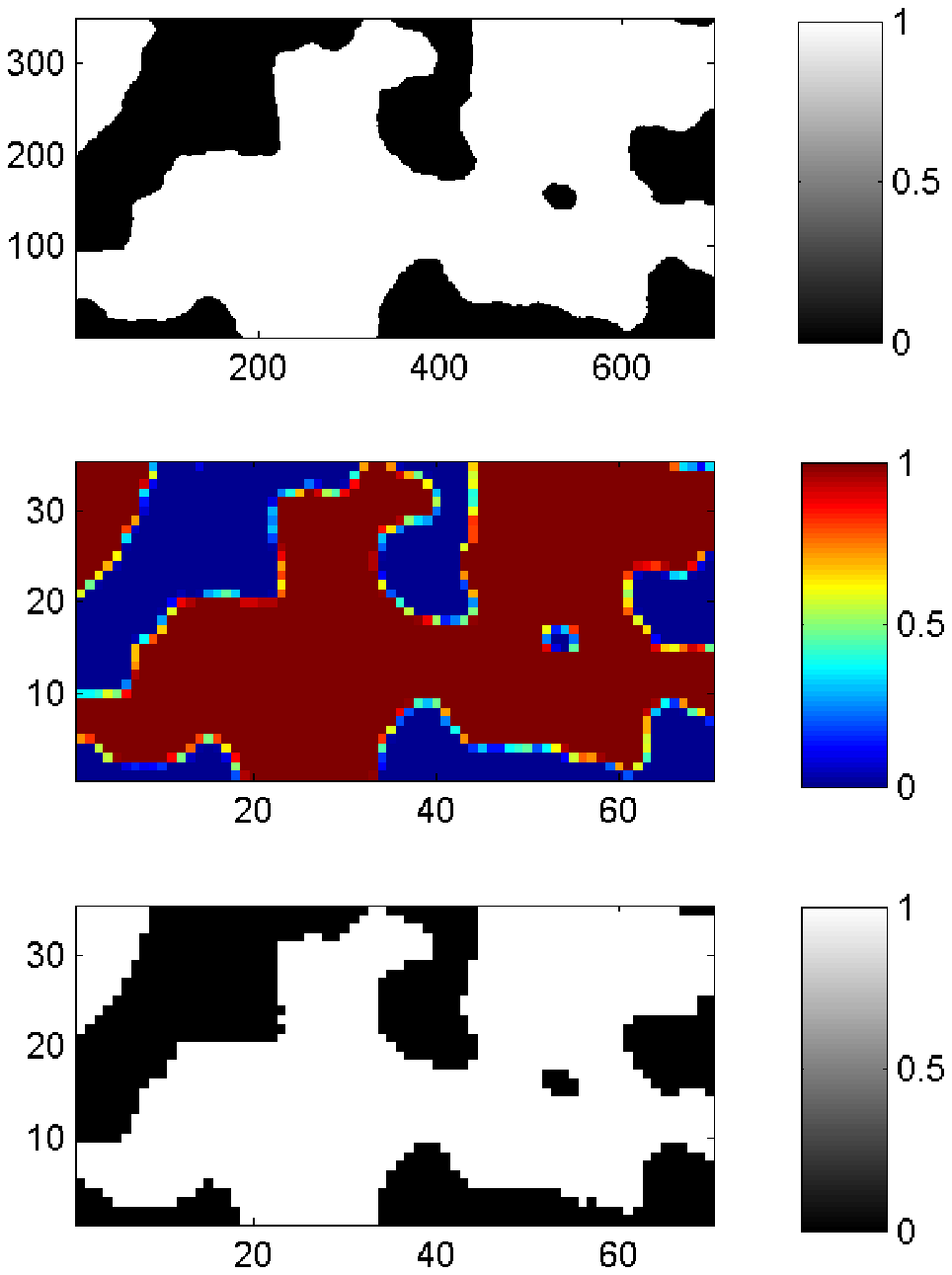}
\includegraphics[scale=0.49]{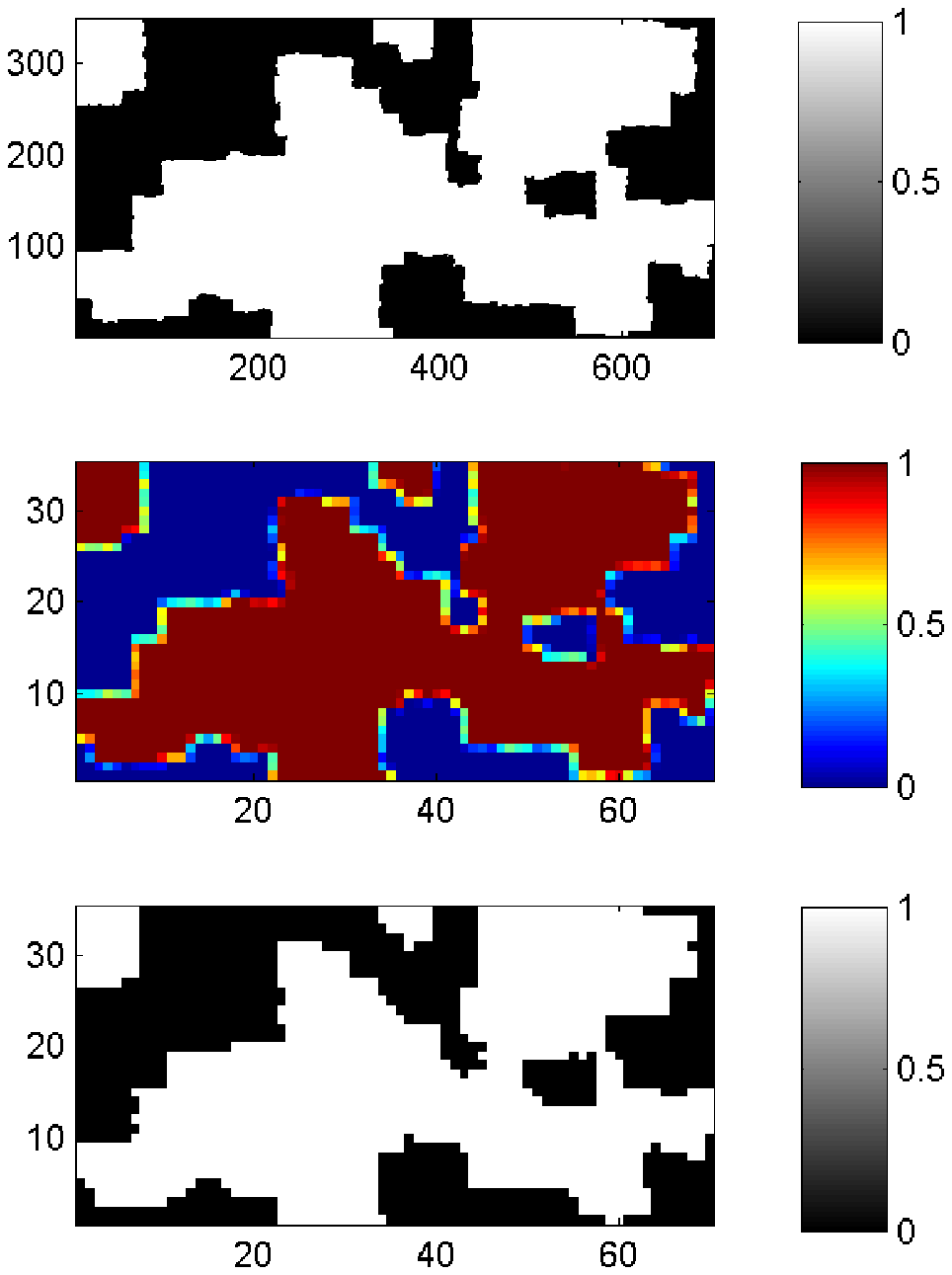}
\caption[The maps at the top represent the high resolution realizations, the ones in the middle are channel proportions in upscaled map and the ones shown in the last row are the final upscaled binary map based on cut off value of 25\%]{\small The maps at the top represent the high resolution realizations, the ones in the middle are ore proportion in upscaled map and the ones shown in the last row are the final upscaled binary map based on cut off value of 25\%. Colored edges indicate the blocks with mixed material. First column refers to upscaling of first scenario, the middle is upscaling of second scenario and the last column refers to third case.}
\label{fig:upscale}
\end{figure}

The constructed high resolution training image is considered to generate high resolution realizations in the following discussion.
\subsection{Simulation with Resolution-Enhanced Training Image}
\nin This subsection conducts an exercise assuming the available training image (top left in Figure~\ref{fig:3case}) does not have the adequate resolution (top right in Figure~\ref{fig:3case}). The coarse resolution training image is constructed by extracting the cells regularly from the channel image at 25$\times$25 cells. As can be observed in Figure~\ref{fig:3case}, the high resolution image is generated using DF kriging (bottom left), and nearest neighbor interpolation (bottom right). Each of these high resolutions images are considered as the training image in three scenarios to construct MPS-based conditional realizations.

To provide conditioning data to this exercise, an unconditional MPS-based realization is generated using SNESIM program \citep{ss:00} with the channel depiction as its training image. Then, 100 samples are regularly extracted from this realization. This is shown in Figure~\ref{fig:locan} at the top right. This unconditional realization consists of 30\% channel. Having 100 samples as conditioning data, SNESIM considers the three high resolution training image of channel, enhance-resolution based on DF kriging, and enhanced-resolution based on nearest neighbor in three scenario of 1 (baseline), 2 and 3, respectively, in order to simulate 100 realizations. One realization of each scenario is shown at the top left, bottom left, and bottom right in Figure~\ref{fig:locan}.

\begin{figure}[t!]\centering
\includegraphics[width=0.65\textwidth,height=0.3\textheight]{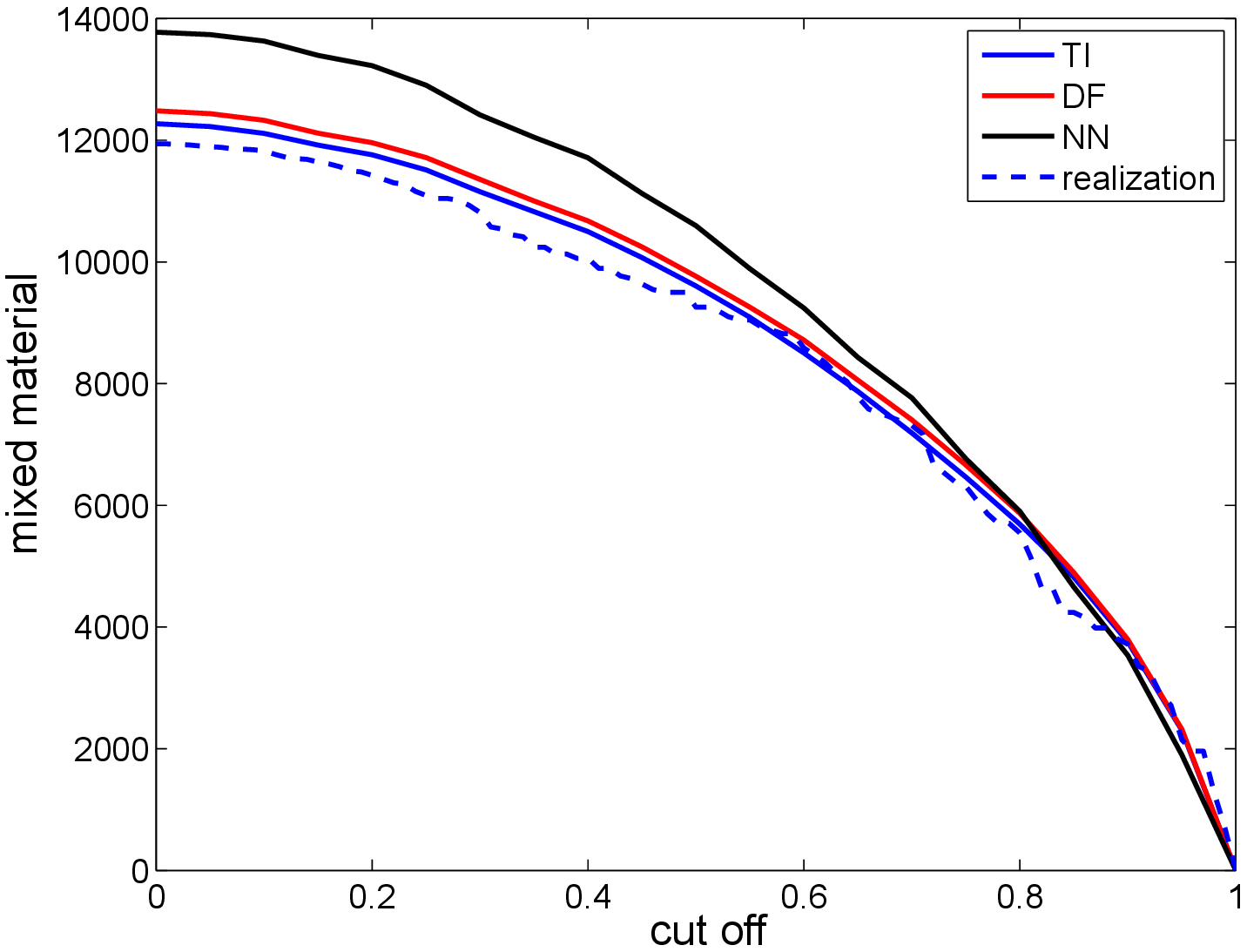}
\includegraphics[width=0.65\textwidth,height=0.3\textheight]{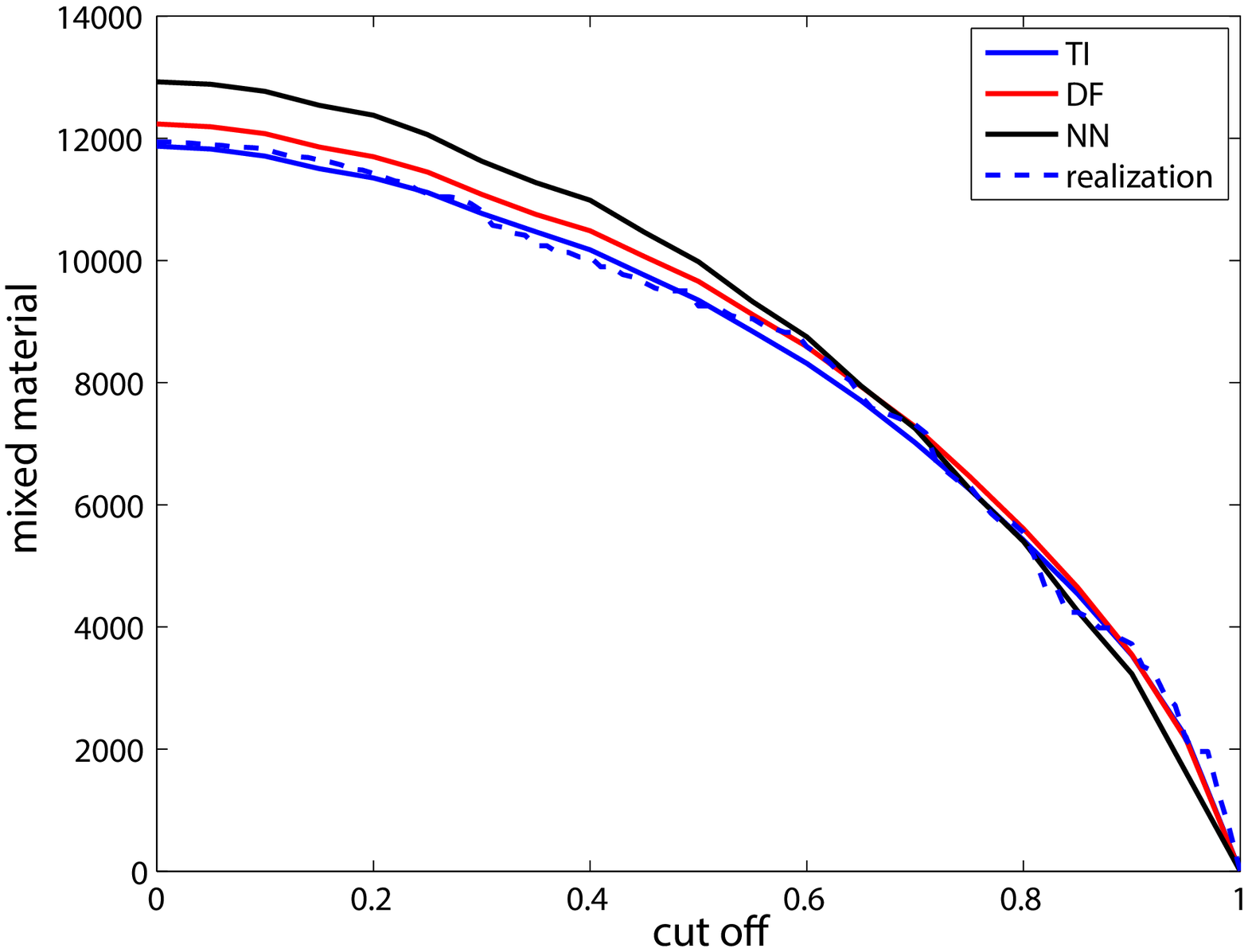}
\caption[Curves for mixed material as a function of cut off values for the three scenarios (upscaled 10$\times$10)]{\small Curves for mixed material as a function of cut off values for the three scenarios (upscaled 10$\times$10). The curves are average value of the amount of mixed material for 100 realizations. The curve at top represents the generated high resolution case conditioned to 100 samples; the one at bottom is generated high resolution case conditioned to 300 samples. With strong conditioning DF scenario performs very closely to original training image and reference case.}
\label{fig:GT}
\end{figure}
\begin{figure}[t!]\centering
\includegraphics[width=0.65\textwidth,height=0.28\textheight]{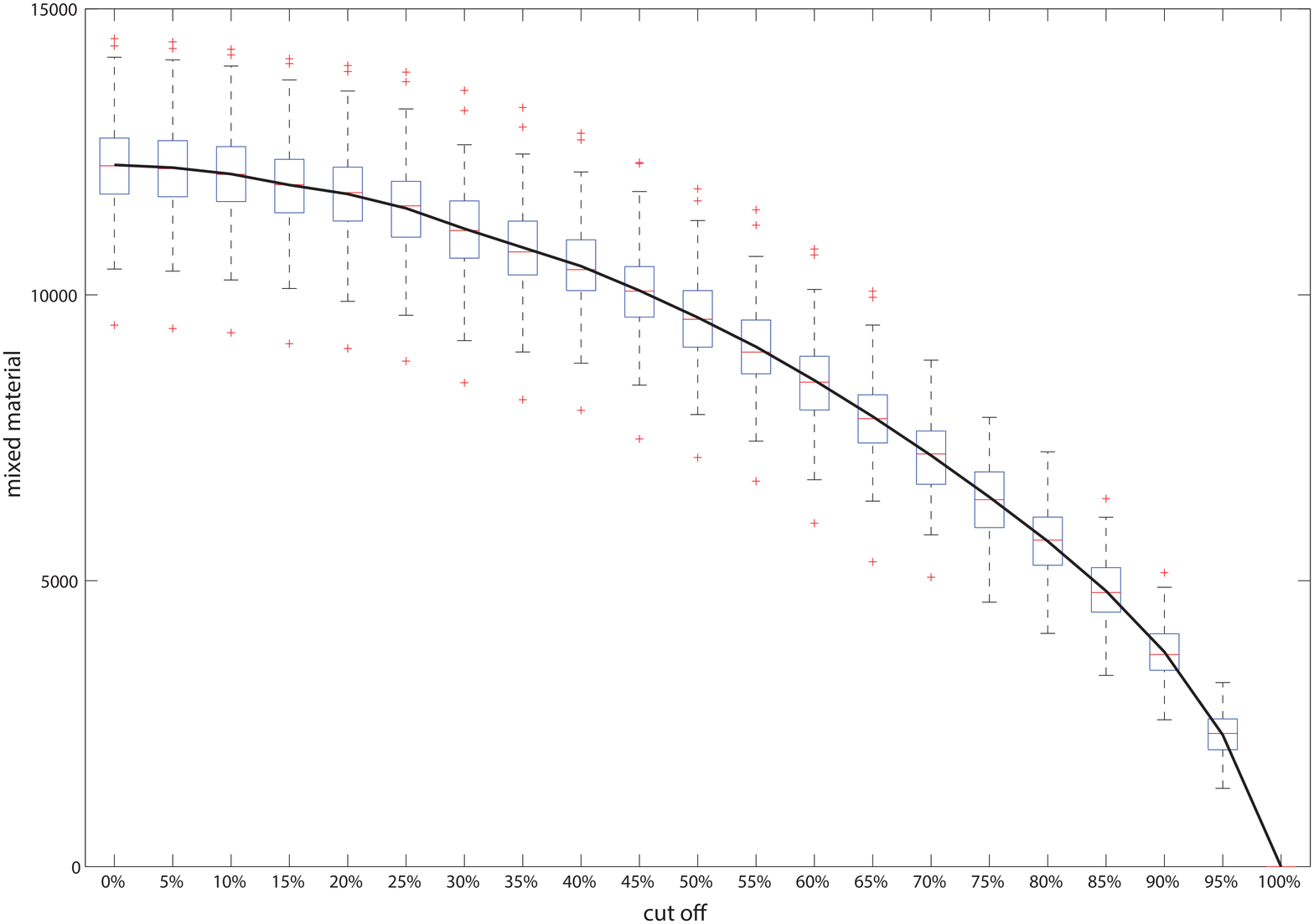}\\
\includegraphics[width=0.65\textwidth,height=0.28\textheight]{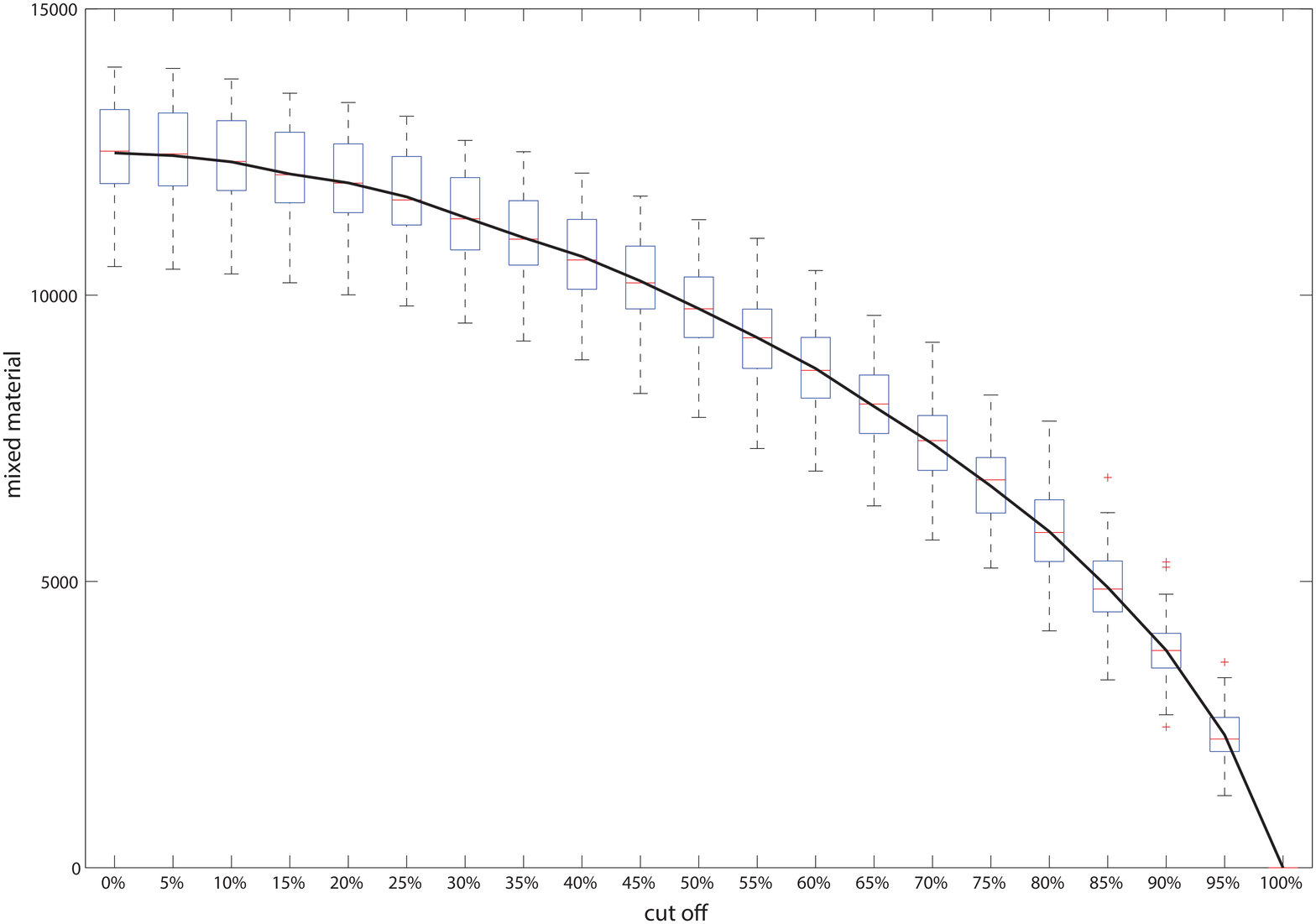}\\
\includegraphics[width=0.65\textwidth,height=0.28\textheight]{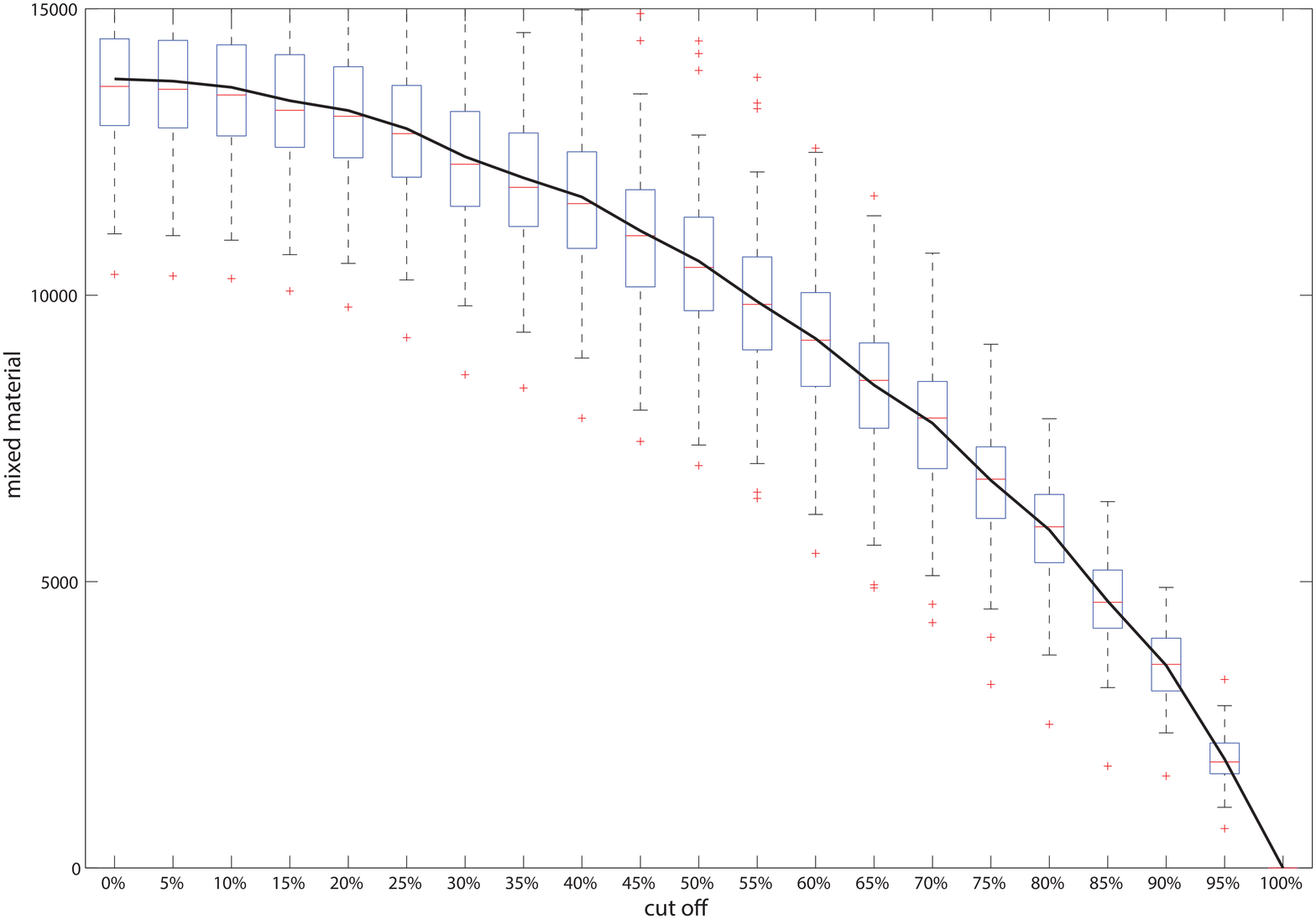}
\caption[Boxplots for amount of mixed material of 100 realizations at every cut off value for all three scenarios]{\small Boxplots for amount of mixed material of 100 realizations at every cut off value for all three scenarios. The one at the top represents the first scenario with original training image, the one in the middle is the second scenario with enhanced-resolution training image generated using distance function kriging, and the last one illustrates the third scenario with enhanced-resolution training image generated using nearest neighbor.}
\label{fig:box}
\end{figure}

The performance of the high resolution realizations is now examined through upscaling. The upscaling cell size of 10$\times$10 is considered as the target block size.
The small scale features in the high resolution realizations are transferred into block features in the upscaled maps, shown in Figure~\ref{fig:upscale}. The block assignments of categories are based on exceeding cut off values. This is mostly due to the exclusive assignment of categories into upscaled blocks. This proportion indicates the amount of high quality reserve present in the block. The performance of three scenarios is compared by measuring the fraction of upscaled blocks that are mixed. Figure~\ref{fig:GT} shows the average amount of mixed material as a function of cut off.

The amount of mixed material for 100 upscaled realizations and the three scenarios are summarized in Figure~\ref{fig:GT}. The plot at the top represents the average amount of mixed material for 100 realizations---conditioned to 100 sample---as a function of cut off. As can be seen, the curves for first and second scenarios perform very closely and also close to a realization sample. The black curve belongs to the realizations generated using enhanced resolution training image with nearest neighbor interpolation. The figure at the bottom demonstrates the curve of mixed material for three scenarios and reference case, conditioned to 300 samples. The strong conditioning takes over in this case and the curves overlap. The third scenario's performance is slightly poor compares to the rest. Figure~\ref{fig:box} shows the boxplots for the three scenarios in terms of variability in amount of mixed material. The boxplots illustrate the variability for every 5\% cut off values. The boxplot for scenario of nearest neighbor interpolation shows relatively larger variance with slightly larger means which indicate the overestimation. This must be due to blocky edges that results in unrealistic transitions. However, in general, there is no significant impact on the dispersion between the three scenarios. The main influence of using different training images is in the mean.

\section{Conclusions}
\nin The smallest scale statistics that can be extracted from a training image is at its native resolution.
This makes the generation of MPS high resolution models challenging since the training image comes at a fixed scale. In an attempt to enhance the model resolution, the frequency of patterns (FOP) for specific configurations are utilized to determine data interactions through high-order statistics. The independence of occurrence of patterns is defined relative to a random map with the same global proportions of categories. The FOP is then re-expressed as the natural logarithm of odds ratio and determined as a function of lag distance for lags changing from 0 to some large value.
The FOP evaluations are then utilized to demonstrate the challenges in generation of high resolution models through extrapolation of such statistics with respect to lag distance. Investigating the behavior of FOP at different scales suggests that prediction at smaller scales is not possible.

At the end, the resolution of a training image is enhanced directly to account for small scale data interactions. To reach data statistics at scales smaller than that of the training image, spatial resolution enhancement interpolation techniques are considered. Different interpolation techniques generate data in between coarse scale grid cells differently. Kriging signed distance function is advocated as a robust approach to construct the training image at finer scale. The paper advocates the generation of high resolution MPS models by directly simulating from the high resolution training image. The examples shown in this paper demonstrate the effectiveness of this technique.

\singlespace
\bibliographystyle{abbrvnat}
\bibliography{MPSbib}
\end{document}